\documentclass [12pt]{article}
\usepackage{amsmath,amssymb,cite}
\usepackage{appendix}
\usepackage{feynmp}
\usepackage[vcentermath,enableskew]{youngtab}
\usepackage{tabularx}
\usepackage[english]{babel}
\usepackage[dvips]{graphicx}
\usepackage{indentfirst}
\usepackage{epsfig}
\usepackage{physics}
\usepackage{pdfpages}
\usepackage{cases}
\usepackage{appendix}
\usepackage{tabularx}
\usepackage[english]{babel}
\usepackage[dvips]{graphicx}
\usepackage{indentfirst}
\usepackage{epsfig}
\usepackage{slashed}
\usepackage{fancyhdr}
\usepackage{cases}
\usepackage{tcolorbox}
\usepackage{amsmath,amssymb,cite}
\usepackage{amsfonts, geometry, hyperref}
\usepackage{booktabs,longtable}
\usepackage{appendix}
\usepackage{feynmp}
\usepackage{float}
\usepackage{subfigure}
\usepackage[vcentermath,enableskew]{youngtab}
\usepackage{tabularx}
\usepackage[english]{babel}
\usepackage{graphicx}
\usepackage{indentfirst}
\usepackage{epsfig}
\usepackage{color}
\usepackage{epstopdf}
\usepackage[]{hyperref}
\usepackage[section]{placeins}
\usepackage[stable]{footmisc}
\usepackage{appendix}
\usepackage{adjustbox}
\usepackage{tabularx}
\usepackage[english]{babel}
\usepackage{graphicx}
\usepackage{indentfirst}
\usepackage{epsfig}
\usepackage{slashed}
\usepackage{fancyhdr}
\usepackage{url}
\usepackage{pgfplots}
\usepackage{amsmath}

\pgfplotsset{compat=1.18}
\numberwithin{equation}{section}

\makeatletter
\patchcmd{\beamer@sectionintoc}
  {\vfill}
  {\vskip\itemsep}
  {}
  {}

\long\def\beamer@section[#1]#2{%
  \beamer@savemode%
  \mode<all>%
  \ifbeamer@inlecture
    \refstepcounter{section}%
    \beamer@ifempty{#2}%
    {\long\def\secname{#1}\long\def\lastsection{#1}}%
    {\global\advance\beamer@tocsectionnumber by 1\relax%
      \long\def\secname{#2}%
      \long\def\lastsection{#1}%
      \addtocontents{toc}{\protect\beamer@sectionintoc{\the\c@section}{#2\hfill\the\c@page}{\the\c@page}{\the\c@part}%
        {\the\beamer@tocsectionnumber}}}%
    {\let\\=\relax\xdef\sectionlink{{Navigation\the\c@page}{\noexpand\secname}}}%
    \beamer@tempcount=\c@page\advance\beamer@tempcount by -1%
    \beamer@ifempty{#1}{}{%
      \addtocontents{nav}{\protect\headcommand{\protect\sectionentry{\the\c@section}{#1}{\the\c@page}{\secname}{\the\c@part}}}%
      \addtocontents{nav}{\protect\headcommand{\protect\beamer@sectionpages{\the\beamer@sectionstartpage}{\the\beamer@tempcount}}}%
      \addtocontents{nav}{\protect\headcommand{\protect\beamer@subsectionpages{\the\beamer@subsectionstartpage}{\the\beamer@tempcount}}}%
    }%
    \beamer@sectionstartpage=\c@page%
    \beamer@subsectionstartpage=\c@page%
    \def\insertsection{\expandafter\hyperlink\sectionlink}%
    \def\insertsubsection{}%
    \def\insertsubsubsection{}%
    \def\insertsectionhead{\hyperlink{Navigation\the\c@page}{#1}}%
    \def\insertsubsectionhead{}%
    \def\insertsubsubsectionhead{}%
    \def\lastsubsection{}%
    \Hy@writebookmark{\the\c@section}{\secname}{Outline\the\c@part.\the\c@section}{2}{toc}%
    \hyper@anchorstart{Outline\the\c@part.\the\c@section}\hyper@anchorend%
    \beamer@ifempty{#2}{\beamer@atbeginsections}{\beamer@atbeginsection}%
  \fi%
  \beamer@resumemode}%

\def\beamer@subsection[#1]#2{%
  \beamer@savemode%
  \mode<all>%
  \ifbeamer@inlecture%
    \refstepcounter{subsection}%
    \beamer@ifempty{#2}{\long\def\subsecname{#1}\long\def\lastsubsection{#1}}
    {%
      \long\def\subsecname{#2}%
      \long\def\lastsubsection{#1}%
      \addtocontents{toc}{\protect\beamer@subsectionintoc{\the\c@section}{\the\c@subsection}{#2\hfill\the\c@page}{\the\c@page}{\the\c@part}{\the\beamer@tocsectionnumber}}%
    }%
    \beamer@tempcount=\c@page\advance\beamer@tempcount by -1%
    \addtocontents{nav}{%
      \protect\headcommand{\protect\beamer@subsectionentry{\the\c@part}{\the\c@section}{\the\c@subsection}{\the\c@page}{\lastsubsection}}%
      \protect\headcommand{\protect\beamer@subsectionpages{\the\beamer@subsectionstartpage}{\the\beamer@tempcount}}%
    }%
    \beamer@subsectionstartpage=\c@page%
    \edef\subsectionlink{{Navigation\the\c@page}{\noexpand\subsecname}}%
    \def\insertsubsection{\expandafter\hyperlink\subsectionlink}%
    \def\insertsubsubsection{}%
    \def\insertsubsectionhead{\hyperlink{Navigation\the\c@page}{#1}}%
    \def\insertsubsubsectionhead{}%
    \Hy@writebookmark{\the\c@subsection}{#2}{Outline\the\c@part.\the\c@section.\the\c@subsection.\the\c@page}{3}{toc}%
    \hyper@anchorstart{Outline\the\c@part.\the\c@section.\the\c@subsection.\the\c@page}\hyper@anchorend%
    \beamer@ifempty{#2}{\beamer@atbeginsubsections}{\beamer@atbeginsubsection}%
  \fi%
  \beamer@resumemode}

\makeatother

\begin{document}

\title{Endpoint Entropy and Emergent Grassmannian Geometry in BFSS/BMN Systems\\
\large\emph{BFSS/BMN Matrix Quantum Mechanics VI}
}

\author{
Badis Ydri\\[2mm]
Department of Physics, Badji Mokhtar Annaba University, Algeria\\
}

\maketitle

\begin{abstract}
  We develop a Gram--Wishart--Stiefel formulation of the \(N=2\),
large--\(d\) BFSS/BMN endpoint theory. After absorbing the large--\(R\) aligned
contribution into a shifted Gaussian kernel, the endpoint model separates into
radial Wishart eigenvalues, relative Stiefel angles, a nonperturbative Bessel
sector, and a transverse Harish--Chandra--Itzykson--Zuber (HCIZ) interaction.
The Bessel kernel both shifts the endpoint mass and supplies an inverse
Wishart--Stiefel prefactor, removing the spurious doubled entropy. In the
stable rank--two regime this gives the symmetric--Wishart saddle and the
universal continuum \(D\)-channel coefficient
\({\cal C}_{2,D}^{\rm W}=-2(d-2)\).

We then analyze the lower-dimensional models directly. In BFSS\(_4\)
(\(d=3\)), the \(O(3)\) angular integral reduces to a \(c\)-resolved HCIZ
problem whose saddle selects a boundary branch; the Bessel dressing gives
\({\cal C}_{2,D}^{\rm bdry,+}=-1\). In BFSS\(_3\) (\(d=2\)), the relative
Grassmannian collapses but the singular Stiefel/Wishart measure produces a
boundary layer, with \({\cal C}_{2,D}^{\rm bdry}=-2\). In BFSS\(_2\)
(\(d=1\)), the rank--two endpoint structure disappears and the Bessel
convolution leaves a purely radial threshold logarithm, again giving
\({\cal C}_{2,D}^{\rm rad}=-2\).

The geometric interpretation is that the rank--two endpoint construction
produces a relative Grassmannian sector \(Gr(2,d)\). Its two principal angles
are only the invariant singular-value coordinates; the full
\(2(d-2)\)-dimensional Grassmannian counts the relative angular fluctuation
modes. Equivalently, the Schur--Ingham--Siegel matrix--ball picture describes
an interior \(2\times2\) matrix-ball saddle for \(d\ge4\), a boundary stratum
at \(d=3\), a singular-value corner at \(d=2\), and rank collapse at \(d=1\).

For \(d\ge4\), the two endpoint two-planes generically span a
four-dimensional subspace. The relative Grassmannian may therefore be
interpreted as the configuration space of a relational four-dimensional
geometry, constituted by these two planes and embedded in \(\mathbb{R}^d\).
The endpoint entropy hierarchy is then governed by the emergence of this stable
relational geometry as \(d\) increases, and by its successive degenerations as
\(d\) is lowered.

\end{abstract}

{\small
\noindent
\textbf{Keywords:} BFSS/BMN matrix models; large-\(d\) expansion; endpoint
entropy; HCIZ/Bessel reduction; Wishart--Stiefel variables; Grassmannian
geometry; principal angles; boundary saddles.
}

\newpage
\tableofcontents

\newpage

\section{Introduction}

\subsection{BFSS/BMN systems}

\medskip
\noindent
The concept of ``matrix quantum gravity'' refers here to three intertwined
notions:
\begin{itemize}
\item Matrix quantum-mechanical models of the BFSS
\cite{BanksFischlerShenkerSusskind1997} and BMN
\cite{BerensteinMaldacenaNastase2002} types.

\item The gauge/gravity duality approach to quantum black-hole dynamics, in
which these matrix models admit emergent gravity duals in the holographic
planar regime~\cite{tHooft1993,Susskind1995,tHooft1974},
\begin{eqnarray}
N\to\infty,
\qquad
g^2\to0,
\qquad
\lambda=g^2N
\quad \hbox{fixed}.
\end{eqnarray}
Since the gauge theory admits a nonperturbative lattice
regularization~\cite{Wilson:1974sk}, these systems provide a concrete
framework for investigating quantum gravity and black-hole thermodynamics.
For Monte Carlo simulations see
\cite{Catterall2008,Anagnostopoulos2008,Hanada2014,Hanada2016b,Filev:2015hiaF},
for analytic approaches see
\cite{Kabat2001,Hanada2009,Hyakutake2014},
and for a pedagogical review see~\cite{Hanada2016}.

\item The noncommutative-geometry/matrix-model approach to emergent geometry
and gravity \cite{Ydri:2022ueu,Ydri:2021cam,Ydri:2020fry}, in which
BFSS/BMN-type matrix models provide the fundamental microscopic degrees of
freedom.
\end{itemize}

\medskip
\noindent
BFSS-type models are obtained by reducing supersymmetric Yang--Mills theory to
one time dimension~\cite{Brink:1976bc}. The allowed dimensions
\begin{eqnarray}
D_{\rm YM}=d+1=10,6,4,3,2,
\end{eqnarray}
are fixed by the corresponding Fierz identities~\cite{Baake:1984ie}. BMN-type
models are the maximally supersymmetric mass deformations of these matrix
quantum-mechanical systems~\cite{Kim:2006,Park:2005}.

\medskip
\noindent
The BFSS class has two complementary physical interpretations. First, it
provides a nonperturbative formulation of M-theory in the infinite-momentum
frame, or equivalently the quantum mechanics of coincident D\(0\)-branes
\cite{Witten1996}. Second, within the framework of Maldacena's
gauge/gravity conjecture
\cite{Maldacena1999,Gubser1998,Witten1998}, strongly coupled large-\(N\)
matrix quantum mechanics is holographically dual, in the appropriate
regime, to type IIA supergravity on the black 0-brane
background~\cite{Itzhaki1998}, which describes the corresponding
D\(0\)-brane bound state in type IIA string theory.
In both interpretations, D\(0\)-branes play a central role, with
the matrices \(X_a\) describing their noncommutative transverse coordinates
and the open-string degrees of freedom connecting them
\cite{Polchinski1995}; see also
\cite{Azeyanagi2009,Zwiebach2009,Becker2006}.

\medskip
\noindent
The BFSS model is also closely related to light-cone supermembranes
\cite{Hoppe1982,Hoppe1988,deWitHoppeNicolai1988} and to superparticle dynamics
in maximally supersymmetric pp-wave backgrounds
\cite{Kowalski-Glikman:1984qtj,Blau:2001ne}. This web of relations follows
from the fact that compactifying eleven-dimensional supergravity
\cite{Cremmer1978} on a circle gives type IIA theory~\cite{Witten1995}, while
reducing the M-wave solution gives the ten-dimensional non-extremal black
0-brane~\cite{Hyakutake:2014maa,Hyakutake:2006aq}.

\medskip
\noindent
At the bosonic level, the actions of the BFSS model and its
mass-deformed BMN counterpart may be written schematically as

\begin{eqnarray}
S_{\rm BFSS}^{\rm E}
=
\frac{1}{g^2}
\int_{0}^{\beta}\! dt~
\Tr\!\left[
\frac{1}{2}(D_t X_a)^2
-
\frac{1}{4}[X_a,X_b]^2
\right]
+
\text{fermionic terms},
\qquad
D_t=\partial_t-i[A_t,\cdot],
\label{BFSSIntro}
\end{eqnarray}
and
\begin{eqnarray}
S_{\rm BMN}^{\rm E}
=
S_{\rm BFSS}^{\rm E}
+
\frac{1}{g^2}
\int_{0}^{\beta}\! dt~
\Tr\!\left[
\mu_1 X_a^2
+
\mu_2 \epsilon_{ijk}X_iX_jX_k
\right]
+
\text{fermionic terms}.
\label{BMNIntro}
\end{eqnarray}
The quadratic mass term lifts the flat directions, while the Myers interaction
\cite{Myers} selects an \(SO(3)\) sector. These deformations describe pp-wave
rather than flat backgrounds. In the maximally supersymmetric case, their
half-BPS sectors are related to LLM bubbling geometries~\cite{Lin:2004nb}, and
their thermal dynamics has been studied numerically, for example in
\cite{Asano:2018nol,Asano:2020yry}. For related recent developments, see also the preprints
\cite{OConnor:2026zlf,Lei:2026fep}.

\subsection{Large-\(d\) saddle and Molien--Weyl integrals}

\medskip
\noindent
In the Gaussian, or large-mass, regime, BFSS/BMN matrix quantum mechanics
reduces to a gauged matrix oscillator whose singlet partition
function is a Molien--Weyl integral
\cite{OConnor:2023mss,OConnor:2024udv}, equivalently the Hilbert series
counting invariant oscillator operators~\cite{CoxLittleOShea2005}. The same
oscillator arises dynamically in the large--\(d\) BFSS systems
\eqref{BFSSIntro}, where the commutator interaction becomes self-consistently
Gaussian and generates an effective mass \(s\sim d^{1/3}\)
\cite{Mandal:2009vzN,Mandal:2011hbN,Kabat:2000zv,Kabat:2001ve}. For the BMN
systems \eqref{BMNIntro}, the corresponding double-scaling regime keeps
\cite{Ydri:2026gsy}
\begin{eqnarray}
\hat m=\kappa^{2/3}
\equiv
\frac{m}{d^{2/3}}
\end{eqnarray}
fixed as both \(m\) and \(d\) become large.

\medskip
\noindent
The effective theory is therefore the gauged matrix harmonic oscillator
\begin{eqnarray}
S_{\rm MHO}[X;\theta]
=
N\int_{0}^{\beta}\! dt\,
\mathrm{Tr}\!\left[
\frac{1}{2}(D_tX_a)^2
+
\frac{s^2}{2}X_a^2
\right]
+
\text{fermionic terms},
\label{MHO_action_for_Ward}
\end{eqnarray}
with self-consistent gap equation
\begin{eqnarray}
s^2=m+k_0,
\qquad
s^3-ms=d.
\end{eqnarray}
The confinement/deconfinement transition is then controlled by the holonomy
effective action
\cite{Kawahara:2007fnF,Aharony:2003sxF,Aharony:2004igF,AlvarezGaume:2005fvF,Gross:1980heF,Wadia:1980cpF},
with critical temperature
\begin{eqnarray}
T_c(\kappa)
=
\frac{s(\kappa)}{\log d}
=
\frac{d^{1/3}}{\log d}
\left(
\kappa^{1/3}
+
\frac{1}{2\kappa^{2/3}}
+\cdots
\right).
\end{eqnarray}
Thus the transition is pushed to parametrically higher temperature as
\(d\to\infty\).

\begin{figure}[htbp]
\begin{center}
   \includegraphics[width=8cm,angle=-0,page=3]{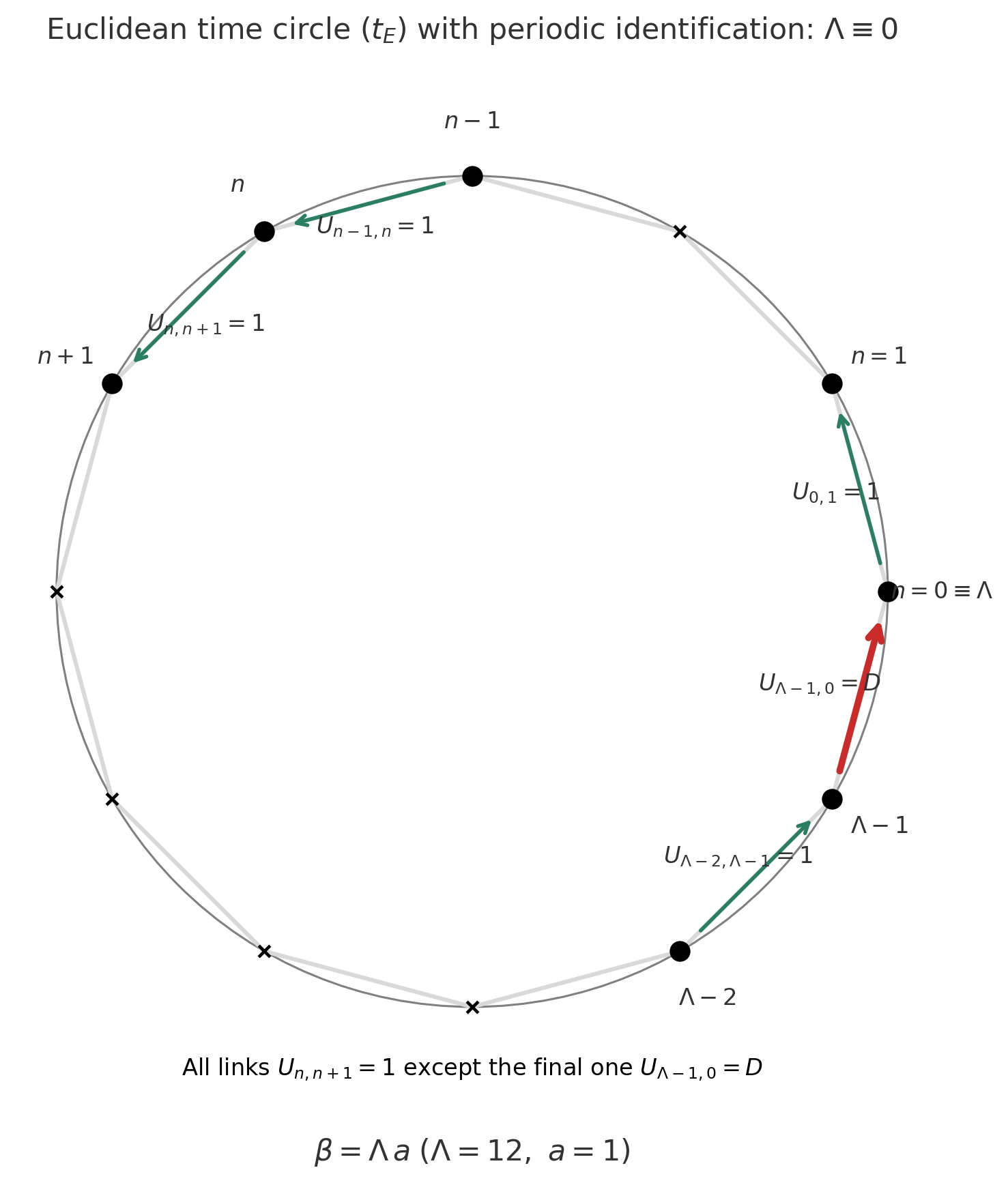}
\end{center}
\caption{The static diagonal (Polyakov) gauge.}\label{GF}
\end{figure}

\medskip
\noindent
On the lattice, the thermal circle is discretized with spacing \(a=\beta/\Lambda\), and the gauge field is represented by link variables \(U_{n,n+1}\). Using gauge invariance and Haar-measure invariance, one may work in the static Polyakov gauge~\cite{Filev:2015hiaF}, where all links are gauged to unity except the closing link. See Fig.~\ref{GF}. The gauge field is then reduced to a single holonomy
\begin{eqnarray}
g=\mathcal{P}\exp\!\left(i\int_0^\beta dt\,A_t\right)\in U(N).
\end{eqnarray}

\medskip
\noindent
After integrating out the Gaussian matrix coordinates, the \(U(N)\) partition
function becomes a single group integral. Since the \(d\) matrices factorize,
the normal-ordered bosonic partition function is equivalently, with
\(x=\exp(-\beta s)\) and \(g=\mathrm{diag}(z_1,\ldots,z_N)\),
\begin{eqnarray}
Z^{U(N)}_{N,d}(x)
&=&
\int_{U(N)} d\mu(g)\;
\frac{1}{\Big[{\bf det}\big(1-x\,g\otimes g^{-1}\big)\Big]^d}
\nonumber\\
&=&
\frac{1}{N!}
\oint
\prod_{i=1}^{N}\frac{dz_i}{2\pi i z_i}\;
\Delta(z)\Delta(z^{-1})
\prod_{i,j=1}^{N}
\frac{1}{\big(1-x\,z_i z_j^{-1}\big)^d}.
\label{eq:MW_BFSSd}
\end{eqnarray}

\subsection{Organization of the paper}

\medskip
\noindent
The present paper focuses on the Gram--Wishart--Stiefel endpoint formulation
of the \(N=2\) planar theory after large-\(d\) Gaussian reduction. Thus the
starting point is the effective gauged matrix harmonic oscillator obtained from
the BFSS/BMN matrix model, rather than the unreduced interacting Yang--Mills
matrix quantum mechanics. The same viewpoint is used in the lower-dimensional
BFSS/BMN cases discussed below: we consider the corresponding harmonic
approximation and subject it to the Wishart--Stiefel endpoint reduction.
BFSS\(_2\) is exceptional only because, with a single matrix, the commutator
interaction is absent and the model is already of harmonic-oscillator type once
the mass scale is specified. The purpose is to understand how the holonomy
dynamics, the shifted Gaussian structure, and the universal
\(D_\Lambda\)-channel contribution are encoded in radial Wishart variables and
relative Stiefel angular data.

\medskip
\noindent
This paper is organized as follows. Section \eqref{section2} develops the Gram--Wishart--Stiefel formulation of the \(N=2\),
large--\(d\) BFSS/BMN endpoint theory. Its purpose is to rewrite the shifted
planar boundary model in radial Wishart variables and relative Stiefel
orientations, isolate the aligned \(-A\) sector, derive the rank--two
Bessel/HCIZ angular kernel, and show how the factorized Bessel/HCIZ ansatz
leads to the symmetric--Wishart saddle and the universal \(D\)-channel
coefficient
\begin{eqnarray}
{\cal C}_{2,D}\longrightarrow -2d.
\end{eqnarray}

\medskip
\noindent
Section \eqref{section3} is devoted to the HCIZ/Bessel reduction of lower--dimensional BFSS
models. We adopt a simplified holonomy potential in which only the quadratic
term is retained, and analyze BFSS\(_4\), BFSS\(_3\), and BFSS\(_2\) separately.
For BFSS\(_4\), the \(O(3)\) angular integral is reduced to a \(c\)-resolved
HCIZ problem, whose saddle dynamically selects a boundary branch before the
Bessel factor is included. For BFSS\(_3\), the \(O(2)\) HCIZ integral is exact,
but the Stiefel/Wishart measure develops an integrable boundary singularity,
leading again to a boundary-layer saddle. For BFSS\(_2\), the rank itself
collapses and the problem becomes purely radial. In all three cases the Bessel
convolution shifts the endpoint mass, while also modifying the logarithmic
entropy coefficient: in BFSS\(_4\) and BFSS\(_3\) this logarithm is tied to
boundary angular or boundary-layer geometry, whereas in BFSS\(_2\) it is purely
radial.

\medskip
\noindent
Section \ref{section4} extracts the geometric meaning of the endpoint and
Schur--Ingham--Siegel (SIS)/Wishart formulations. We first show that the
rank--two endpoint construction naturally produces a relative Grassmannian
sector \(Gr(2,d)\), whose tangent-space dimension \(2(d-2)\) controls the
large--\(d\) angular entropy. The two principal angles are identified as the
invariant singular-value coordinates on this relative Grassmannian, while the
remaining variables describe their angular degeneracies. We then explain how
this geometry degenerates in low dimensions: BFSS\(_4\) retains only one
nontrivial principal angle, BFSS\(_3\) has a collapsed Grassmannian but a
nontrivial boundary-layer remnant, and BFSS\(_2\) reduces to a purely radial
rank--one endpoint problem. The same hierarchy is reinterpreted from the
Schur--Ingham--Siegel matrix--ball perspective, where the stable \(d\ge4\)
regime is an interior \(2\times2\) matrix-ball saddle, \(d=3\) is a boundary
stratum, \(d=2\) is a singular-value corner, and \(d=1\) is a rank collapse.
Finally, we interpret the relative Grassmannian as the configuration space of a
relational four-dimensional geometry generated by two endpoint two-planes: the
generic span \(P_V+P_W\) is four-dimensional, while \(Gr(2,d)\) describes the
ways in which its two two-plane constituents move, orient, and embed themselves
inside \(\mathbb{R}^d\).

\medskip
\noindent
Section~\ref{conclusion} summarizes the endpoint entropy hierarchy,
the HCIZ/Bessel reductions, and the emergent Grassmannian interpretation.
The paper closes with a historical note in Section~\ref{history},
which situates the present work within our broader matrix quantum
gravity programme and traces its origins in our earlier investigations
of emergent geometry, noncommutative field theory and fuzzy spaces.
Appendices~\ref{appendix1} and~\ref{appendix2} collect the technical
material used in the main text, including details of the angular
reductions, endpoint asymptotics, and auxiliary Wishart/Stiefel and SIS
matrix--ball formulae.

\section{Gram--Wishart--Stiefel formulation of the (N=2), large--(d) BFSS/BMN QM}\label{section2}

In this section we follow closely the analysis of~\cite{YdriUnpublished1}. See also \cite{Ydri:2026rke} and \cite{Ydri:2026gsy,Ydri:2026smo,YdriUnpublished0}.  

\subsection{Planar boundary model and shifted Gaussian structure}

\medskip
\noindent
We start by recasting the \emph{\(N=2\), large--\(d\) BFSS/BMN matrix quantum mechanics on the lattice} as a planar endpoint, or boundary, model.

\medskip
\noindent
After the bulk fluctuations, the gauge field, and the longitudinal endpoint modes have been integrated out, the remaining degrees of freedom are two families of transverse two-dimensional endpoint vectors,

\begin{eqnarray}
V_a=(V_a^1,V_a^2),
\qquad
W_a=(W_a^1,W_a^2),
\qquad
a=1,\ldots,d.
\end{eqnarray}
The natural collective variables are the aligned and transverse invariants

\begin{eqnarray}
A=\lambda\sum_{a=1}^{d}V_a\cdot W_a,
\qquad
B=\lambda\sum_{a=1}^{d}V_a\times W_a,
\qquad
R=\sqrt{A^2+B^2},
\qquad
\lambda=\frac{2N}{a}.
\end{eqnarray}
The corresponding planar endpoint action is then written as

\begin{eqnarray}
S_{N=2,\rm eff}^{\rm BFSS_{d+1}}
=
m\sum_{a=1}^{d}
\Big(
V_a^2+W_a^2
\Big)
-2\beta_\Lambda A
+
V_{\rm hol}(A,R),
\qquad
m=\lambda\alpha_\Lambda.
\end{eqnarray}
The holonomy-induced potential is explicitly given by the expression 
\begin{eqnarray}
V_{\rm hol}(A,R)
=
-\log\Big(I_0(R)-\frac{A}{R}I_1(R)\Big).
\end{eqnarray}

\medskip
\noindent 
The endpoint coefficients \(\alpha_\Lambda\) and \(\beta_\Lambda\) are expressed explicitly in terms of the oscillator mass \(s\), the lattice spacing \(a\), and the fugacity \(x=e^{-\beta s}\) as

\begin{eqnarray}
\alpha_\Lambda
=
\frac12+\frac{as}{2}\frac{1+x^2}{1-x^2}+O(a^2),
\qquad 
\beta_\Lambda
=
as\,\frac{x}{1-x^2}+O(a^2),
\qquad
x=e^{-\beta s}.
\label{alphabeta_x_intro}
\end{eqnarray}

\medskip
\noindent
The physical domain is bounded by the inequality \(R^2\ge A^2\). Hence the saddle relevant to the endpoint theory is not an unconstrained critical point in the ambient \((A,R)\)-space. It is instead a constrained boundary saddle supported on the aligned branch
\begin{eqnarray}
B=0,\qquad A=R.
\end{eqnarray}
Restricting the holonomy potential to this branch gives
\begin{eqnarray}
V_{\rm b}(A)
=
-\log\Big(I_0(A)-I_1(A)\Big),
\end{eqnarray}
whose constrained saddle
\begin{eqnarray}
  A=A_*=R_*\label{ARstar}
\end{eqnarray}
is determined by
\begin{eqnarray}
I_0(R_*)
=
\left(1+\frac{1}{R_*}\right)I_1(R_*),
\qquad
R_*\simeq 1.545.
\end{eqnarray}
Expanding around this saddle separates the longitudinal displacement \(A^2-R_*^2\) from the transverse variable \(B^2\). At fixed \(A=A_*=R_*\), the transverse expansion is

\begin{eqnarray}
V_{\rm hol}(A=R_*,B)
=
V_*
-\frac{1}{2R_*}B^2
+
\frac{3-R_*}{8R_*^3}B^4
+
O(B^6).
\end{eqnarray}
This potential exhibits the standard local Landau structure: the negative quadratic term signals a transverse instability of the aligned branch, while the positive quartic term provides the leading stabilizing correction.

\medskip
\noindent
In the low-temperature continuum regime the invariants are driven into the large-\(R\) region. On the constrained aligned boundary, the holonomy potential has the asymptotic behavior

\begin{eqnarray}
V_{\rm b}(R)
=
-\log\Big(I_0(R)-I_1(R)\Big)
=
-R+O(\log R).
\end{eqnarray}
Since \(A=R\) on this branch, the holonomy potential is dominated by the linear aligned contribution \(-A\). Thus the appropriate continuum Gaussian is not the bare Gaussian, but a shifted Gaussian in which the \(-A\) term is absorbed into the Gaussian sector:

\begin{eqnarray}
S_{N=2,\rm eff}^{\rm BFSS_{d+1}}
=
\Big[
m\sum_{a=1}^{d}
\big(
V_a^2+W_a^2
\big)
-A
\Big]
+
\widetilde V_{\rm hol}(A,R),
\qquad
\widetilde V_{\rm hol}(A,R)
=
A+V_{\rm hol}(A,R).
\end{eqnarray}

\subsection{Continuum limit}

\medskip
\noindent
In the full endpoint theory, the anisotropic coupling \(\beta_\Lambda\) enters the reduced formulation through two separate structures:

\begin{eqnarray}
D_\Lambda=\alpha_\Lambda^2-\beta_\Lambda^2,
\qquad
\text{and}
\qquad
-2\beta_\Lambda A.
\end{eqnarray}
The first structure defines the determinant-like \(D_\Lambda\)-channel, whereas the second is a direct or explicit aligned coupling to the longitudinal invariant \(A\).

\medskip
\noindent
In a pure \(B\)-theory, by contrast, the longitudinal variable is frozen, and the only remaining dynamical variable is the transverse invariant \(B\). Once the constrained saddle has been fixed, the only parameter which remains sensitive to the continuum limit is therefore

\begin{eqnarray}
D_\Lambda=\alpha_\Lambda^2-\beta_\Lambda^2.
\end{eqnarray}
Thus the pure \(B\)-theory isolates the \(D_\Lambda\)-channel. Its continuum behavior is controlled by the approach of \(D_\Lambda\) to the critical value \(1/4\). Since \(\beta_\Lambda\to0\) in the continuum limit, the anisotropic coupling does not survive as an independent dynamical coupling in the pure \(B\)-theory. Hence, for the purpose of analyzing the continuum singular structure, one may effectively set \(\beta_\Lambda=0\).

\medskip
\noindent
Let us now recall the continuum structure of the exact endpoint theory in a form adapted to the present discussion. The object of interest is the quadratic coefficient in the low-\(x\) expansion of the normalized full endpoint ratio

\begin{eqnarray}
\frac{\widetilde Z_\perp(x)}{\widetilde Z_\perp(0)}.
\end{eqnarray}
In the exact theory, the \(x\)-dependence enters through the expansion of the two endpoint parameters,

\begin{eqnarray}
D_\Lambda(x)=D_0+D_2x^2+O(x^4),
\qquad
\beta_\Lambda(x)=\beta_1x+\beta_3x^3+O(x^5),
\end{eqnarray}
and through the exact holonomy kernel
\begin{eqnarray}
\Phi(A,B)
=
I_0(R)-\frac{A}{R}I_1(R),
\qquad
R^2=A^2+B^2.
\end{eqnarray}
The Gaussian average of this kernel defines the exact holonomy correction

\begin{eqnarray}
G_{\rm ex}(D_\Lambda,\beta_\Lambda)=
\frac{\widetilde Z_\perp(x)}{Z_{\perp,0}(x)}
=
\left\langle \Phi(A,B)\right\rangle_0.
\end{eqnarray}

\medskip
\noindent
The normalized endpoint ratio can therefore be written as

\begin{eqnarray}
\frac{\widetilde Z_\perp(x)}{\widetilde Z_\perp(0)}
=
\frac{Z_{\perp,0}(x)}{Z_{\perp,0}(0)}
\widehat Z_\perp(x)=\frac{Z_{\perp,0}(x)}{Z_{\perp,0}(0)}
\frac{G_{\rm ex}(D_\Lambda(x),\beta_\Lambda(x))}
{G_{\rm ex}(D_0,0)}.
\end{eqnarray}
It follows that the quadratic coefficient in its low-\(x\) expansion separates into three contributions:

\begin{eqnarray}
{\cal C}_2^{\rm ex}
=
-\frac{dD_2}{D_0}
+
D_2
\frac{\partial_D G_{\rm ex}(D_0,0)}
{G_{\rm ex}(D_0,0)}
+
\frac{\beta_1^2}{2}
\frac{\partial_\beta^2G_{\rm ex}(D_0,0)}
{G_{\rm ex}(D_0,0)}.
\end{eqnarray}
The first term is the universal Gaussian contribution, the second is the \(D\)-channel contribution, and the third is the \(\beta\)-channel contribution. In the strict continuum limit, the Gaussian term drops out, while the two exact holonomy channels remain finite:

\begin{eqnarray}
{\cal C}_{2,D}^{\rm ex}\longrightarrow -2d,
\qquad
{\cal C}_{2,\beta}^{\rm ex}\longrightarrow d(d+1).
\end{eqnarray}
The exact continuum coefficient is therefore reconstructed as

\begin{eqnarray}
{\cal C}_2^{\rm ex}\longrightarrow d(d-1).
\end{eqnarray}

\medskip
\noindent
Because a pure transverse model freezes the longitudinal variable \(A\), it cannot reproduce the \(\beta\)-channel. It can nevertheless reproduce the \(D\)-channel, provided that its Gaussian average develops the same singular dependence on \(D_\Lambda\) as the exact holonomy kernel. This is exactly the point at which finite polynomial truncations break down. Although the transverse expansion captures the local Landau-type structure near the constrained boundary saddle, any finite polynomial \(B\)-theory produces, after Gaussian averaging, a function which is regular at the continuum point \(D_\Lambda=1/4\). Its continuum limit is therefore trivial.

\subsection{Toy model and non-polynomial completion}

\medskip
\noindent
The pure transverse model must reproduce the \(D_\Lambda\)-channel singularity while depending only on \(B\). Finite polynomial truncations cannot do this, since their Gaussian averages are regular at \(D_\Lambda=1/4\). We therefore introduce the non-polynomial toy completion
\begin{eqnarray}
V_{\rm toy}(B)
=
-\log\cosh B,
\qquad
e^{-V_{\rm toy}(B)}=\cosh B.
\end{eqnarray}
Its local expansion is
\begin{eqnarray}
V_{\rm toy}(B)
=
-\frac12 B^2
+
\frac1{12}B^4
-\frac1{45}B^6
+
\frac{17}{2520}B^8
+
O(B^{10}),
\end{eqnarray}
so it has the same qualitative Landau structure as the exact transverse holonomy potential: a negative quadratic term followed by a positive quartic correction. Globally, however, it is better behaved than any finite polynomial truncation, since it does not generate artificial off-center minima.

\medskip
\noindent
The essential point is the Gaussian average. Since the Gaussian moments of \(\cosh B\) resum exactly, one obtains
\begin{eqnarray}
G_{\rm toy}(D_\Lambda)
=
\big\langle \cosh B\big\rangle_0
=
\left(
1-\frac{1}{4D_\Lambda}
\right)^{-d}.
\end{eqnarray}
Thus the toy model has precisely the continuum singularity at \(D_\Lambda=1/4\), and hence reproduces the full \(D\)-channel contribution
\begin{eqnarray}
{\cal C}_{2,D}^{\rm toy}
\longrightarrow
-2d.
\end{eqnarray}
It does not reproduce the \(\beta\)-channel, as expected, because the longitudinal invariant \(A\) has been frozen.

\medskip
\noindent
The same toy interaction is also related to the large-\(R\) aligned branch. The Gaussian master identity
\begin{eqnarray}
\big\langle e^{uA}\cosh(vB)\big\rangle_0
=
\left(
\frac{D_\Lambda}
{D_\Lambda-\beta_\Lambda u-\frac{u^2+v^2}{4}}
\right)^d
\end{eqnarray}
shows that, after Gaussian averaging, a transverse \(B\)-insertion may be transmuted into an effective aligned \(A\)-insertion. In particular, for small \(\beta_\Lambda\), the pure toy insertion is equivalent to the large-\(R\) branch governed by \(-A\), up to small \(\beta_\Lambda\)-corrections:
\begin{eqnarray}
\cosh B
\;\sim\;
e^{A+O(\beta_\Lambda)A}.
\end{eqnarray}
Equivalently,
\begin{eqnarray}
\big\langle e^{-A}\cosh B\big\rangle_0=1,
\label{toy_trivialization_intro}
\end{eqnarray}
or
\begin{eqnarray}
\int dV\,dW\;
e^{-\alpha_\Lambda(V^2+W^2)}\,
\cosh B
=
\int dV\,dW\;
e^{-\alpha_\Lambda(V^2+W^2)+A}.
\label{remarkable_intro}
\end{eqnarray}

\medskip
\noindent
Relative to the shifted Gaussian sector, the completed toy potential is therefore
\begin{eqnarray}
\widetilde V_{\rm toy}(B)
=
V_{\rm toy}(B)+A
=
-\log\cosh B+\sqrt{R_*^2-B^2},
\label{Vtoy_modified_intro}
\end{eqnarray}
where \(A\) has been pulled back to the fixed-radius shell \(R=R_*\). Its small-\(B\) expansion is
\begin{eqnarray}
\widetilde V_{\rm toy}(B)
&=&
R_*+\sum_{n\ge1}(-1)^n c'_{2n}B^{2n}\nonumber\\
&=&
R_*
-\left(\frac12+\frac{1}{2R_*}\right)B^2
+\left(\frac1{12}-\frac{1}{8R_*^3}\right)B^4
-O(B^{6}).
\label{Vtoy_modified_series_intro}
\end{eqnarray}
Thus the toy model should be viewed as a non-polynomial completion of the transverse expansion: locally approximate, globally better behaved than a polynomial truncation, and exact in its reproduction of the singular \(D_\Lambda\)-channel.

\subsection{Gram--Wishart--Stiefel variables}

\medskip
\noindent
The Wishart--Stiefel decomposition used below is a standard tool in
multivariate statistics and invariant integration on matrix spaces. It is rooted
in Wishart's covariance distribution, Stiefel's theory of frame manifolds, and
the work of James and Herz on orthogonal invariance, latent roots, and
matrix-variate special functions
\cite{Wishart:1928,BStiefel:1935,James:1954AMS,Herz:1955,James:1960AMS,James:1961AMS,James:1964AMS,Muirhead:1982}.

\medskip
\noindent
The planar endpoint theory admits a natural Gram--Wishart--Stiefel formulation. We assemble the two endpoint configurations into \(2\times d\) matrices
\begin{eqnarray}
(X_0)_{\mu a}=V_a^\mu,
\qquad
(X_1)_{\mu a}=W_a^\mu,
\qquad
\mu=1,2,
\qquad
a=1,\ldots,d.
\end{eqnarray}
The corresponding endpoint Gram blocks and cross--Gram block are
\begin{eqnarray}
Q^V=X_0^T X_0,
\qquad
Q^W=X_1^T X_1,
\qquad
\Phi=X_1^T X_0.
\end{eqnarray}
Since \(X_0\) and \(X_1\) have only two rows, \(Q^V\) and \(Q^W\) are positive semidefinite matrices of rank at most two.

\medskip
\noindent
With \(K=X_1X_0^T\), the aligned invariant is
\begin{eqnarray}
A_0={\rm tr}\,K={\rm Tr}\,\Phi,
\end{eqnarray}
while the radial invariant decomposes as
\begin{eqnarray}
R_0^2
=
A_0^2+B_0^2
=
{\rm Tr}(Q^WQ^V)
+
\Big[
({\rm Tr}\,\Phi)^2-{\rm Tr}(\Phi^2)
\Big].
\end{eqnarray}
Equivalently, the transverse invariant is
\begin{eqnarray}
B_0^2
=
{\rm Tr}(Q^WQ^V)-{\rm Tr}(\Phi^2).
\label{B0}
\end{eqnarray}
Thus the holonomy invariants split into a Frobenius sector, controlled by \(Q^V\) and \(Q^W\), and a minor sector, controlled by the cross--Gram block \(\Phi\).

\medskip
\noindent
The rank--two structure suggests the Wishart--Stiefel decomposition
\begin{eqnarray}
X_0
=
U_0\,{\rm diag}(\sqrt{v_1},\sqrt{v_2})\,R_V^T,
\qquad
X_1
=
U_1\,{\rm diag}(\sqrt{w_1},\sqrt{w_2})\,R_W^T,
\end{eqnarray}
where \(v_i,w_i\ge0\) are Wishart eigenvalues. Here \(U_0,U_1\in O(2)\), while \(R_V,R_W\in{\rm St}(2,d)\subset{\mathbb R}^{d\times2}\) are \(d\times2\) Stiefel matrices with orthonormal columns; hence \(R_V^T\) and \(R_W^T\) are \(2\times d\). Writing
\begin{eqnarray}
R_V=(r_1\ r_2),
\qquad
R_W=(s_1\ s_2),
\qquad
r_i\cdot r_j=s_i\cdot s_j=\delta_{ij},
\end{eqnarray}
the two endpoint configurations determine two two-planes
\begin{eqnarray}
\Pi_V=\mathrm{span}(r_1,r_2),
\qquad
\Pi_W=\mathrm{span}(s_1,s_2),
\qquad
\Pi_V,\Pi_W\subset\mathbb R^d.
\end{eqnarray}
Choosing the canonical frame
\begin{eqnarray}
E=\binom{\mathbf 1_2}{0}\in\mathbb R^{d\times2},
\end{eqnarray}
we may write
\begin{eqnarray}
R_V=O_VE,
\qquad
R_W=O_WE,
\qquad
O_V,O_W\in O(d).
\end{eqnarray}
The common \(O(d)\) rotation is redundant, so the angular dependence is carried by the relative orientation
\begin{eqnarray}
O=O_WO_V^T\in O(d),
\end{eqnarray}
and, more precisely, by its upper \(2\times2\) block
\begin{eqnarray}
\Omega=\big[O_{ij}\big]_{i,j=1,2}.
\end{eqnarray}

\medskip
\noindent
In these variables,
\begin{eqnarray}
Q^V
=
R_V\,{\rm diag}(v_1,v_2)\,R_V^T,
\qquad
Q^W
=
R_W\,{\rm diag}(w_1,w_2)\,R_W^T,
\end{eqnarray}
and
\begin{eqnarray}
\Phi
=
R_W\,{\rm diag}(\sqrt{w_1},\sqrt{w_2})\,
S\,
{\rm diag}(\sqrt{v_1},\sqrt{v_2})\,R_V^T,
\qquad
S:=U_1^T U_0\in O(2).
\end{eqnarray}
Thus the action depends only on the relative \(O(d)\) orientation \(O\) and the residual relative \(O(2)\) variable \(S\).

\medskip
\noindent
Up to an overall normalization, the endpoint measure becomes
\begin{eqnarray}
\prod_{a=1}^d d^2V_a
&\propto&
d\mu(R_V)\,
(v_1v_2)^{\frac{d-3}{2}}
|v_1-v_2|\,dv_1\,dv_2\,d\mu(U_0),
\nonumber\\
\prod_{a=1}^d d^2W_a
&\propto&
d\mu(R_W)\,
(w_1w_2)^{\frac{d-3}{2}}
|w_1-w_2|\,dw_1\,dw_2\,d\mu(U_1).
\end{eqnarray}

\medskip
\noindent
The radial sector is therefore Wishart, while the angular sector is Stiefel. The mass term depends only on the radial variables:
\begin{eqnarray}
S_{\rm mass}
=
m\big(\mathrm{Tr}Q^V+\mathrm{Tr}Q^W\big)
=
m(v_1+v_2+w_1+w_2).
\end{eqnarray}
The shifted aligned coupling is controlled by the effective \(2\times2\) overlap matrix
\begin{eqnarray}
\phi:=\sqrt{w}\,S\,\sqrt{v}\,\Omega^T,
\end{eqnarray}
where \(\sqrt v={\rm diag}(\sqrt{v_1},\sqrt{v_2})\) and
\(\sqrt w={\rm diag}(\sqrt{w_1},\sqrt{w_2})\). Since
\begin{eqnarray}
A=\lambda A_0,
\qquad
A_0={\rm tr}\,\phi,
\end{eqnarray}
the shifted linear part of the action is
\begin{eqnarray}
-2\beta_\Lambda A-A
=
-\left(1+2\beta_\Lambda\right)\lambda\,{\rm tr}\,\phi
=
-\ell_\Lambda\,{\rm tr}\,\phi\quad ,\quad
\ell_\Lambda:=
\left(1+2\beta_\Lambda\right)\lambda .
\end{eqnarray}
Thus the shifted Gaussian sector contains the large-\(R\) aligned contribution \(-A\) from the outset, together with the anisotropic aligned coupling \(-2\beta_\Lambda A\), and separates into a radial Wishart mass term and an aligned cross term.

\medskip
\noindent
The two basic angular invariants are
\begin{eqnarray}
{\cal T}
:=
\sum_{i,j=1}^{2}w_i v_j(\Omega_{ij})^2,
\qquad
\delta
:=
\sqrt{v_1v_2w_1w_2}\,\det\Omega.
\end{eqnarray}
Since
\begin{eqnarray}
\det\phi=(\det S)\,\delta,
\end{eqnarray}
the exact transverse invariant may be written as
\begin{eqnarray}
B_0^2
=
{\cal T}
+
2\det\phi
-
({\rm tr}\,\phi)^2
=
{\cal T}
+
2(\det S)\delta
-
({\rm tr}\,\phi)^2.
\end{eqnarray}
Equivalently,
\begin{eqnarray}
R_0^2
=
A_0^2+B_0^2
=
{\cal T}
+
2(\det S)\delta .
\end{eqnarray}
Thus the full rank--two Gram structure is retained: the invariant \(A_0\) is carried by \({\rm tr}\,\phi\), while the radial invariant \(R_0\) depends on both the Frobenius variable \({\cal T}\) and the minor variable \(\delta\).

\medskip
\noindent
The shifted endpoint partition function therefore takes the Wishart--Stiefel form
\begin{eqnarray}
Z_\perp
&\propto&
\int dv_1\,dv_2\;
(v_1v_2)^{\frac{d-3}{2}}|v_1-v_2|\,
e^{-m(v_1+v_2)}
\int dw_1\,dw_2\;
(w_1w_2)^{\frac{d-3}{2}}|w_1-w_2|\,
e^{-m(w_1+w_2)}
\nonumber\\
&&\times
\int_{O(d)}d\mu(O)
\int_{O(2)}d\mu(S)\;
\exp\!\Big[
\ell_\Lambda\,{\rm tr}\,\phi
-
\widetilde V_{\rm hol}
\Big(
\lambda\,{\rm tr}\,\phi,
\lambda\sqrt{{\cal T}+2(\det S)\delta}
\Big)
\Big],
\end{eqnarray}
where
\begin{eqnarray}
\widetilde V_{\rm hol}(A,R)
=
A+V_{\rm hol}(A,R).
\end{eqnarray}
The shifted aligned source is therefore included through
\(\exp(\ell_\Lambda{\rm tr}\,\phi)\), while the remaining holonomy dependence is carried by the shifted potential \(\widetilde V_{\rm hol}\). At this stage no transverse constraint has been imposed, and both angular integrations, over \(O(d)\) and \(O(2)\), are kept intact.

\medskip
\noindent
We now introduce the transverse expansion. The prescription has three separate parts.

\begin{itemize}
\item[(i)] \emph{The genuine holonomy potential.}
The genuine holonomy piece is expanded around the aligned boundary saddle
\begin{eqnarray}
B=0,
\qquad
A=\lambda A_0=\lambda\,{\rm tr}\phi=R_*.
\label{ARstar0}
\end{eqnarray}
The longitudinal fluctuation is then projected out, with \(A\) held fixed. Thus the holonomy expansion becomes a purely transverse expansion in powers of \(B^2\).

In Gram variables, the same transverse invariant contains the trace of \(\phi\), since this trace is precisely the longitudinal component. Hence the background-field projection freezes only
\begin{eqnarray}
{\rm tr}\phi=\frac{R_*}{\lambda},
\end{eqnarray}
while leaving the Frobenius and determinant parts dynamical. Using the angular invariants \({\cal T}\) and \(\delta\) already introduced above, the background-field projection gives
\begin{eqnarray}
B_0^2
=
{\cal T}
+
2(\det S)\delta
-
\frac{R_*^2}{\lambda^2}.
\end{eqnarray}
Thus \({\cal T}\) carries the Frobenius contribution, while the residual minor contribution is the determinant term \(2\det\phi=2(\det S)\delta\).

\item[(ii)] \emph{The compensating \(+A\) term.}
The compensating \(+A\) term is not part of the genuine local holonomy expansion. It only compensates the linear \(-A\) contribution absorbed into the shifted Gaussian sector. Its transverse effect is therefore defined geometrically by pulling it back to the fixed-radius shell
\begin{eqnarray}
R^2=A^2+B^2,
\qquad
R=R_*,
\end{eqnarray}
so that
\begin{eqnarray}
+A=+\sqrt{R_*^2-B^2}.
\end{eqnarray}
\medskip
\noindent
The transverse holonomy interaction is then written as
\begin{eqnarray}
\widetilde V_{\rm hol}(B_0)
=
-\kappa_2 B_0^2
+\kappa_4 B_0^4
-\kappa_6 B_0^6
+\cdots,
\qquad
\kappa_{2n}
=
c'_{2n}\left(\frac{2N}{a}\right)^n.
\end{eqnarray}
Consequently the coefficients \(c'_{2n}\), and hence the couplings \(\kappa_{2n}\), are those of the completed transverse potential: the genuine holonomy piece expanded at fixed \(A=R_*\), together with the compensating \(+A\) term pulled back at fixed \(R=R_*\).

\item[(iii)] \emph{The aligned source.}
The anisotropic coupling \(-2\beta_\Lambda A\) is frozen to its background value in the pure \(B\)-sector and may be set to zero as an independent dynamical coupling. By contrast, the subtracted large-\(R\) contribution \(-A\) is kept explicitly as a shifted source:
\begin{eqnarray}
-A=-\lambda A_0=-\lambda\,{\rm tr}\phi.
\end{eqnarray}
Thus this term remains inside the angular kernel as \(\ell_\Lambda{\rm tr}\phi\). In particular, the \(O(2)\) variable \(S\) enters through \({\rm tr}\phi\) in the shifted source, and through \(\det\phi\) in the transverse invariant \(B_0^2\).
\end{itemize}

\medskip
\noindent
The corresponding shifted transverse partition function is therefore of Wishart--Stiefel form:
\begin{eqnarray}
Z_\perp
&\propto&
\int dv_1\,dv_2\;
(v_1v_2)^{\frac{d-3}{2}}|v_1-v_2|\,
e^{-m(v_1+v_2)}
\int dw_1\,dw_2\;
(w_1w_2)^{\frac{d-3}{2}}|w_1-w_2|\,
e^{-m(w_1+w_2)}
\nonumber\\
&&\times
\int_{O(d)}d\mu(O)
\int_{O(2)}d\mu(S)\;
\exp\!\Big[
\ell_\Lambda\,{\rm tr}\phi
+
\kappa_2B_0^2
-\kappa_4B_0^4
+\kappa_6B_0^6
+\cdots
\Big].
\end{eqnarray}
Here \(\ell_\Lambda=(1+2\beta_\Lambda)\lambda\), and in the pure \(B\)-sector one may set \(\beta_\Lambda=0\), so that \(\ell_\Lambda\to\ell=\lambda\).

The essential point is that the shifted source \(\ell_\Lambda{\rm tr}\phi\), coming from the large-\(R\) contribution \(-A\), is kept inside the angular integral. Thus the \(O(2)\) variable \(S\) enters both through \({\rm tr}\phi\) and through the determinant term in \(B_0^2\).

\medskip
\noindent
Equivalently, the non-polynomial transverse interaction may be generated from the shifted quadratic kernel
\begin{eqnarray}
Z_\perp^{(2)}(\kappa,\ell_\Lambda)
&\propto&
\int dv_1\,dv_2\;
(v_1v_2)^{\frac{d-3}{2}}|v_1-v_2|\,
e^{-m(v_1+v_2)}
\int dw_1\,dw_2\;
(w_1w_2)^{\frac{d-3}{2}}|w_1-w_2|\,
e^{-m(w_1+w_2)}
\nonumber\\
&&\times
\int_{O(d)}d\mu(O)
\int_{O(2)}d\mu(S)\;
\exp\!\Big[
\ell_\Lambda\,{\rm tr}\phi
+
\kappa
\left(
{\cal T}
+
2(\det S)\delta
-
\frac{R_*^2}{\lambda^2}
\right)
\Big],
\end{eqnarray}
by the differential rule
\begin{eqnarray}
Z_\perp
=
\left.
\exp\!\left(
\kappa_2\frac{\partial}{\partial\kappa}
-\kappa_4\frac{\partial^2}{\partial\kappa^2}
+\kappa_6\frac{\partial^3}{\partial\kappa^3}
+\cdots
\right)
Z_\perp^{(2)}(\kappa,\ell_\Lambda)
\right|_{\kappa=0}.
\label{qy6}
\end{eqnarray}
Thus the \(O(2)\) integral is already nontrivial at quadratic order: the shifted source \(\ell_\Lambda{\rm tr}\phi\) comes from the retained \(-A\) term, while the determinant branch \(2(\det S)\delta\) comes from the transverse invariant \(B_0^2\).

\subsection{The minor approximation and emergent alignment}

\medskip
\noindent
The detailed angular calculation was carried out in the preceding paper \cite{YdriUnpublished1}. We recall here only the structure needed for the present discussion. After the transverse background-field reduction, the shifted Wishart--Stiefel formulation contains two angular integrations: a residual \(O(2)\) integral over \(S\), and a remaining \(O(d)\) integral over the relative Stiefel orientation \(O\). The shifted quadratic kernel has the form
\begin{eqnarray}
Z_\perp^{(2)}(\kappa,\ell)
\propto
\int d\mu_{\rm W}(v,w)
\int_{O(d)}d\mu(O)
\int_{O(2)}d\mu(S)\;
\exp\!\Big[
\ell\,{\rm tr}\phi
+
\kappa
\Big(
{\cal T}
+
2(\det S)\delta
-
c
\Big)
\Big],
\end{eqnarray}
where
\begin{eqnarray}
d\mu_{\rm W}(v,w)
&:=&
dv_1\,dv_2\;
(v_1v_2)^{\frac{d-3}{2}}
|v_1-v_2|\,
e^{-m(v_1+v_2)}
\nonumber\\
&&\times
dw_1\,dw_2\;
(w_1w_2)^{\frac{d-3}{2}}
|w_1-w_2|\,
e^{-m(w_1+w_2)},
\qquad
c=\frac{R_*^2}{\lambda^2}.
\end{eqnarray}

\medskip
\noindent
The \(O(2)\) integral can be performed exactly. Splitting \(O(2)\) into its two connected components gives
\begin{eqnarray}
K(\kappa,\ell;v,w,\Omega)
&:=&
\int_{O(2)}d\mu(S)\;
\exp\!\Big[
\ell\,{\rm tr}\phi
+
\kappa
\Big(
{\cal T}
+
2(\det S)\delta
-
c
\Big)
\Big]
\nonumber\\
&=&
\frac12\,e^{\kappa({\cal T}-c)}
\Big[
e^{2\kappa\delta}I_0(\ell\rho_+)
+
e^{-2\kappa\delta}I_0(\ell\rho_-)
\Big],
\qquad
\rho_\pm^2={\cal T}\pm2\delta.
\end{eqnarray}

\medskip
\noindent
Thus the shifted aligned source \(-A\) is converted into a rank--two Bessel kernel, while the determinant, or minor, variable \(\delta\) splits the two \(O(2)\) branches.

\medskip
\noindent
There are now two natural ways of treating the minor sector. The rougher one, which we call simply the \emph{minor approximation}, freezes the shifted aligned source \(-A\) in the same way as the anisotropic term. Equivalently, one sets
\begin{eqnarray}
\ell=0.
\end{eqnarray}
Since \(I_0(0)=1\), the \(O(2)\)-reduced kernel becomes
\begin{eqnarray}
K(\kappa,0;v,w,\Omega)
&=&
\frac12\,e^{\kappa({\cal T}-c)}
\Big[
e^{2\kappa\delta}
+
e^{-2\kappa\delta}
\Big]=
e^{\kappa({\cal T}-c)}
\cosh(2\kappa\delta).
\end{eqnarray}
The leading minor truncation is then obtained by suppressing the determinant variable,
\begin{eqnarray}
\delta\longrightarrow0,
\end{eqnarray}
which gives
\begin{eqnarray}
K(\kappa,0;v,w,\Omega)
\longrightarrow
e^{\kappa({\cal T}-c)}.
\end{eqnarray}
Thus the remaining angular problem is the pure rank--two orthogonal Harish-Chandra--Itzykson--Zuber (HCIZ) integral 
\begin{eqnarray}
{\cal K}_d(\kappa;v,w)
=
e^{-\kappa c}
\int_{O(d)}d\mu(O)\;
e^{\kappa{\cal T}}.
\end{eqnarray}
This approximation keeps only the Frobenius part of the transverse invariant and drops the determinant sector at leading order.

\medskip
\noindent
The orthogonal HCIZ integral provides the natural angular generating function
for invariant integrations over \(O(d)\)
\cite{Harish-Chandra:1957dhy,Itzykson:1979fi,Hikami:2006wya,Brezin:2002mathph,Duistermaat:1982vw,Guhr:2000mathph,Guhr:1997ve}.
Its systematic expansion is given by zonal polynomials, equivalently Jack
polynomials at parameter \(\alpha=2\)
\cite{James:1960AMS,James:1961AMS,James:1965AMS}. This expansion is the
orthogonal analogue of the character expansion of the unitary HCIZ integral,
but the orthogonal case has no simple determinantal formula and does not enjoy
the same WKB-exact localization structure. For this reason, explicit
Haar-moment or low-order zonal expansions are often the most practical route in
applications.

\medskip
\noindent
The second, more refined route is the \emph{improved weak--minor approximation}. The determinant, or minor, sector enters the \(O(2)\)-reduced kernel in two ways: through the explicit factors \(e^{\pm2\kappa\delta}\), and through the Bessel radii \(\rho_\pm\). In the improved minor approximation one suppresses only the explicit \(\kappa\)-dependent determinant splitting,
\begin{eqnarray}
e^{\pm2\kappa\delta}\longrightarrow 1,
\end{eqnarray}
while keeping the two Bessel radii
\begin{eqnarray}
\rho_\pm^2={\cal T}\pm2\delta.
\end{eqnarray}
Thus the minor sector is not completely discarded: it remains encoded in the two-branch Bessel structure.

\medskip
\noindent
The large--argument behavior of the Bessel kernel explains the emergence of alignment, but also shows why the aligned configuration alone is not the full answer. For a large positive argument one has

\begin{eqnarray}
I_0(\ell\rho_\pm)
\sim
\frac{e^{\ell\rho_\pm}}{\sqrt{2\pi\ell\rho_\pm}},
\qquad
\ell\rho_\pm\gg1.
\end{eqnarray}
At fixed Wishart eigenvalues, the leading angular saddle is therefore obtained by maximizing \(\rho_+(\Omega)\). Since the \(2\times2\) block \(\Omega\) is the overlap of two Stiefel frames, it is a contraction, and its singular values are bounded by one. The maximum is reached when the two endpoint two-planes are aligned. After fixing orientation and ordering conventions, this gives
\begin{eqnarray}
\Omega_2=\mathbf 1_2.
\end{eqnarray}
At this aligned point,
\begin{eqnarray}
\rho_+
=
\sqrt{w_1v_1}+\sqrt{w_2v_2},
\qquad
\rho_-
=
\left|\sqrt{w_1v_1}-\sqrt{w_2v_2}\right|.
\end{eqnarray}
Hence the \(\rho_+\) branch is exponentially dominant at an interior radial saddle. In particular, at the symmetric radial saddle \(w_i=v_i\), one has
\begin{eqnarray}
\rho_+=v_1+v_2,
\qquad
\rho_-=|v_1-v_2|.
\end{eqnarray}
For positive interior eigenvalues, \(\rho_+\geq \rho_-\), with strict inequality away from the boundary, so the minus branch is subdominant in the leading large-\(\ell\) exponential.

\medskip
\noindent
Thus the large-\(\ell\) Bessel sector selects the aligned branch. A priori, this is only the leading exponential statement. The exact angular kernel still contains the logarithmic Bessel prefactors, the subleading angular fluctuations, and the \(\rho_-\) branch. As it turns out, these effects do not change the leading aligned/symmetric saddle; rather, they only dress it by a nontrivial angular prefactor.

\subsection{Factorized matrix--Bessel/orthogonal--HCIZ ansatz for the full transverse theory}

\medskip
\noindent
The full rank--two orthogonal matrix--Bessel/orthogonal--HCIZ kernel is obtained after the remaining \(O(d)\) convolution:
\begin{eqnarray}
{\cal K}_d(v,w;\ell,\kappa)
&:=&
\int_{O(d)}d\mu(O)\;
K(\kappa,\ell;v,w,\Omega)
\nonumber\\
&=&
\int_{O(d)}d\mu(O)\;
\frac12\,e^{\kappa({\cal T}-c)}
\Big[
e^{2\kappa\delta}I_0(\ell\rho_+)
+
e^{-2\kappa\delta}I_0(\ell\rho_-)
\Big].
\end{eqnarray}
Thus \(K\) is the exact \(O(2)\)-reduced Bessel kernel at fixed relative Stiefel orientation, whereas \({\cal K}_d\) is the convolved orthogonal Bessel/HCIZ kernel. In terms of this kernel, the shifted quadratic partition function becomes simply
\begin{eqnarray}
Z_\perp^{(2)}(\kappa,\ell)
\propto
\int d\mu_{\rm W}(v,w)\;
{\cal K}_d(v,w; \ell,\kappa).
\end{eqnarray}

\medskip
\noindent
The role of this kernel is clarified by the pure \(-A\) theory. In this sector one sets
\begin{eqnarray}
\kappa=0,
\qquad
\ell=\lambda,
\end{eqnarray}
so that the matrix--Bessel/orthogonal--HCIZ kernel reduces to
\begin{eqnarray}
{\cal K}_d(v,w;\lambda)
=
\int_{O(d)}d\mu(O)\;
\frac12
\Big[
I_0(\lambda\rho_+)
+
I_0(\lambda\rho_-)
\Big].
\end{eqnarray}
This case is exactly solvable in the original Cartesian endpoint variables. Holding \(X_0\) fixed, the Gaussian integral over \(X_1\) gives
\begin{eqnarray}
\int dX_1\;
\exp\!\Big[
-m\,{\rm Tr}(X_1^TX_1)
+\lambda\,{\rm Tr}(X_1^TX_0)
\Big]
=
\left(\frac{\pi}{m}\right)^d
\exp\!\left[
\frac{\lambda^2}{4m}(v_1+v_2)
\right].
\end{eqnarray}
Equivalently, in Wishart--Stiefel variables this implies the exact identity
\begin{eqnarray}
\left(\frac{\pi}{m}\right)^d
\exp\!\left[
\frac{\lambda^2}{4m}(v_1+v_2)
\right]
=
C_d
\int_0^\infty dw_1\,dw_2\;
(w_1w_2)^{\frac{d-3}{2}}
|w_1-w_2|\,
e^{-m(w_1+w_2)}
{\cal K}_d(v,w;\lambda).
\end{eqnarray}

\medskip
\noindent
This identity fixes the leading structure of the orthogonal Bessel/HCIZ kernel. It shows that the angular integral cannot be replaced by the value of the integrand at the aligned configuration alone. The exact kernel must also contain a nontrivial inverse Wishart prefactor, whose role is to cancel the spurious doubled Wishart entropy. Thus, at leading order, the kernel has the structural form
\begin{eqnarray}
{\cal K}_d(v,w;\lambda)
\sim
\frac{
\exp\!\Big[
\lambda(\sqrt{w_1v_1}+\sqrt{w_2v_2})
\Big]
}{
(v_1v_2)^{\frac{d-3}{4}}|v_1-v_2|^{1/2}
(w_1w_2)^{\frac{d-3}{4}}|w_1-w_2|^{1/2}
}
\,
{\cal P}(v,w;\lambda),\label{fpor}
\end{eqnarray}
where \({\cal P}\) contains subleading corrections. Thus the exponential selects the aligned Stiefel configuration, while the denominator supplies the inverse Wishart factor required by the exact Cartesian identity. More precisely, it provides the inverse square root of the two endpoint Wishart blocks, so that near the symmetric locus one full Wishart entropy block is removed.

\medskip
\noindent
We now return to the full transverse holonomy theory. The lesson of the pure \(-A\) model is that the \(O(d)\) angular kernel is not obtained by the naive aligned substitution alone. Instead, the shifted source generates a matrix--Bessel/orthogonal--HCIZ kernel whose leading form contains both the aligned exponential and the inverse Wishart prefactor.

\medskip
\noindent
The exact angular factor is reorganized as follows:
\begin{eqnarray}
&&\int_{O(d)} d\mu(O)\; \int_{O(2)} d\mu(S)\;
\exp\!\Big[\ell\,{\rm tr}\phi+2\kappa \det\phi+\kappa{\cal T}-\kappa c\Big]
\nonumber\\
&=&
\int_{O(d)} d\mu(O)\;e^{\kappa({\cal T}-c)}\;
\frac12\Big[
e^{2\kappa\delta}\,I_0\!\big(\ell\,\rho_+\big)
+
e^{-2\kappa\delta}\,I_0\!\big(\ell\,\rho_-\big)
\Big]
\nonumber\\
&\simeq&
\int_{O(d)} d\mu(O)\;e^{\kappa({\cal T}-c)}\;
\frac12\Big[
\,I_0\!\big(\ell\,\rho_+\big)
+
\,I_0\!\big(\ell\,\rho_-\big)
\Big]\qquad \delta\longrightarrow 0
\nonumber\\
&\simeq&
\Big[\exp\!\big(\kappa({\cal T}-c)\big)\Big]_{\Omega_2=\mathbf 1_2}
\,\mathcal K_d(v,w;\ell).\label{ctcal0}
\end{eqnarray}
In the last line we have employed the minor approximation: the determinant variable has been neglected only in the explicit factors \(e^{\pm2\kappa\delta}\). It is still retained inside the Bessel radii
\begin{eqnarray}
\rho_\pm^2(\Omega)
=
{\cal T}(\Omega)\pm2\delta(\Omega)
=
\sum_{i,j=1}^{2}w_i v_j\,\Omega_{ij}^2
\pm
2\sqrt{w_1w_2v_1v_2}\,\det\Omega .
\end{eqnarray}
Thus the \(\kappa\)-dependent determinant splitting is suppressed, while the nonperturbative Bessel sector still retains the two-branch memory of the minor sector.

\medskip
\noindent
In \eqref{ctcal0}, the \(\kappa\)-dependent HCIZ factor is evaluated at the same aligned saddle selected by the nonperturbative Bessel sector. The latter is kept inside the \(O(d)\) integral and collected into the \(-A\) kernel \({\cal K}_d(v,w;\ell)\) given in \eqref{fpor}. This resums the aligned \(-A\) sector nonperturbatively: the exponential gives the aligned source contribution, while the denominator supplies the inverse Wishart factor fixed by the pure \(-A\) identity.

\medskip
\noindent
All in all, we arrive at the factorized Bessel/HCIZ ansatz
\begin{eqnarray}
&&
\int_{O(d)} d\mu(O)\;
\int_{O(2)} d\mu(S)\;
\exp\!\Big[
\ell\,{\rm tr}\phi
+
2\kappa\det\phi
+
\kappa{\cal T}
-
\kappa c
\Big]
\nonumber\\
&\propto&
e^{-\kappa c}\,
{\cal I}_{\rm HCIZ}(\kappa;v,w)\,
\frac{
\exp\!\Big[
\ell(\sqrt{w_1v_1}+\sqrt{w_2v_2})
\Big]
}{
(v_1v_2)^{\frac{d-3}{4}}|v_1-v_2|^{1/2}
(w_1w_2)^{\frac{d-3}{4}}|w_1-w_2|^{1/2}
},
\end{eqnarray}
where the residual transverse HCIZ factor is
\begin{eqnarray}
{\cal I}_{\rm HCIZ}(\kappa;v,w)
:=
\int_{O(d)}d\mu(O)\;e^{\kappa{\cal T}(\Omega)}
\sim
{\cal P}_{\rm HCIZ}(\kappa;v,w)\,
e^{\kappa{\cal T}_*}.
\label{HCIZ_recalled}
\end{eqnarray}
This is the factorized Bessel/HCIZ ansatz: the Bessel kernel supplies the aligned \(-A\) exponential together with the inverse Wishart--Stiefel prefactor, while \({\cal I}_{\rm HCIZ}\) retains the Haar-resummed transverse \(\kappa\)-interaction.

\subsection{Symmetric--Wishart saddle and the universal \texorpdfstring{\(-2d\)}{-2d} law}

\medskip
\noindent
The factorized Bessel/HCIZ ansatz separates the angular problem into two parts. The nonperturbative Bessel kernel carries the aligned \(-A\) sector and supplies the inverse Wishart--Stiefel prefactor, while the residual HCIZ factor carries the transverse \(\kappa\)-interaction. In the symmetric endpoint sector
\begin{eqnarray}
w_i=v_i=:z_i,
\end{eqnarray}
the aligned Bessel exponential combines with the Gaussian weight to produce the shifted mass
\begin{eqnarray}
m_{\rm eff}
=
m-\frac{\lambda}{2},
\qquad
(\alpha_\Lambda)_{\rm eff}
=
\alpha_\Lambda-\frac12.
\end{eqnarray}
At the same time, the inverse Wishart prefactor removes the spurious doubled entropy and leaves the standard rank--two Wishart/Vandermonde structure. Thus the symmetric saddle is the natural large--\(d\) saddle selected by the resummed angular kernel.

\medskip
\noindent
The resulting symmetric effective action has the schematic form
\begin{eqnarray}
S_{\rm eff}^{\rm sym}(u)
=
2(\alpha_\Lambda)_{\rm eff}u
-(d-2)\log u
+
\sum_{n\ge1}\frac{a_{2n}}{d^n}u^{2n},
\qquad
u=\lambda(z_1+z_2),
\qquad
a_{2n}=O(1).
\end{eqnarray}
A finite truncation of the transverse angular tower suggests the apparent condition
\begin{eqnarray}
(\alpha_\Lambda)_{\rm eff}\gtrsim d^{1/2}.
\end{eqnarray}
However, this is not a physical continuum scaling law. It is only a signal that a finite polynomial expansion in \(B^2\) is inadequate in the continuum regime.

\medskip
\noindent
The correct continuum object is the summed local completion
\begin{eqnarray}
V_{\rm comp}(B)
=
-\log\cosh B
+
\sqrt{R_*^2-B^2}.
\end{eqnarray}
On the Wishart branch this completion drives the reduced dynamics toward a flat cancellation region, rather than toward the small-\(B\) Taylor neighborhood. The detailed construction of this flat region, including the balanced compensating split of the \(+A\) term, was given in the previous paper \cite{YdriUnpublished1}. We only use its consequence here: after this split is imposed, the residual slope at the flat point is cancelled, and the completed transverse sector no longer renormalizes the continuum mass.

\medskip
\noindent
The leading reduced problem therefore returns to the pure Wishart form
\begin{eqnarray}
F_{\rm W}(u)
=
2M u
-
(d-2)\log u,
\qquad
M=\alpha_\Lambda-\frac12.
\end{eqnarray}
Its saddle is
\begin{eqnarray}
u_{\rm W}
=
\frac{d-2}{2M},
\end{eqnarray}
and the \(M\)-dependent part of the saddle free energy is
\begin{eqnarray}
F_{\rm W}(u_{\rm W})
=
(d-2)\log(2M)
+
\hbox{\(M\)-independent terms}.
\end{eqnarray}
Consequently,
\begin{eqnarray}
\log\frac{Z_{\rm W}(x)}{Z_{\rm W}(0)}
=
-(d-2)\log\frac{M(x)}{M(0)}.
\end{eqnarray}
Using
\begin{eqnarray}
M(x)
=
\frac{\mu}{2}
+
\mu x^2
+
O(x^4),
\qquad
\mu=as,
\end{eqnarray}
one obtains
\begin{eqnarray}
\log\frac{Z_{\rm W}(x)}{Z_{\rm W}(0)}
=
-2(d-2)x^2
+
O(x^4).
\end{eqnarray}
Thus the continuum \(D\)-channel entropy coefficient is
\begin{eqnarray}
{\cal C}_{2,D}
=
-2(d-2).
\end{eqnarray}

\medskip
\noindent
The final interpretation is that the symmetric saddle is the correct continuum saddle, while the universal entropy is controlled by the surviving Wishart logarithm. The apparent perturbativity bound is not fundamental; it is reinterpreted as the requirement that the flat-region radius scale with the typical large-\(R\) endpoint scale,
\begin{eqnarray}
R_*
\sim
R_{\rm typ}
\sim
\frac{d}{as}.
\end{eqnarray}

\section{HCIZ/Bessel reductions for lower--dimensional BFSS models}
\label{section3}

\medskip
\noindent
In this section we study lower-dimensional BFSS models within the same
large-\(d\) reduced framework used above. Thus the starting point is not the
full interacting Yang--Mills matrix quantum mechanics, but the corresponding
gauged matrix harmonic oscillator obtained after Gaussian reduction.

\medskip
\noindent
For the lower-dimensional models considered here, we also adopt a controlled
simplification of the holonomy potential by retaining only its quadratic term.
This approximation makes the saddle structure transparent and isolates the
leading mechanism. Higher holonomy corrections are expected to dress this
structure, but not to alter it qualitatively.

\medskip
\noindent For each model we first perform the corresponding HCIZ reduction, then study the
resulting saddle-point structure, and finally incorporate the Bessel convolution
factor. This order of operations is important, because in low dimensions the
HCIZ sector and the endpoint measure may dynamically select boundary saddles
before the Bessel dressing is included.

\subsection{HCIZ/Bessel reduction for BFSS$_4$}
\subsubsection{HCIZ reduction for BFSS$_4$}
\medskip
\noindent
We now treat the \(d=3\) case, corresponding to BFSS\(_4\) in our notation,
directly. In this section we reduce only the angular HCIZ factor; the effect of
the Bessel dressing will be incorporated separately in a later section.

\medskip
\noindent
This case is special because
the endpoint vectors are planar. Hence the endpoint Gram matrices have rank at
most two, and we may write
\begin{eqnarray}
\Lambda_V=\mathrm{diag}(v_1,v_2,0),
\qquad
\Lambda_W=\mathrm{diag}(w_1,w_2,0),
\qquad
v_i,w_i\ge 0.
\end{eqnarray}
The angular invariant is
\begin{eqnarray}
{\cal T}
:=
\mathrm{Tr}\!\big(\Lambda_W O\Lambda_VO^T\big)
=
\sum_{i=1}^{2}\sum_{j=1}^{2}
w_i v_j\,O_{ij}^2,
\qquad
O\in O(3).
\label{T_O3_main}
\end{eqnarray}
Thus only the upper--left \(2\times2\) minor of the orthogonal matrix enters.
This is the basic structural simplification of BFSS\(_4\).

\medskip
\noindent
Since \({\cal T}\) depends only on the squared matrix elements \(O_{ij}^2\), the
integrand is insensitive to the disconnected component of \(O(3)\). Up to an
overall eigenvalue-independent normalization, one may therefore replace the
\(O(3)\) integral by an \(SO(3)\) integral:
\begin{eqnarray}
{\cal I}_{d=3}(\Lambda_W,\Lambda_V)
\propto
\int_{SO(3)}d\mu(O)\;
\exp\!\Big(\kappa_2{\cal T}\Big).
\end{eqnarray}
Using the Euler parametrization
\begin{eqnarray}
O(\alpha,\beta,\gamma)
=
R_z(\alpha)R_y(\beta)R_z(\gamma),
\qquad
0\leq\alpha,\gamma<2\pi,
\qquad
0\leq\beta\leq\pi,
\end{eqnarray}
and setting
\begin{eqnarray}
c:=\cos\beta,
\end{eqnarray}
the upper--left \(2\times2\) block of \(O\) factorizes as
\begin{eqnarray}
A(\alpha,\beta,\gamma)
=
R(\alpha)
\left(\begin{array}{cc}
c&0\\
0&1
\end{array}\right)
R(\gamma).
\label{A_minor_factor_main}
\end{eqnarray}
Equivalently, the two singular values of this minor are
\begin{eqnarray}
1,\qquad |c|.
\end{eqnarray}
The angular variable \(c\) therefore measures the loss of rank of the projected
minor. In particular, at \(c=0\) the minor becomes rank one.

\medskip
\noindent
Writing
\begin{eqnarray}
D_V=\mathrm{diag}(v_1,v_2),
\qquad
D_W=\mathrm{diag}(w_1,w_2),
\end{eqnarray}
the invariant becomes
\begin{eqnarray}
{\cal T}
=
\mathrm{Tr}
\Big(
D_W A D_V A^T
\Big).
\label{T_minor_trace_main}
\end{eqnarray}
It is useful to introduce the combinations
\begin{eqnarray}
\varphi_1=\alpha+\gamma,
\qquad
\varphi_2=\alpha-\gamma.
\end{eqnarray}
Then the invariant takes the form
\begin{eqnarray}
{\cal T}
=
T_0(c)
+
T_1(c)\cos(2\varphi_1)
+
T_2(c)\cos(2\varphi_2)
+
T_-(c)\cos(\varphi_1-\varphi_2)
+
T_+(c)\cos(\varphi_1+\varphi_2),
\label{T_c_resolved_main}
\end{eqnarray}
where, with
\begin{eqnarray}
v_\pm=v_1\pm v_2,
\qquad
w_\pm=w_1\pm w_2,
\end{eqnarray}
one finds
\begin{eqnarray}
T_0(c)
&=&
\frac{w_+v_+}{4}(1+c^2),
\nonumber\\[1mm]
T_1(c)
&=&
\frac{w_-v_-}{8}(1+c)^2,
\qquad
T_2(c)
=
\frac{w_-v_-}{8}(1-c)^2,
\nonumber\\[1mm]
T_-(c)
&=&
-\frac{w_+v_-}{4}(1-c^2),
\qquad
T_+(c)
=
-\frac{w_-v_+}{4}(1-c^2).
\label{T_coefficients_main}
\end{eqnarray}
The derivation of these formulae is elementary but somewhat lengthy, and is
given in Appendix \ref{appendix1}.

\medskip
\noindent
We therefore arrive at the exact \(c\)-resolved HCIZ representation
\begin{eqnarray}
{\cal I}_{d=3}(\Lambda_W,\Lambda_V)
=
\int_{-1}^{1}\frac{dc}{2}\;
{\cal Z}_{\rm ang}(c;\Lambda_W,\Lambda_V),
\label{Id3_c_resolved_main}
\end{eqnarray}
where \({\cal Z}_{\rm ang}\) is the exact two--torus average
\begin{eqnarray}
{\cal Z}_{\rm ang}(c;\Lambda_W,\Lambda_V)
:=
\int_{0}^{2\pi}\frac{d\varphi_1}{2\pi}
\int_{0}^{2\pi}\frac{d\varphi_2}{2\pi}\;
\exp\!\Big(\kappa_2\,{\cal T}(c;\varphi_1,\varphi_2)\Big).
\label{Zang_c_main}
\end{eqnarray}
This formula is the main result of the \(O(3)\) HCIZ reduction. It rewrites the
BFSS\(_4\) angular integral as an exact average over one residual variable
\(c=\cos\beta\), together with a coupled two--torus integral over the angular
variables \(\varphi_1\) and \(\varphi_2\).

\medskip
\noindent
The important point is that the angular sector does not collapse to a single
ordinary Bessel function. The two torus angles remain coupled through the mixed
terms
\begin{eqnarray}
\cos(\varphi_1-\varphi_2),
\qquad
\cos(\varphi_1+\varphi_2).
\end{eqnarray}
Equivalently, applying the Jacobi--Anger expansion to the four angular terms
gives an exact discrete Bessel sum, but not a factorized product of independent
one-angle integrals. This is why the \(d=3\) angular problem has to be treated
separately from the higher-dimensional symmetric--Wishart analysis.

\subsubsection{Exact \(c\)-resolved saddle equations and the BFSS$_4$ boundary branch}

\medskip
\noindent
In the reduced endpoint formulation, the planar endpoint partition function
factorizes into a Wishart--type eigenvalue measure times the holonomy factor.
For general \(d\), we have
\begin{eqnarray}
Z_\perp
&\propto&
\int dv_1\,dv_2\;
e^{-m(v_1+v_2)}
(v_1v_2)^{\frac{d-3}{2}}
|v_1-v_2|
\int dw_1\,dw_2\;
e^{-m(w_1+w_2)}
(w_1w_2)^{\frac{d-3}{2}}
|w_1-w_2|
\nonumber\\
&&\times
{\cal I}_{d}(\Lambda_W,\Lambda_V).
\label{Zperp_general_form}
\end{eqnarray}
Specializing to \(d=3\) removes the factors
\((v_1v_2)^{(d-3)/2}\) and \((w_1w_2)^{(d-3)/2}\). Thus
\begin{eqnarray}
Z_\perp^{(d=3)}
\propto
\int dv_1\,dv_2\;
e^{-m(v_1+v_2)}
|v_1-v_2|
\int dw_1\,dw_2\;
e^{-m(w_1+w_2)}
|w_1-w_2|\;
{\cal I}_{d=3}(\Lambda_W,\Lambda_V).
\label{Zperp_d3_form}
\end{eqnarray}

\medskip
\noindent At this point one may be tempted to integrate out the residual angular variable \(c=\cos\beta\) and
work with the four-dimensional reduced potential
\begin{eqnarray}
V_{\rm eff}^{(4d)}(v,w)
=
m(v_1+v_2)-\log|v_1-v_2|
+
m(w_1+w_2)-\log|w_1-w_2|
-\log{\cal I}_{d=3}(\Lambda_W,\Lambda_V).\nonumber\\
\label{Veff_4d_reduced_main}
\end{eqnarray}
However, as explained in Appendix
\ref{appendix1}, this potential is not a reliable
saddle diagnostic in the weak-\(\kappa_2\) expansion. After \(c\) is integrated
out perturbatively, the splitting directions are non-confining and no finite
interior saddle is obtained.

\medskip
\noindent
Thus the BFSS\(_4\) saddle must be studied in the \(c\)-resolved formulation.
At fixed \(c=\cos\beta\), define the exact angular partition function
\begin{eqnarray}
{\cal Z}_{\rm ang}(c;v,w)
=
\int_{0}^{2\pi}\frac{d\varphi_1}{2\pi}
\int_{0}^{2\pi}\frac{d\varphi_2}{2\pi}\;
\exp\!\big(\kappa_2 {\cal T}(c;v,w,\varphi_1,\varphi_2)\big),
\end{eqnarray}
and the corresponding \(c\)-resolved holonomy potential
\begin{eqnarray}
V_{\rm hol}(c;v,w)
:=
-\log{\cal Z}_{\rm ang}(c;v,w).
\end{eqnarray}
The transverse partition function is then written as the five-dimensional
integral
\begin{eqnarray}
Z_\perp^{(d=3)}
\propto
\int dv_1\,dv_2\,dw_1\,dw_2
\int_{-1}^{1}\frac{dc}{2}\;
\exp\!\Big[-V_{\rm eff}(v,w,c)\Big],
\label{Z5d_main}
\end{eqnarray}
with
\begin{eqnarray}
V_{\rm eff}(v,w,c)
=
m(v_1+v_2)-\log|v_1-v_2|
+
m(w_1+w_2)-\log|w_1-w_2|
+
V_{\rm hol}(c;v,w).
\label{Veff_5d_main}
\end{eqnarray}
Equivalently,
\begin{eqnarray}
V_{\rm eff}(v,w,c)
=
m(v_1+v_2)-\log|v_1-v_2|
+
m(w_1+w_2)-\log|w_1-w_2|
-\log{\cal Z}_{\rm ang}(c;v,w).\label{5D3d}\nonumber\\
\end{eqnarray}

\medskip
\noindent
Introduce the exact angular Gibbs average
\begin{eqnarray}
\langle F\rangle_{c;v,w}
:=
\frac{1}{{\cal Z}_{\rm ang}(c;v,w)}
\int\frac{d\varphi_1}{2\pi}\frac{d\varphi_2}{2\pi}\;
F(\varphi_1,\varphi_2)\,e^{\kappa_2{\cal T}(c;v,w,\varphi_1,\varphi_2)}.
\end{eqnarray}
Then the key identity
\begin{eqnarray}
\partial_\xi V_{\rm hol}(c;v,w)
=
-\kappa_2
\Big\langle
\partial_\xi {\cal T}(c;v,w,\varphi_1,\varphi_2)
\Big\rangle_{c;v,w},
\qquad
\xi\in\{v_1,v_2,w_1,w_2,c\},
\end{eqnarray}
gives the exact saddle equations:

\begin{eqnarray}
0&=&
m-\frac{1}{v_1-v_2}
-\kappa_2
\Big\langle \partial_{v_1}{\cal T}\Big\rangle,\label{saddle_v1_exact}
\\
0&=&
m+\frac{1}{v_1-v_2}
-\kappa_2
\Big\langle \partial_{v_2}{\cal T}\Big\rangle,\label{saddle_v2_exact}
\\
0&=&
m-\frac{1}{w_1-w_2}
-\kappa_2
\Big\langle \partial_{w_1}{\cal T}\Big\rangle,\label{saddle_w1_exact}
\\
0&=&
m+\frac{1}{w_1-w_2}
-\kappa_2
\Big\langle \partial_{w_2}{\cal T}\Big\rangle,\label{saddle_w2_exact}
\\
0&=&
-\kappa_2
\Big\langle \partial_c{\cal T}\Big\rangle.\label{saddle_c_exact}
\end{eqnarray}

\medskip
\noindent Recall that expansion \eqref{T_c_resolved_main} where $T_0$, $T_{i}$ and $T_{\pm}$ are given by equations \eqref{T0}--\eqref{Tuv}--\eqref{T-+} or \eqref{T_coefficients_main}. Then for example

\begin{eqnarray}
\partial_{v_1}{\cal T}
&=&
\frac{w_+}{4}(1+c^2)
+\frac{w_-}{8}(1+c)^2\cos(2\varphi_1)
+\frac{w_-}{8}(1-c)^2\cos(2\varphi_2)
\nonumber\\
&&
-\frac{w_+}{4}(1-c^2)\cos(\varphi_1-\varphi_2)
-\frac{w_-}{4}(1-c^2)\cos(\varphi_1+\varphi_2),
\end{eqnarray}
and similarly for the other derivatives.

\medskip
\noindent
Two remarks are now in order.

\medskip
\noindent
(i) \emph{Exchange symmetry.}
Because ${\cal T}$ is symmetric under exchanging the endpoint spectra,
\begin{eqnarray}
{\cal T}(c;v,w,\varphi_1,\varphi_2)={\cal T}(c;w,v,\varphi_1,\varphi_2),
\end{eqnarray}
the full potential satisfies $V_{\rm eff}(v,w,c)=V_{\rm eff}(w,v,c)$. Hence the saddle system
\eqref{saddle_v1_exact}--\eqref{saddle_c_exact} is invariant under $v\leftrightarrow w$.
Therefore the \emph{diagonal branch}
\begin{eqnarray}
w_i=v_i
\label{diagonal_branch}
\end{eqnarray}
is a consistent exact solution manifold of the stationarity equations (if a saddle exists, its
$v\leftrightarrow w$ image is also a saddle; restricting to the fixed-point set \eqref{diagonal_branch}
is consistent).  This statement is exact and does not use small $\kappa_2$.

\medskip
\noindent However, the exchange symmetry alone does \emph{not} guarantee that the {global} saddle (or dominant region)
lies on the diagonal branch; it only guarantees that saddles come in $v\leftrightarrow w$ pairs,
so one may consistently analyze the diagonal sector as a candidate.

\medskip
\noindent
(ii) \emph{Boundary saddle.}

\medskip
\noindent In $d=2$ the Vandermonde term enforces a boundary saddle at $v_2=0$,
while for $d>3$ it suppresses such configurations.
The case $d=3$ is marginal.
Approaching the boundary with $v_1=x>0$ fixed and $v_2\to0^+$, the Vandermonde piece gives

\begin{eqnarray}
-\log|v_1-v_2|=-\log(x-v_2),
\quad\Rightarrow\quad
\partial_{v_2}\big[-\log(x-v_2)\big]
=
\frac{1}{x-v_2}
\longrightarrow
\frac{1}{x},
\label{vandermonde_derivative_boundary}
\end{eqnarray}
which remains finite.
Because the endpoint measure in $d=3$ carries no $(v_1v_2)^{(d-3)/2}$ prefactor,
there is no $1/v_2$ divergence that would automatically pin the saddle at $v_2=0$.

\medskip
\noindent 
Therefore the boundary ansatz
\begin{eqnarray}
v_2=w_2=0,\qquad v_1=w_1=:x>0
\label{boundary_saddle_ansatz}
\end{eqnarray}
must be analyzed dynamically.
It is a possible saddle (or edge-dominant configuration),
not a consequence of log--repulsion alone.

\medskip
\noindent  Thus, by exchange symmetry we restrict to
\begin{eqnarray}
w_i=v_i,
\end{eqnarray}
and we first analyze the boundary configuration
\begin{eqnarray}
v_2=w_2=0,
\qquad
v_1=w_1=:x>0.
\label{boundary_ansatz_smallk}
\end{eqnarray}

\medskip
\noindent From \eqref{I_O3_expanded_integrand} we have the mall--$\kappa_2$ expansion:

\begin{eqnarray}
{\cal Z}_{\rm ang}
=
1+\kappa_2 T_0
+\kappa_2^2\left(
\frac{T_0^2}{2}
+\frac{T_1^2+T_2^2+T_-^2+T_+^2}{4}
\right)
+O(\kappa_2^3).
\end{eqnarray}
One gets then the corresponding holonomy potential
\begin{eqnarray}
V_{\rm hol}(c;v,w):=-\log{\cal Z}_{\rm ang}(c;v,w)
=
-\kappa_2 T_0
-\frac{\kappa_2^2}{4}\Big(T_1^2+T_2^2+T_-^2+T_+^2\Big)
+O(\kappa_2^3).
\label{Vhol_from_Zang}
\end{eqnarray}

\medskip
\noindent
On the diagonal+boundary ansatz \eqref{boundary_ansatz_smallk} (so $v_\pm=w_\pm=x$),
the coefficients \eqref{T0}--\eqref{Tuv}--\eqref{T-+} become
\begin{eqnarray}
T_0=\frac{x^2}{4}(1+c^2),\quad
T_1=\frac{x^2}{8}(1+c)^2,\quad
T_2=\frac{x^2}{8}(1-c)^2,\quad
T_-=-\frac{x^2}{4}(1-c^2),\quad
T_+=-\frac{x^2}{4}(1-c^2).\nonumber\\
\label{Tcoeff_boundary_diag}
\end{eqnarray}
Therefore
\begin{eqnarray}
V_{\rm hol}(c)\Big|_{\rm bdry}
=
-\frac{\kappa_2 x^2}{4}(1+c^2)
-\frac{\kappa_2^2 x^4}{128}\Big(5-2c^2+5c^4\Big)
+O(\kappa_2^3),
\label{Vhol_boundary_general_c}
\end{eqnarray}
and the five-dimensional potential \eqref{Veff_5d_main} restricted to the diagonal+boundary sector reads
\begin{eqnarray}
V_{\rm eff}(x,c)\Big|_{\rm bdry}
=
2mx-2\log x
-\frac{\kappa_2 x^2}{4}(1+c^2)
-\frac{\kappa_2^2 x^4}{128}\Big(5-2c^2+5c^4\Big)
+O(\kappa_2^3).
\label{Veff_boundary_general_c}
\end{eqnarray}

\medskip
\noindent We now minimize in the $c$ direction. The $c$--dependence of the full
effective potential is
\begin{eqnarray}
V_{\rm eff}(c)\;=\;\text{const}
\;-\;\frac{\kappa_2 x^2}{4}\,c^2
\;+\;\frac{\kappa_2^2 x^4}{64}\,c^2
\;-\;\frac{5\kappa_2^2 x^4}{128}\,c^4
\;+\;O(\kappa_2^3).
\label{Veff_c_expansion}
\end{eqnarray}

\medskip
\noindent
Differentiating gives the stationarity equation
\begin{eqnarray}
\frac{\partial V_{\rm eff}}{\partial c}
=
-\frac{\kappa_2 x^2}{2}\,c
-\frac{\kappa_2^2 x^4}{32}\,c(-1+5c^2)
+O(\kappa_2^3),
\label{dVdc}
\end{eqnarray}
Thus, the interior critical point is $c=0$ (the other algebraic branch
requires $5c^2=1-\frac{16}{\kappa_2 x^2}$ and is absent for small $\kappa_2 x^2$).

\medskip
\noindent
However, for weak coupling (more precisely $\kappa_2 x^2<16$),
the point $c=0$ is an interior \emph{maximum}
in the $c$ direction, not a minimum.

\medskip
\noindent
Since $c\in[-1,1]$ is compact, the global minimum must then lie on the boundary
of the interval, i.e.
\begin{eqnarray}
c_\star=\pm1.
\label{cstar_pm1}
\end{eqnarray}
Equivalently, we can compare values directly:
\begin{eqnarray}
V_{\rm eff}(\pm1)-V_{\rm eff}(0)
=
-\frac{\kappa_2 x^2}{4}
-\frac{\kappa_2^2 x^4}{16}
+O(\kappa_2^3)
\;<\;0,
\label{deltaV_endpoint}
\end{eqnarray}
so the endpoints are energetically preferred at small $\kappa_2$.

\medskip
\noindent
Indeed, when minimizing a potential $V_{\rm eff}(c)$ on a closed interval $c\in[-1,1]$,
an interior minimum satisfies $V'(c_\star)=0$, but a boundary minimum
need not.

\medskip
\noindent
The appropriate necessary conditions are the one--sided
Karush–Kuhn–Tucker (KKT) inequalities:

\begin{eqnarray}
c_\star=1 \ \Rightarrow\ V'(1)\le 0,
\qquad\qquad
c_\star=-1 \ \Rightarrow\ V'(-1)\ge 0.
\label{KKT_1d_interval}
\end{eqnarray}
Equivalently, for any small $\epsilon>0$,
\begin{eqnarray}
V(1)-V(1-\epsilon)\le 0,
\qquad
V(-1)-V(-1+\epsilon)\le 0.
\end{eqnarray}

\medskip
\noindent
Evaluating the derivative at the endpoints gives
\begin{eqnarray}
V'(1)=
-\frac{\kappa_2 x^2}{2}
-\frac{\kappa_2^2 x^4}{8}
+O(\kappa_2^3)\;<\;0,
\qquad
V'(-1)=
+\frac{\kappa_2 x^2}{2}
+\frac{\kappa_2^2 x^4}{8}
+O(\kappa_2^3)\;>\;0,\nonumber\\
\label{endpoint_slopes}
\end{eqnarray}
which is \emph{exactly} the one--sided optimality condition \eqref{KKT_1d_interval}.
Thus $c=\pm1$ are global minimizers on $[-1,1]$ even though $V'(\pm1)\neq 0$.

\medskip
\noindent We can now compute the saddle in $x$ at $c_\star=\pm1$ and the value of the potential. Setting $c^2=1$ in \eqref{Veff_boundary_general_c} gives
\begin{eqnarray}
V_{\rm eff}(x,c_\star)\Big|_{\rm bdry}
=
2mx-2\log x
-\frac{\kappa_2}{2}x^2
-\frac{\kappa_2^2}{16}x^4
+O(\kappa_2^3).
\label{Veff_boundary_c1}
\end{eqnarray}
The $x$--stationarity equation is therefore
\begin{eqnarray}
0=\frac{\partial V_{\rm eff}}{\partial x}
=
2m-\frac{2}{x}-\kappa_2 x-\frac{\kappa_2^2}{4}x^3+O(\kappa_2^3).
\label{x_saddle_eq_c1}
\end{eqnarray}
Solving perturbatively with $x_\star=x_0+\kappa_2 x_1+\kappa_2^2 x_2+\cdots$ yields
\begin{eqnarray}
x_0=\frac{1}{m},\qquad
x_1=\frac{1}{2m^3},\qquad
x_2=\frac{5}{8m^5},
\end{eqnarray}
i.e.
\begin{eqnarray}
x_\star
=
\frac{1}{m}
+\frac{\kappa_2}{2m^3}
+\frac{5\kappa_2^2}{8m^5}
+O(\kappa_2^3).
\label{xstar_series}
\end{eqnarray}
\medskip
\noindent The curvature in $x$ is
\begin{eqnarray}
\frac{\partial^2 V_{\rm eff}}{\partial x^2}
=
\frac{2}{x^2}-\kappa_2-\frac{3\kappa_2^2}{4}x^2+O(\kappa_2^3),
\end{eqnarray}
which is positive at weak coupling when evaluated at $x_\star\sim 1/m$.
Thus the diagonal+boundary saddle is locally stable in $x$ within this ansatz.

\medskip
\noindent Finally, inserting \eqref{xstar_series} into \eqref{Veff_boundary_c1} gives the saddle value of the potential, viz.
\begin{eqnarray}
V_{\rm eff}^\star\Big|_{\rm bdry}
=
2+2\log m
-\frac{\kappa_2}{2m^2}
-\frac{5\kappa_2^2}{16m^4}
+O(\kappa_2^3).
\label{Vstar_series}
\end{eqnarray}

\subsubsection{Bessel dressing of the BFSS$_4$ boundary saddle}

\medskip
\noindent
We now incorporate the Bessel dressing. The important point is that the
\(c\)-resolved HCIZ reduction is structurally unchanged by this dressing. At
the level of the angular HCIZ integrand, the full Bessel factor is
\begin{eqnarray}
{\cal B}_\ell(O;\Lambda_W,\Lambda_V)
=
\frac{1}{2}
\left[
e^{\kappa_2\delta(O)}
I_0\!\Big(\ell\sqrt{\rho_+(O)}\Big)
+
e^{-\kappa_2\delta(O)}
I_0\!\Big(\ell\sqrt{\rho_-(O)}\Big)
\right],
\label{Bessel_factor_full}
\end{eqnarray}
where
\begin{eqnarray}
\rho_\pm(O)
=
{\cal T}(O)\pm 2\delta(O),
\qquad
\delta(O)
=
\sqrt{v_1v_2w_1w_2}\,\det\Omega .
\label{rho_pm_delta_def}
\end{eqnarray}
Here \(\Omega\) is the upper-left \(2\times2\) block of \(O\). In the
\(d=3\) Euler parametrization used above, \(\det\Omega=c\), and hence
\begin{eqnarray}
\delta(c)
=
\sqrt{v_1v_2w_1w_2}\,c.
\end{eqnarray}

\medskip
\noindent
Applying the same \(O(3)\) reduction as in the pure HCIZ case gives
\begin{eqnarray}
{\cal I}_{d=3}^{(\ell)}(\Lambda_W,\Lambda_V)
=
\int_{-1}^{1}\frac{dc}{2}\;
{\cal Z}_{\rm ang}^{(\ell)}(c;v,w),
\end{eqnarray}
with
\begin{eqnarray}
{\cal Z}_{\rm ang}^{(\ell)}(c;v,w)
&=&
\frac{1}{2}
\int_0^{2\pi}\frac{d\varphi_1}{2\pi}
\int_0^{2\pi}\frac{d\varphi_2}{2\pi}\;
\exp\!\Big(\kappa_2{\cal T}(c;v,w,\varphi_1,\varphi_2)\Big)
\nonumber\\
&\times&
\left[
e^{\kappa_2\delta(c)}
I_0\!\Big(\ell\sqrt{\rho_+(c;v,w,\varphi_1,\varphi_2)}\Big)
+
e^{-\kappa_2\delta(c)}
I_0\!\Big(\ell\sqrt{\rho_-(c;v,w,\varphi_1,\varphi_2)}\Big)
\right].\nonumber\\
\label{Zang_Bessel_dressed_full}
\end{eqnarray}
Thus the Bessel factor remains a Bessel factor at the level of the
\(c\)-resolved integrand. It is not replaced by a new HCIZ structure; it is
simply carried along inside the same two-angle angular integral.

\medskip
\noindent
We now specialize to the boundary branch selected by the pure HCIZ analysis,
\begin{eqnarray}
v_2=w_2=0,
\qquad
v_1=w_1=x,
\qquad
c_\star=\pm1.
\label{boundary_branch_Bessel}
\end{eqnarray}
On this branch,
\begin{eqnarray}
\delta=0,
\qquad
\rho_+=\rho_-={\cal T}.
\end{eqnarray}
Therefore the two-branch factor \eqref{Bessel_factor_full} collapses exactly to
a single Bessel factor,
\begin{eqnarray}
{\cal B}_\ell
=
I_0\!\Big(\ell\sqrt{{\cal T}}\Big).
\end{eqnarray}
Thus, for the boundary saddle and entropy calculation, it is sufficient to keep
one representative branch, say \(\rho_+\), with the understanding that
\(\delta=0\) and the \(\rho_-\) branch gives the same contribution.

\medskip
\noindent
On the rank-one boundary branch, the angular invariant simplifies
substantially. Since only \(v_1\) and \(w_1\) are nonzero, the HCIZ invariant is
\begin{eqnarray}
{\cal T}
=
x^2 O_{11}^2.
\end{eqnarray}
Equivalently, after the \(O(3)\) reduction, the two angular variables
\(\varphi_1,\varphi_2\) enter only through the single matrix element
\(O_{11}\). We therefore write
\begin{eqnarray}
t:=O_{11},
\qquad
-1\leq t\leq 1.
\end{eqnarray}
Then
\begin{eqnarray}
{\cal T}=x^2t^2,
\qquad
\sqrt{\rho_+}=x|t|.
\end{eqnarray}
For Haar measure on \(O(3)\), the matrix element \(t=O_{11}\) is uniformly
distributed on \([-1,1]\). Hence the two-angle angular average may be rewritten
as a one-dimensional integral over \(t\):
\begin{eqnarray}
{\cal Z}_{\rm bdry}^{(\ell)}(x)
&=&
\int_{-1}^{1}\frac{dt}{2}\;
\exp\!\big(\kappa_2 x^2t^2\big)
I_0(\ell x|t|)=
\int_0^1 dt\;
\exp\!\big(\kappa_2 x^2t^2\big)
I_0(\ell xt).
\label{Zbdry_Bessel_exact}
\end{eqnarray}
This is the invariant boundary-layer form of the dressed angular integral. The
logarithmic prefactor obtained below comes from this residual Haar angular
average; it would be missed if one froze the angular variable too early.

\medskip
\noindent
For large \(\ell x\), we use the standard asymptotic form
\begin{eqnarray}
I_0(\ell xt)
\sim
\frac{\exp(\ell xt)}{\sqrt{2\pi \ell xt}}.
\end{eqnarray}
Thus
\begin{eqnarray}
{\cal Z}_{\rm bdry}^{(\ell)}(x)
\sim
\int_0^1 dt\;
\frac{
\exp\!\big(\ell xt+\kappa_2x^2t^2\big)
}{
\sqrt{2\pi \ell xt}
}.
\end{eqnarray}
The exponent is increasing on the interval, so the integral is dominated by the
endpoint \(t=1\).

\medskip
\noindent
We shall use the endpoint form of Laplace's method. If
\begin{eqnarray}
{\cal J}
=
\int_a^b dt\;g(t)\,e^{F(t)},
\end{eqnarray}
where \(F(t)\) has its maximum at the endpoint \(b\), with \(F'(b)>0\), then
\begin{eqnarray}
{\cal J}
\sim
\frac{g(b)\,e^{F(b)}}{F'(b)}.
\label{endpoint_Laplace_formula}
\end{eqnarray}
In the present case,
\begin{eqnarray}
F(t)=\ell xt+\kappa_2x^2t^2,
\qquad
g(t)=\frac{1}{\sqrt{2\pi \ell xt}}.
\end{eqnarray}
Therefore
\begin{eqnarray}
F(1)=\ell x+\kappa_2x^2,
\qquad
F'(1)=\ell x+2\kappa_2x^2,
\qquad
g(1)=\frac{1}{\sqrt{2\pi \ell x}}.
\end{eqnarray}
Applying \eqref{endpoint_Laplace_formula}, we obtain
\begin{eqnarray}
{\cal Z}_{\rm bdry}^{(\ell)}(x)
\sim
\frac{
\exp\!\big(\ell x+\kappa_2x^2\big)
}{
\sqrt{2\pi \ell x}\,
\big(\ell x+2\kappa_2x^2\big)
}.
\label{Zbdry_Bessel_asymptotic}
\end{eqnarray}

\medskip
\noindent
Therefore the dressed boundary contribution to the endpoint integral is
\begin{eqnarray}
e^{-2mx}\,x^2\,{\cal Z}_{\rm bdry}^{(\ell)}(x)
\sim
\frac{x^2}{
\sqrt{2\pi \ell x}\,
\big(\ell x+2\kappa_2x^2\big)
}
\exp\!\Big[-(2m-\ell)x+\kappa_2x^2\Big].
\label{boundary_measure_Bessel_combined}
\end{eqnarray}
Here the factor \(x^2\) is precisely the Vandermonde contribution. Indeed, on
the boundary branch \(v_2=w_2=0\), \(v_1=w_1=x\), one has
\begin{eqnarray}
|v_1-v_2|\,|w_1-w_2|
=
x^2.
\label{vandBFSS4}
\end{eqnarray}

\medskip
\noindent
Equation \eqref{boundary_measure_Bessel_combined} displays the two effects of
the Bessel dressing. First, the exponential factor shifts the endpoint mass:
\begin{eqnarray}
e^{-2mx}e^{\ell x}
=
e^{-2m_{\rm eff}x},
\qquad
m_{\rm eff}:=m-\frac{\ell}{2}.
\label{meff_Bessel_shift}
\end{eqnarray}
Second, the Bessel endpoint prefactor changes the logarithmic entropy. In the
source-dominated regime
\begin{eqnarray}
\ell x\gg \kappa_2x^2,
\end{eqnarray}
we have
\begin{eqnarray}
\ell x+2\kappa_2x^2\simeq \ell x,
\end{eqnarray}
and therefore
\begin{eqnarray}
e^{-2mx}\,x^2\,{\cal Z}_{\rm bdry}^{(\ell)}(x)
\sim
{\rm const}\;
x^{1/2}
\exp\!\Big[-2m_{\rm eff}x+\kappa_2x^2\Big].
\label{boundary_Wishart_power}
\end{eqnarray}
Thus the naive Vandermonde power \(x^2\) is reduced by the Bessel endpoint
prefactor to the effective boundary power \(x^{1/2}\). Equivalently, the
dressed boundary potential contains
\begin{eqnarray}
V_{\rm bdry,+}^{(\ell)}(x)
=
2m_{\rm eff}x
-\kappa_2x^2
-\frac{1}{2}\log x
+\cdots .
\label{Vbdry_Bessel_log}
\end{eqnarray}
Hence the dressed boundary measure takes the form
\begin{eqnarray}
d\mu_{\rm bdry,+}^{(\ell)}(x)
\sim
dx\;x^{1/2}
\exp\!\Big[-2m_{\rm eff}x+\kappa_2x^2\Big].
\label{dmu_bdry_Bessel_final}
\end{eqnarray}
Thus the logarithmic coefficient of the dressed boundary measure is
\begin{eqnarray}
q_{\rm bdry,+}=\frac{1}{2}.
\end{eqnarray}

\medskip
\noindent
We finally translate this logarithmic coefficient into the continuum entropy. In
the following ratio, \(x\) denotes the thermal fugacity
\begin{eqnarray}
x=e^{-\beta s},
\end{eqnarray}
where \(s\) is the oscillator frequency. This fugacity should not be confused
with the boundary radial variable \(v_1=w_1=x\) used above.

\medskip
\noindent
Since the boundary measure carries the power
\(q_{\rm bdry,+}=1/2\), its normalized mass dependence is
\begin{eqnarray}
\log\frac{Z_{\rm bdry,+}(x)}{Z_{\rm bdry,+}(0)}
=
-\frac{1}{2}
\log\frac{m_{\rm eff}(x)}{m_{\rm eff}(0)}
+\cdots .
\label{Zbdry_mass_ratio}
\end{eqnarray}
Using
\begin{eqnarray}
m_{\rm eff}(x)
=
m_{\rm eff}(0)\Big(1+2x^2+O(x^4)\Big),
\end{eqnarray}
we obtain
\begin{eqnarray}
\log\frac{Z_{\rm bdry,+}(x)}{Z_{\rm bdry,+}(0)}
=
-x^2+O(x^4).
\label{bdry_entropy_coefficient}
\end{eqnarray}

\medskip
\noindent
This should be compared with the interior symmetric--Wishart saddle. There the
logarithmic coefficient is \(d-2\), so that
\begin{eqnarray}
\log\frac{Z_{\rm W}(x)}{Z_{\rm W}(0)}
=
-(d-2)
\log\frac{M(x)}{M(0)},
\qquad
M(x)=M(0)\Big(1+2x^2+O(x^4)\Big).
\end{eqnarray}
Therefore
\begin{eqnarray}
\log\frac{Z_{\rm W}(x)}{Z_{\rm W}(0)}
=
-2(d-2)x^2+O(x^4),
\qquad
{\cal C}_{2,D}^{\rm W}=-2(d-2).
\label{Wishart_entropy_coefficient}
\end{eqnarray}

\medskip
\noindent
There is no contradiction between the boundary result
\eqref{bdry_entropy_coefficient} and the symmetric--Wishart result
\eqref{Wishart_entropy_coefficient}. The coefficient \(-2(d-2)\) is the entropy
of the interior rank--two symmetric--Wishart saddle, where the two Stiefel
vectors carry \(2(d-2)\) angular directions. In the present BFSS\(_4\) branch,
the HCIZ sector has already driven the endpoint spectrum to rank one. The
angular problem then contains only one nontrivial principal angle, and the
dressed boundary entropy is correspondingly reduced:
\begin{eqnarray}
{\cal C}_{2,D}^{\rm bdry,+}
=
-1.
\end{eqnarray}

\medskip
\noindent
Thus \(d=3\) is exceptional. For \(d>3\), the dominant continuum saddle is the
interior symmetric--Wishart saddle selected by the nonperturbative Bessel
sector, with universal entropy \(-2(d-2)\). In BFSS\(_4\), by contrast, the
HCIZ endpoint geometry fixes the boundary branch first. The Bessel factor still
produces the mass shift \(m\to m_{\rm eff}\), but the entropy is that of a
rank--one boundary angular problem rather than that of the full interior
symmetric--Wishart saddle.

\subsubsection{The Schur--Ingham--Siegel formulation for  BFSS\(_4\)}

\medskip
\noindent
In the Schur--Ingham--Siegel (SIS) reduction, discussed in detail elsewhere
\cite{Ydri2025}, we exploit the fact that for rank--\(2\) endpoint
spectra the invariant

\begin{eqnarray}
{\cal T}
=\mathrm{Tr}(\Lambda_W O\Lambda_V O^T)
\end{eqnarray}
depends only on the $2\times2$ principal block  of $O\in O(d)$,

\begin{eqnarray}
X_{ia}:=O_{ia},\qquad i,a=1,2.
\end{eqnarray}
Explicitly:
\begin{eqnarray}
{\cal T}=\sum_{i,a=1}^2 w_i v_a\,X_{ia}^2
=\mathrm{Tr}(W X V X^T),
\qquad
W=\mathrm{diag}(w_1,w_2),\ \ V=\mathrm{diag}(v_1,v_2).
\label{T_block_form_honest}
\end{eqnarray}
Thus the relevant object is not the full Haar geometry,
but the pushforward of Haar measure to the block $X$:
\begin{eqnarray}
O(d)
\;\longrightarrow\;
X \in \mathbb{R}^{2\times2}.
\end{eqnarray}
\medskip
\noindent
For $d\ge 4$, this pushforward measure is supported on the interior
of the $2\times2$ matrix ball,
\begin{eqnarray}
XX^T < \mathbf 1_2,
\end{eqnarray}
with an induced (marginal) density given by the matrix--ball law
\begin{eqnarray}
\rho_{d,2}(X)\ \propto\ \mathbf 1_{\,XX^T<\mathbf 1_2}\;
\det(\mathbf 1_2-XX^T)^{\frac{d-5}{2}}.
\label{rho_general_d2}
\end{eqnarray}
Therefore
\begin{eqnarray}
{\cal I}_{O(d)}
=
\int dX\;\rho_{d,2}(X)\;
\exp\!\Big(\kappa_2\,\mathrm{Tr}(W X V X^T)\Big).
\label{I_block_marginal_start}
\end{eqnarray}

\medskip
\noindent 
In this regime, the two singular values of $X$,
\begin{eqnarray}
s_1, s_2 \in (0,1),
\end{eqnarray}
are independent invariants.  Geometrically, they are
\begin{eqnarray}
s_i=\cos\theta_i,
\end{eqnarray}
where $(\theta_1,\theta_2)$ are the two principal angles
between the two $2$--planes in $\mathbb R^d$.
Hence for $d\ge4$ there are two independent principal angles,
and correspondingly two independent singular values.

Thus, for $d\ge 4$, the pushforward of Haar to $X$ is absolutely continuous on the
interior of the matrix ball $XX^T<\mathbf 1_2$. In particular, both singular values vary in $(0,1)$ and correspond to two
nontrivial principal angles.

\medskip
\noindent The case $d=3$ is qualitatively different.  For $d\ge4$ the pushforward
is absolutely continuous on the interior $XX^T<\mathbf 1_2$ with density
\eqref{rho_general_d2}.  Formally setting $d=3$ in \eqref{rho_general_d2}
would give an interior weight $\det(\mathbf 1_2-XX^T)^{-1}$, but this is
\emph{not} the correct pushforward measure at $d=3$.

\medskip
\noindent
Indeed, at $d=3$ the image of the map $O(3)\to X$ lies on a boundary stratum.
Using Euler angles on $SO(3)$ one finds
\begin{eqnarray}
X(\alpha,\beta,\gamma)
=
R(\alpha)\,
\left(\!\begin{array}{cc}
c&0\\[1mm]
0&1
\end{array}\!\right)\,
R(\gamma),
\qquad c:=\cos\beta,
\end{eqnarray}
so that
\begin{eqnarray}
\mathrm{singval}(X)=\{1,|c|\},
\qquad
\mathrm{spec}(XX^T)=\{1,c^2\},
\qquad
\det(\mathbf 1_2-XX^T)\equiv0.
\end{eqnarray}
Therefore the pushforward of Haar to $X$ at $d=3$ is a \emph{singular measure}
supported on $s_1=1$ (equivalently $\theta_1=0$), reflecting the geometric fact
that any two $2$--planes in $\mathbb R^3$ intersect along a line:
\begin{eqnarray}
(\theta_1,\theta_2)=(0,\theta),
\qquad
(s_1,s_2)=(1,\cos\theta).
\end{eqnarray}

\medskip
\noindent
Moreover, with $c=\cos\beta$ we have $\sin\beta\,d\beta=-dc$, hence the Haar measure becomes
\begin{eqnarray}
\frac{1}{8\pi^2}\sin\beta\,d\alpha\,d\beta\,d\gamma
=
\frac{1}{8\pi^2}\,d\alpha\,d\gamma\,dc,
\qquad c\in[-1,1].
\label{Haar_to_c}
\end{eqnarray}
Thus, 
\begin{eqnarray}
{\cal I}_{d=3}=
\int_{-1}^{1}\frac{dc}{2}\int_0^{2\pi}\frac{d\alpha}{2\pi}\int_0^{2\pi}\frac{d\gamma}{2\pi}\;
\exp\!\Big(\kappa_2\,\mathrm{Tr}\big(W X(\alpha,c,\gamma)V X(\alpha,c,\gamma)^T\big)\Big).
\end{eqnarray}
This expression, evaluated in \eqref{I_O3_double_Bessel_series_symmetric}
in terms of Bessel functions, is the explicit low--rank reduction at $d=3$:
it isolates the single genuine principal variable $c=\cos\beta$,
together with two in--plane rotations.
It therefore replaces the $2\times2$ auxiliary--matrix formulation in this case.

In fact, the Schur--Ingham--Siegel (SIS) construction does not genuinely
apply at $d=3$, as we explain in Appendix \eqref{appendix2}.

\subsection{HCIZ/Bessel reduction for BFSS$_3$}

\medskip
\noindent
The large--$d$ saddle that produced a term $-2(d-2)x^2$ relied crucially on the
Wishart/Stiefel eigenvalue weight
\begin{eqnarray}
(v_1v_2)^{\frac{d-3}{2}}\;|v_1-v_2|
\end{eqnarray}
with \emph{positive} exponent $(d-3)/2>0$.
In the full large--$d$ endpoint problem this favors the interior region and makes the
standard near--coincident scaling (in particular $v_-=v_1-v_2\to0$ and similarly $w_-\to0$)
self--consistent at the saddle.

\medskip
\noindent
The case \(d=3\), corresponding to BFSS\(_4\), is marginal: the exponent
\((d-3)/2\) vanishes, so the measure neither suppresses nor enhances the
boundary \(v_1v_2=0\). As shown above, this marginal case must be treated by
keeping the full \(c\)-resolved HCIZ integral. The result is that the HCIZ
sector dynamically selects a boundary saddle, which is then dressed by the
Bessel factor.

\medskip
\noindent
At $d=2$ the exponent is negative,
\begin{eqnarray}
\frac{d-3}{2}=-\frac12,
\end{eqnarray}
so the integrand develops an \emph{integrable} singularity on the boundary stratum
where one eigenvalue approaches $0$, i.e. $v_1v_2\to0$ (equivalently $v_+^2-v_-^2\to0$,
or $|v_-|\to v_+$), and likewise for $w$.
This changes the dominant region of integration and invalidates the naive
``$v_-=w_-=0$ saddle'' logic.
In particular, one must compute the $O(2)$ angular integral \emph{directly} and only then
build the effective potential.

\subsubsection{HCIZ reduction for BFSS$_3$}

\medskip
\noindent For $d=2$ there are no zero eigenvalues:
\begin{eqnarray}
\Lambda_V=\mathrm{diag}(v_1,v_2),\qquad 
\Lambda_W=\mathrm{diag}(w_1,w_2),\qquad v_i,w_i\ge 0.
\end{eqnarray}
Write $O(2)=SO(2)\rtimes\mathbb Z_2$.  For diagonal $\Lambda_{V,W}$, the reflection
component produces the same invariant ${\cal T}:=\mathrm{Tr}(\Lambda_W O\Lambda_V O^T)$
as the rotation component (because ${\cal T}$ depends only on squared entries of $O$),
so the Haar integral reduces (up to an overall $v,w$--independent factor) to the
$SO(2)$ average.  Parameterize
\begin{eqnarray}
O(\phi)=
\left(
\begin{array}{cc}
\cos\phi & -\sin\phi\\
\sin\phi & \cos\phi
\end{array}
\right),
\qquad 0\le\phi<2\pi,
\end{eqnarray}
and compute
\begin{eqnarray}
{\cal T}(\phi)
&=&
\mathrm{Tr}\Big(\Lambda_W O(\phi)\Lambda_V O(\phi)^T\Big)\nonumber\\
&=&
\frac12(w_1+w_2)(v_1+v_2)\;+\;\frac12(w_1-w_2)(v_1-v_2)\cos(2\phi).
\label{Tphi_d2}
\end{eqnarray}
Therefore the $d=2$ HCIZ integral is \emph{exactly}
\begin{eqnarray}
{\cal I}_{d=2}(\Lambda_W,\Lambda_V)
\;\propto\;
\int_0^{2\pi}\frac{d\phi}{2\pi}\;
\exp\!\Big(\kappa_2\,{\cal T}(\phi)\Big)
=
\exp\!\Big(\frac{\kappa_2}{2}(w_1+w_2)(v_1+v_2)\Big)\;
I_0\!\Big(\frac{\kappa_2}{2}(w_1-w_2)(v_1-v_2)\Big).\nonumber\\
\label{I_O2_exact_again}
\end{eqnarray}

\medskip
\noindent
Introduce $v_\pm:=v_1\pm v_2$ and $w_\pm:=w_1\pm w_2$.  Then
\begin{eqnarray}
{\cal I}_{d=2}\;\propto\;
\exp\!\Big(\frac{\kappa_2}{2}w_+v_+\Big)\;
I_0\!\Big(\frac{\kappa_2}{2}w_-v_-\Big).
\label{I_O2_vpmwpm}
\end{eqnarray}

\medskip
\noindent
Geometrically, this reflects the fact that for \(d=2\) the rank--\(2\)
endpoint support fills the whole ambient space. Thus there is no nontrivial
Grassmannian principal angle between two endpoint two-planes:
\(Gr(2,2)\) is a point. The remaining angular variable \(\phi\) is instead the
relative \(O(2)\) eigenbasis angle between \(\Lambda_V\) and \(\Lambda_W\), and
the entire angular average collapses to a single Bessel function \(I_0\).

\medskip
\noindent Let us now expand \(I_0\) to quadratic order:

\begin{eqnarray}
I_0(x)=1+\frac{x^2}{4}+O(x^4),
\qquad x=\frac{\kappa_2}{2}\,w_-v_-.
\end{eqnarray}
Then
\begin{eqnarray}
{\cal I}_{d=2}
&\propto&
\Big(1+\frac{\kappa_2}{2}w_+v_+ +\frac{\kappa_2^2}{8}w_+^2v_+^2+O(\kappa_2^3)\Big)
\Big(1+\frac{\kappa_2^2}{16}w_-^2v_-^2+O(\kappa_2^4)\Big)\nonumber\\
&=&
1+\frac{\kappa_2}{2}\,w_+v_+
+\kappa_2^2\Big(\frac{1}{8}w_+^2v_+^2+\frac{1}{16}w_-^2v_-^2\Big)
+O(\kappa_2^3).
\label{IO2_expand_vpm}
\end{eqnarray}

\medskip
\noindent
On the other hand, the general zonal-moment formula obtained elsewhere
\cite{Ydri2025} has the following structure:

\begin{eqnarray}
{\cal I}_{d}
=
1+\frac{\kappa_2}{d}\,S_1^W S_1^V
+\frac{\kappa_2^2}{2}\;
\frac{
(d+1)(S_1^W)^2(S_1^V)^2
-2(S_1^W)^2 S_2^V
-2(S_1^V)^2 S_2^W
+2d\,S_2^W S_2^V
}{
d(d-1)(d+2)
}
+O(\kappa_2^3).\nonumber\\
\label{zonal_prime_struct}
\end{eqnarray}
Specializing to $d=2$ gives $d(d-1)(d+2)=8$, $(d+1)=3$, $2d=4$, hence
\begin{eqnarray}
{\cal I}_{d=2}
=
1+\frac{\kappa_2}{2}\,S_1^W S_1^V
+\frac{\kappa_2^2}{16}\Big[
3(S_1^W)^2(S_1^V)^2
-2(S_1^W)^2 S_2^V
-2(S_1^V)^2 S_2^W
+4\,S_2^W S_2^V
\Big]
+O(\kappa_2^3).\nonumber\\
\label{zonal_prime_d2}
\end{eqnarray}
The spectral invariants are defined by 
\begin{eqnarray}
S_1^V=v_1+v_2=v_+,\qquad S_2^V=v_1^2+v_2^2=\frac{v_+^2+v_-^2}{2},
\qquad
S_1^W=w_+,\qquad S_2^W=\frac{w_+^2+w_-^2}{2}.\nonumber\\
\end{eqnarray}
Inserting these definitions into the bracket yields the identity
\begin{eqnarray}
3(S_1^W)^2(S_1^V)^2
-2(S_1^W)^2 S_2^V
-2(S_1^V)^2 S_2^W
+4\,S_2^W S_2^V
=
2w_+^2v_+^2+w_-^2v_-^2.
\label{d2_bracket_identity}
\end{eqnarray}
Therefore \eqref{zonal_prime_d2} becomes
\begin{eqnarray}
{\cal I}_{d=2}
=
1+\frac{\kappa_2}{2}\,w_+v_+
+\kappa_2^2\Big(\frac{1}{8}w_+^2v_+^2+\frac{1}{16}w_-^2v_-^2\Big)
+O(\kappa_2^3),
\end{eqnarray}
which matches the direct expansion \eqref{IO2_expand_vpm} of the \emph{exact} $O(2)$ integral.

\subsubsection{Effective potential and boundary saddle for BFSS$_3$}

\medskip
\noindent 
At \(d=2\), the Stiefel/Wishart measure for each Gram block is, up to constants,
\begin{eqnarray}
d\mu_V 
\;\propto\; 
(v_1v_2)^{-1/2}\,|v_1-v_2|\,dv_1\,dv_2,
\qquad
d\mu_W 
\;\propto\; 
(w_1w_2)^{-1/2}\,|w_1-w_2|\,dw_1\,dw_2 .
\end{eqnarray}
In the variables \(v_\pm=v_1\pm v_2\) and \(w_\pm:=w_1\pm w_2\), this becomes
\begin{eqnarray}
d\mu_V 
\;\propto\; 
\frac{|v_-|}{\sqrt{v_+^2-v_-^2}}\;dv_+\,dv_-,
\qquad 
{\cal D}_v:\quad v_+\ge 0,\quad |v_-|\le v_+,
\end{eqnarray}
and
\begin{eqnarray}
d\mu_W 
\;\propto\; 
\frac{|w_-|}{\sqrt{w_+^2-w_-^2}}\;dw_+\,dw_-,
\qquad
{\cal D}_w:\quad w_+\ge 0,\quad |w_-|\le w_+ .
\end{eqnarray}
Including the Gaussian endpoint factor \(e^{-m(v_+ + w_+)}\) and the exact
\(O(2)\) angular integral, the reduced endpoint integral reads
\begin{eqnarray}
Z_\perp^{(d=2)}
\;\propto\;
\int_{{\cal D}_v} dv_+\,dv_-\;
\int_{{\cal D}_w} dw_+\,dw_-\;
\exp\!\Big[-V_{\rm eff}^{(2)}(v_\pm,w_\pm)\Big],
\end{eqnarray}
with effective potential
\begin{eqnarray}
V_{\rm eff}^{(2)}(v_\pm,w_\pm)
&=&
m(v_+ + w_+)
+\frac12\log(v_+^2-v_-^2)-\log|v_-|
+\frac12\log(w_+^2-w_-^2)-\log|w_-|
\nonumber\\
&-&
\frac{\kappa_2}{2}\,w_+v_+
-\log I_0\!\Big(\frac{\kappa_2}{2}\,w_-v_-\Big).
\label{Veff_d2_full_exact}
\end{eqnarray}
Here all additive constants independent of \((v_\pm,w_\pm)\) have been dropped,
and the domain is \(v_+\ge0\), \(|v_-|\le v_+\), and similarly for \(w_\pm\).

\medskip
\noindent
In the low--temperature regime, the Bessel argument
\begin{eqnarray}
z:=\frac{\kappa_2}{2}w_-v_-
\end{eqnarray}
is small, and one may use
\begin{eqnarray}
\log I_0(z)=\frac{z^2}{4}+O(z^4).
\end{eqnarray}
This gives
\begin{eqnarray}
V_{\rm eff}^{(2)}(v_\pm,w_\pm)
&=&
m(v_+ + w_+)
+\frac12\log(v_+^2-v_-^2)-\log|v_-|
+\frac12\log(w_+^2-w_-^2)-\log|w_-|
\nonumber\\
&-&
\frac{\kappa_2}{2}w_+v_+
-\frac{\kappa_2^2}{16}\,w_-^2v_-^2
+O(z^4).
\label{Veff_d2_smallz}
\end{eqnarray}

\medskip
\noindent
For $d=2$ the eigenvalue domain which is
\begin{eqnarray}
v_+\ge0,\qquad |v_-|\le v_+,
\qquad
w_+\ge0,\qquad |w_-|\le w_+,
\end{eqnarray}
corresponds to
\begin{eqnarray}
v_1=\frac{v_+ + v_-}{2}\ge0,
\qquad
v_2=\frac{v_+ - v_-}{2}\ge0,
\end{eqnarray}
and similarly for $w$.

\medskip
\noindent
The boundary of the domain corresponds to collapse of one eigenvalue:
\begin{eqnarray}
v_+^2-v_-^2=4v_1v_2\to0
\qquad\Longleftrightarrow\qquad
|v_-|\to v_+,
\end{eqnarray}
and likewise for $w$.

\medskip
\noindent
At $d=2$ the Stiefel/Wishart measure contains the factor
\begin{eqnarray}
  (v_1v_2)^{-1/2}
\propto
\frac{1}{\sqrt{v_+^2-v_-^2}},
\end{eqnarray}
which produces an \emph{integrable singularity} when
$v_+^2-v_-^2\to0$.
Thus the path integral is strongly enhanced near the boundary where
one eigenvalue becomes small.

\medskip
\noindent
Thus, because the measure develops an integrable singularity as
$v_2\to0$ and $w_2\to0$, the path integral remains finite but is strongly
enhanced near the boundary. As a result, the dominant contribution arises from
a boundary layer near
\begin{eqnarray}
|v_-|\to v_+,\qquad |w_-|\to w_+.
\end{eqnarray}
This is therefore a boundary saddle, driven by entropic enhancement of the measure rather than by a conventional
interior stationary point.

\medskip
\noindent
We pick the branch
\begin{eqnarray}
v_2\simeq0,\qquad w_2\simeq0,
\qquad\Longleftrightarrow\qquad
v_-=v_+,\qquad w_-=w_+,
\label{d2_boundary_branch}
\end{eqnarray}
(the other branch $v_1\simeq0$ corresponds to $v_-=-v_+$ and is treated analogously).
However, one must keep the transverse boundary-layer integrations rather than
simply set the boundary values from the outset.

\medskip
\noindent
Write
\begin{eqnarray}
v_-=v_+-\varepsilon_v,\qquad
w_-=w_+-\varepsilon_w,
\qquad
\varepsilon_v,\varepsilon_w\ge0.
\end{eqnarray}
Then, for \(\varepsilon_v,\varepsilon_w\to0^+\),
\begin{eqnarray}
v_+^2-v_-^2
=
\varepsilon_v(2v_+-\varepsilon_v)
\sim
2v_+\varepsilon_v,
\qquad
w_+^2-w_-^2
=
\varepsilon_w(2w_+-\varepsilon_w)
\sim
2w_+\varepsilon_w,
\end{eqnarray}
and therefore
\begin{eqnarray}
\frac{|v_-|}{\sqrt{v_+^2-v_-^2}}\,dv_-
\sim
\sqrt{\frac{v_+}{2\varepsilon_v}}\,d\varepsilon_v,
\qquad
\frac{|w_-|}{\sqrt{w_+^2-w_-^2}}\,dw_-
\sim
\sqrt{\frac{w_+}{2\varepsilon_w}}\,d\varepsilon_w.
\end{eqnarray}
Taking the boundary layer to have fixed fractional thickness,
\begin{eqnarray}
0\le \varepsilon_v \le \eta_v v_+,
\qquad
0\le \varepsilon_w \le \eta_w w_+,
\qquad
\eta_v,\eta_w\ll1,
\end{eqnarray}
with \(\eta_v,\eta_w\) independent of \((v_+,w_+)\), gives
\begin{eqnarray}
\int_0^{\eta_v v_+}
d\varepsilon_v\,
\sqrt{\frac{v_+}{2\varepsilon_v}}
\propto v_+,
\qquad
\int_0^{\eta_w w_+}
d\varepsilon_w\,
\sqrt{\frac{w_+}{2\varepsilon_w}}
\propto w_+.
\end{eqnarray}
Thus, after absorbing the \((v_+,w_+)\)-independent constants into the normalization,
the boundary layer induces the effective measure
\begin{eqnarray}
d\mu_{\rm bdry}\propto v_+\,w_+\;dv_+\,dw_+.
\end{eqnarray}

\medskip
\noindent
On \eqref{d2_boundary_branch} the \(I_0\) argument becomes
\begin{eqnarray}
z:=\frac{\kappa_2}{2}\,w_-v_-
=
\frac{\kappa_2}{2}\,w_+v_+,
\end{eqnarray}
and the effective boundary problem is therefore
\begin{eqnarray}
V_{\rm bdry}(v_+,w_+)
=
m(v_+ + w_+)
-\log v_+
-\log w_+
-\frac{\kappa_2}{2}\,w_+v_+
-\log I_0\!\Big(\frac{\kappa_2}{2}\,w_+v_+\Big),
\label{Vbdry2_exact}
\end{eqnarray}
up to \((v_+,w_+)\)--independent constants.

\medskip
\noindent
The exact saddle equations are obtained using
\begin{eqnarray}
\frac{d}{dz}\log I_0(z)=\frac{I_1(z)}{I_0(z)}.
\end{eqnarray}
Differentiating \eqref{Vbdry2_exact} gives
\begin{eqnarray}
0=\partial_{v_+}V_{\rm bdry}:\qquad
m-\frac{1}{v_+}
-\frac{\kappa_2}{2}\,w_+
-\frac{\kappa_2}{2}\,w_+\,
\frac{I_1(z)}{I_0(z)}=0,
\label{d2_exact_saddle_v}
\\
0=\partial_{w_+}V_{\rm bdry}:\qquad
m-\frac{1}{w_+}
-\frac{\kappa_2}{2}\,v_+
-\frac{\kappa_2}{2}\,v_+\,
\frac{I_1(z)}{I_0(z)}=0.
\label{d2_exact_saddle_w}
\end{eqnarray}

\medskip
\noindent
By symmetry the saddle lies on the diagonal \(v_+=w_+=u\), for which
\begin{eqnarray}
z=\frac{\kappa_2}{2}\,u^2,
\end{eqnarray}
and the exact saddle equation becomes
\begin{eqnarray}
m-\frac{1}{u}
-\frac{\kappa_2}{2}\,u\,
\Big(1+\frac{I_1(z)}{I_0(z)}\Big)=0,
\qquad z=\frac{\kappa_2}{2}\,u^2.
\label{d2_exact_diagonal_saddle}
\end{eqnarray}
This determines \(u_\star(m,\kappa_2)\) implicitly and is valid for all
\(\kappa_2\ge0\) for which the boundary saddle exists.

\medskip
\noindent
The on--shell boundary free energy is likewise exact:
\begin{eqnarray}
V_{\rm bdry}(u_\star,u_\star)
=
2mu_\star-2\log u_\star
-\frac{\kappa_2}{2}\,u_\star^2
-\log I_0\!\Big(\frac{\kappa_2}{2}\,u_\star^2\Big)
\;+\;(\text{\(m,\kappa_2\)--indep. const}).
\label{d2_exact_onshell}
\end{eqnarray}

\medskip
\noindent
We now work in the low--temperature regime and expand the Bessel ratio
and logarithm at small argument. We use
\begin{eqnarray}
\frac{I_1(z)}{I_0(z)}
=
\frac{z}{2}+O(z^3),
\qquad
\log I_0(z)=\frac{z^2}{4}+O(z^4).
\label{Bessel_smallz_identities}
\end{eqnarray}
Inserting \eqref{Bessel_smallz_identities} into the exact diagonal saddle
equation \eqref{d2_exact_diagonal_saddle} gives, up to \(O(\kappa_2^2)\),
\begin{eqnarray}
mu-1-\frac{\kappa_2}{2} u^2-\frac{\kappa_2^2}{8}u^4
+O(\kappa_2^3)=0.
\label{d2_smallk_poly}
\end{eqnarray}

\medskip
\noindent
Solving perturbatively with
\begin{eqnarray}
u_\star=u_0+\kappa_2 u_1+\kappa_2^2 u_2+O(\kappa_2^3),
\end{eqnarray}
one finds
\begin{eqnarray}
u_\star
=
\frac{1}{m}
+\frac{\kappa_2}{2m^3}
+\frac{5\kappa_2^2}{8m^5}
+O(\kappa_2^3).
\label{d2_u_star_series}
\end{eqnarray}

\bigskip
\noindent
Also, expanding \eqref{Vbdry2_exact} on the diagonal \(v_+=w_+=u\) using
\eqref{Bessel_smallz_identities} yields
\begin{eqnarray}
V_{\rm bdry}(u,u)
&=&
2mu-2\log u
-\frac{\kappa_2}{2}u^2
-\frac{\kappa_2^2}{16}u^4
+O(\kappa_2^4)
+(\text{const}).
\label{d2_Vbdry_smallk}
\end{eqnarray}

\medskip
\noindent
Evaluating this at the saddle \eqref{d2_u_star_series} gives
\begin{eqnarray}
V_{\rm bdry}^\star
&=&
2+2\log m
-\frac{\kappa_2}{2m^2}
-\frac{5\kappa_2^2}{16m^4}
+O(\kappa_2^3)
+(\text{const}),
\label{d2_Vstar_series}
\end{eqnarray}
where the additive constant is independent of \((m,\kappa_2)\) and may be dropped.

\subsubsection{Bessel convolution, mass shift, and logarithmic dressing for BFSS$_3$}

\medskip
\noindent
We now include the Bessel convolution factor. In the rank--\(2\) problem it has
the schematic form
\begin{eqnarray}
{\cal B}_\ell(O;\Lambda_W,\Lambda_V)
=
\frac12\left[
e^{\kappa_2\delta(O)}I_0\!\Big(\ell\sqrt{\rho_+(O)}\Big)
+
e^{-\kappa_2\delta(O)}I_0\!\Big(\ell\sqrt{\rho_-(O)}\Big)
\right],
\end{eqnarray}
where
\begin{eqnarray}
\rho_\pm(O)={\cal T}(O)\pm2\delta(O),
\qquad
\delta(O)=\sqrt{v_1v_2w_1w_2}\,\det O .
\end{eqnarray}
For the \(d=2\) boundary branch
\begin{eqnarray}
v_2=w_2=0,
\qquad
v_1=v_+,\qquad
w_1=w_+,
\end{eqnarray}
one has
\begin{eqnarray}
\delta(O)=0,
\qquad
\rho_+(O)=\rho_-(O)={\cal T}(O).
\end{eqnarray}
Thus the two Bessel branches collapse to a single factor
\begin{eqnarray}
{\cal B}_\ell(O;\Lambda_W,\Lambda_V)
=
I_0\!\Big(\ell\sqrt{{\cal T}(O)}\Big).
\end{eqnarray}

\medskip
\noindent
On the boundary branch the exact \(O(2)\) invariant is
\begin{eqnarray}
{\cal T}(\phi)=w_+v_+\cos^2\phi .
\end{eqnarray}
Therefore the Bessel-dressed boundary angular integral is
\begin{eqnarray}
{\cal Z}_{\rm bdry}^{(\ell)}(v_+,w_+)
=
\int_0^{2\pi}\frac{d\phi}{2\pi}\;
\exp\!\Big(\kappa_2 w_+v_+\cos^2\phi\Big)\,
I_0\!\Big(\ell\sqrt{w_+v_+}\,|\cos\phi|\Big).
\label{Zbdry_d2_ell_exact}
\end{eqnarray}
This is the \(d=2\) analogue of the BFSS\(_4\) \(c\)-resolved Bessel dressing.
The difference is that in BFSS\(_3\) the angular variable is the relative
\(O(2)\) eigenbasis angle, not a Grassmannian principal angle.

\medskip
\noindent
For large Bessel source, or more precisely for
\begin{eqnarray}
\ell\sqrt{w_+v_+}\gg1,
\end{eqnarray}
the integral is dominated by the aligned angular endpoints
\begin{eqnarray}
|\cos\phi|=1.
\end{eqnarray}
Writing
\begin{eqnarray}
a:=\sqrt{w_+v_+},
\end{eqnarray}
and using
\begin{eqnarray}
I_0(\ell a|\cos\phi|)
\sim
\frac{\exp(\ell a|\cos\phi|)}
{\sqrt{2\pi\ell a|\cos\phi|}},
\end{eqnarray}
one obtains, up to \(a\)-independent constants,
\begin{eqnarray}
{\cal Z}_{\rm bdry}^{(\ell)}(v_+,w_+)
\sim
\frac{
\exp\!\Big(\ell a+\kappa_2 a^2\Big)
}{
\sqrt{\ell a}\,
\sqrt{\ell a+2\kappa_2 a^2}
},
\qquad
a=\sqrt{w_+v_+}.
\label{Zbdry_d2_ell_asympt}
\end{eqnarray}

\medskip
\noindent
The most important consequence is the linear term
\begin{eqnarray}
\exp(\ell a).
\end{eqnarray}
Near the symmetric boundary saddle \(v_+=w_+=u\), one has \(a=u\), and hence
\begin{eqnarray}
e^{-m(v_+ + w_+)}\,e^{\ell\sqrt{w_+v_+}}
\;\longrightarrow\;
e^{-2mu+\ell u}
=
e^{-2m_{\rm eff}u},
\qquad
m_{\rm eff}=m-\frac{\ell}{2}.
\label{d2_mass_shift}
\end{eqnarray}
Thus the Bessel convolution shifts the effective endpoint mass exactly as in
the BFSS\(_4\) boundary branch.

\medskip
\noindent
There is, however, an additional logarithmic effect. Before the Bessel
convolution, the boundary measure gave
\begin{eqnarray}
d\mu_{\rm bdry}\propto v_+w_+\,dv_+\,dw_+,
\end{eqnarray}
which on the symmetric diagonal contributes the logarithmic term
\begin{eqnarray}
-2\log u
\end{eqnarray}
to the boundary potential. The endpoint asymptotics
\eqref{Zbdry_d2_ell_asympt} adds the prefactor
\begin{eqnarray}
\frac{1}{
\sqrt{\ell u}\,
\sqrt{\ell u+2\kappa_2u^2}
}
\end{eqnarray}
on the diagonal. Therefore the Bessel-dressed diagonal boundary potential is
\begin{eqnarray}
V_{\rm bdry}^{(\ell)}(u)
&\simeq&
2mu-\ell u
-2\log u
-\kappa_2u^2
+\frac12\log(\ell u)
+\frac12\log(\ell u+2\kappa_2u^2)
+(\text{const}) .
\label{Vbdry_d2_ell_general}
\end{eqnarray}

\medskip
\noindent
In the source-dominated regime
\begin{eqnarray}
\ell u\gg \kappa_2u^2,
\end{eqnarray}
this reduces to
\begin{eqnarray}
V_{\rm bdry}^{(\ell)}(u)
\simeq
2m_{\rm eff}u
-\log u
-\kappa_2u^2
+(\text{const}),
\qquad
m_{\rm eff}=m-\frac{\ell}{2}.
\label{Vbdry_d2_ell_source_dominated}
\end{eqnarray}

\medskip
\noindent
Thus the leading Bessel effect is the mass shift
\begin{eqnarray}
m\longrightarrow m_{\rm eff}=m-\frac{\ell}{2}.
\end{eqnarray}
At the same time, the endpoint Bessel prefactor dresses the logarithmic
coefficient. In the source-dominated endpoint regime one finds
\begin{eqnarray}
-2\log u
\quad\longrightarrow\quad
-\log u .
\end{eqnarray}
Equivalently, the Bessel-dressed boundary measure behaves as
\begin{eqnarray}
d\mu_{\rm bdry}^{(\ell)}
\sim
du\;u\,
\exp\!\Big(-2m_{\rm eff}u+\kappa_2u^2\Big),
\qquad
\ell u\gg\kappa_2u^2.
\label{d2_bessel_dressed_boundary_measure}
\end{eqnarray}

\medskip
\noindent
This should be compared with the large--\(d\) Wishart result
\begin{eqnarray}
\log\frac{Z_W(x)}{Z_W(0)}
=
-2(d-2)x^2+O(x^4),
\label{large_d_Wishart_entropy_again}
\end{eqnarray}
where \(x=e^{-\beta s}\) is the thermal fugacity. For \(d=3\), corresponding
to BFSS\(_4\), the naive continuation of \eqref{large_d_Wishart_entropy_again}
would give \(-2x^2\), whereas the boundary HCIZ/Bessel analysis gave instead
\(-x^2\). The coefficient was therefore halved by the boundary/Bessel mechanism.

\medskip
\noindent
The BFSS\(_3\) case is even more singular, since the analytic continuation of
\eqref{large_d_Wishart_entropy_again} to \(d=2\) gives zero. Thus BFSS\(_3\)
is not described by the interior Wishart entropy. Rather, it is controlled by
the boundary layer of the \(d=2\) measure. Before the Bessel convolution, the
boundary measure produces
\begin{eqnarray}
V_{\rm bdry}(u)\supset -2\log u .
\end{eqnarray}

\medskip
\noindent The endpoint Bessel prefactor dresses this logarithmic term as
\begin{eqnarray}
-2\log u
\quad\longrightarrow\quad
-\log u
\end{eqnarray}
in the source-dominated regime. Evaluated at the boundary saddle, one has
\(u_\star\sim 1/m_{\rm eff}\), up to numerical constants independent of the
thermal fugacity. Hence the logarithmic term becomes
\begin{eqnarray}
-\log u_\star
=
\log m_{\rm eff}
+
(\hbox{\(x\)-independent constant}).
\end{eqnarray}
Therefore the normalized boundary partition function contains
\begin{eqnarray}
\log\frac{Z_{\rm bdry}^{(\ell)}(x)}{Z_{\rm bdry}^{(\ell)}(0)}
=
-\log\frac{m_{\rm eff}(x)}{m_{\rm eff}(0)}.
\end{eqnarray}
Using
\begin{eqnarray}
m_{\rm eff}(x)=m_{\rm eff}(0)\Big(1+2x^2+O(x^4)\Big),
\end{eqnarray}
we obtain
\begin{eqnarray}
\log\frac{Z_{\rm bdry}^{(\ell)}(x)}{Z_{\rm bdry}^{(\ell)}(0)}
=
-2x^2+O(x^4),
\end{eqnarray}
and therefore
\begin{eqnarray}
{\cal C}_{2,D}^{\rm bdry}=-2.
\end{eqnarray}
This coefficient is halved relative to the undressed \(d=2\) boundary-layer
measure, where the logarithmic term would have been \(-2\log u\). Hence the BFSS\(_3\) boundary entropy is also
halved, but now relative to its own boundary-layer measure rather than relative
to the smooth large--\(d\) Wishart formula.

\subsection{HCIZ/Bessel reduction for BFSS$_2$ (\(d=1\))}

\medskip
\noindent
At \(d=1\) we have rank collapse, since there is only a single endpoint
species index \(a=1\). Hence each endpoint Gram matrix has only one nonzero
eigenvalue:
\begin{eqnarray}
\Lambda_V=(v),
\qquad
\Lambda_W=(w),
\qquad
v,w\ge0.
\end{eqnarray}
Thus the rank--\(2\) discussion, with two eigenvalues, a Vandermonde
\(|v_1-v_2|\), and principal angles between two-planes, no longer applies.
Geometrically, there is no nontrivial two-plane geometry in
\(\mathbb{R}^1\); the relative Grassmannian and principal-angle structure have
collapsed completely.

\medskip
\noindent
Since
\begin{eqnarray}
O(1)=\{+1,-1\},
\qquad
d\mu(O)=\frac12\sum_{O=\pm1},
\end{eqnarray}
and since the invariant depends only on \(O_{11}^2=1\), one has
\begin{eqnarray}
{\cal T}
=
\mathrm{Tr}\!\big(\Lambda_W O\Lambda_V O^T\big)
=
wv\,O_{11}^2
=
wv.
\end{eqnarray}
Therefore the HCIZ factor is exactly
\begin{eqnarray}
{\cal I}_{O(1)}(\Lambda_W,\Lambda_V)
=
\int_{O(1)}d\mu(O)\;e^{\kappa_2{\cal T}}
=
e^{\kappa_2wv}.
\label{I_O1_exact}
\end{eqnarray}
Its small--\(\kappa_2\) expansion is simply
\begin{eqnarray}
{\cal I}_{O(1)}
=
1+\kappa_2wv+\frac{\kappa_2^2}{2}(wv)^2+O(\kappa_2^3).
\end{eqnarray}
There is no meaningful specialization of the rank--\(2\) zonal-moment formula,
obtained in \cite{Ydri2025}, to \(d=1\). At this point the
representation-theoretic structure changes type: the rank itself has collapsed.

\medskip
\noindent
The endpoint measure is thus purely radial. The endpoint data are two real
two-vectors
\begin{eqnarray}
V^\mu,\;W^\mu\in\mathbb{R}^2,
\qquad
\mu=1,2,
\end{eqnarray}
with scalar Gram invariants
\begin{eqnarray}
v:=|V|^2,
\qquad
w:=|W|^2.
\end{eqnarray}
Thus, in polar coordinates,
\begin{eqnarray}
d^2V=r\,dr\,d\theta=\frac12\,dv\,d\theta,
\qquad
d^2W=\rho\,d\rho\,d\phi=\frac12\,dw\,d\phi.
\end{eqnarray}
After integrating out the angular variables, and omitting overall constants,
\begin{eqnarray}
d\mu_V\propto dv,
\qquad
d\mu_W\propto dw,
\qquad
v,w\in[0,\infty).
\end{eqnarray}

\medskip
\noindent
Inserting the exact HCIZ factor gives the reduced endpoint integral
\begin{eqnarray}
Z_\perp^{(d=1)}
\;\propto\;
\int_0^\infty dv\int_0^\infty dw\;
\exp\!\Big[-m(v+w)+\kappa_2vw\Big].
\label{Zperp_d1_start}
\end{eqnarray}
\medskip
\noindent
For fixed $w$, the $v$--integral converges whenever $\Re(m-\kappa_2 w)>0$, and formally gives
\begin{eqnarray}
\int_0^\infty dv\;e^{-(m-\kappa_2 w)v}=\frac{1}{m-\kappa_2 w}.
\end{eqnarray}
Thus
\begin{eqnarray}
Z_\perp^{(d=1)}
\;\propto\;
\int_0^\infty dw\;\frac{e^{-mw}}{m-\kappa_2 w}.
\label{Zperp_d1_exact_form}
\end{eqnarray}

\medskip
\noindent
For positive \(\kappa_2\), the above integral \eqref{Zperp_d1_start} should be understood in the same
low--temperature perturbative sense as before: the bilinear term is expanded
before the large-field instability becomes relevant. Thus, for \(m^2\gg
\kappa_2\),
\begin{eqnarray}
Z_\perp^{(d=1)}
&\propto&
\int_0^\infty dv\int_0^\infty dw\;
e^{-m(v+w)}
\left[
1+\kappa_2vw+\frac{\kappa_2^2}{2}v^2w^2+\cdots
\right]
\nonumber\\
&=&
\frac{1}{m^2}
\left[
1+\frac{\kappa_2}{m^2}
+2\frac{\kappa_2^2}{m^4}
+O\!\left(\frac{\kappa_2^3}{m^6}\right)
\right].
\label{Zperp_d1_series}
\end{eqnarray}
Hence
\begin{eqnarray}
-\log Z_\perp^{(d=1)}
=
2\log m+\text{const}
+O\!\left(\frac{\kappa_2}{m^2}\right).
\end{eqnarray}

\medskip
\noindent
We now include the Bessel convolution. Since the rank has collapsed to one,
there is no determinant branch and no pair of \(\rho_\pm\) invariants. The
Bessel factor reduces directly to
\begin{eqnarray}
{\cal B}_\ell(v,w)
=
I_0\!\Big(\ell\sqrt{vw}\Big).
\end{eqnarray}
Thus, setting \(\kappa_2=0\) momentarily in order to isolate the Bessel effect,
the dressed radial integral is
\begin{eqnarray}
Z_\perp^{(d=1,\ell)}
\;\propto\;
\int_0^\infty dv\int_0^\infty dw\;
e^{-m(v+w)}\,
I_0\!\Big(\ell\sqrt{vw}\Big).
\label{Zperp_d1_bessel_kappa0}
\end{eqnarray}
Using
\begin{eqnarray}
I_0\!\Big(\ell\sqrt{vw}\Big)
=
\sum_{n\ge0}
\frac{1}{(n!)^2}
\left(\frac{\ell^2vw}{4}\right)^n,
\end{eqnarray}
one obtains
\begin{eqnarray}
Z_\perp^{(d=1,\ell)}
&\propto&
\sum_{n\ge0}
\frac{1}{(n!)^2}
\left(\frac{\ell^2}{4}\right)^n
\left(\int_0^\infty dv\,v^n e^{-mv}\right)
\left(\int_0^\infty dw\,w^n e^{-mw}\right)
\nonumber\\
&=&
\frac{1}{m^2}
\sum_{n\ge0}
\left(\frac{\ell^2}{4m^2}\right)^n
=
\frac{1}{m^2-\ell^2/4},
\qquad |\ell|<2m.
\label{Zperp_d1_bessel_exact}
\end{eqnarray}
Therefore
\begin{eqnarray}
-\log Z_\perp^{(d=1,\ell)}
=
\log\!\left(m^2-\frac{\ell^2}{4}\right)
+\text{const}
=
\log\!\left(m-\frac{\ell}{2}\right)
+
\log\!\left(m+\frac{\ell}{2}\right)
+\text{const}.
\label{d1_log_mass_threshold}
\end{eqnarray}
Thus the Bessel convolution replaces the undressed radial logarithm
\(2\log m\) by a threshold logarithm. Defining
\begin{eqnarray}
m_{\rm eff}:=m-\frac{\ell}{2},
\end{eqnarray}
the singular source-dominated part is
\begin{eqnarray}
-\log Z_\perp^{(d=1,\ell)}
=
\log m_{\rm eff}
+
\text{smooth terms}.
\end{eqnarray}
The second factor, \(\log(m+\ell/2)\), is smooth at the threshold and may be
absorbed into the normalization when one focuses on the endpoint singularity.

Thus the leading Bessel effect is again the endpoint mass shift
\begin{eqnarray}
m\longrightarrow m_{\rm eff}=m-\frac{\ell}{2}.
\end{eqnarray}
At the same time, the Bessel convolution also halves the logarithmic mass
coefficient. Indeed, without the Bessel factor one finds
\begin{eqnarray}
-\log Z_\perp^{(d=1)}
=
2\log m+\text{const}
+O\!\left(\frac{\kappa_2}{m^2}\right),
\end{eqnarray}
where the coefficient \(2\) comes from the two radial endpoint integrations.
After the Bessel convolution is included, the exact rank--one radial integral
at \(\kappa_2=0\) gives
\begin{eqnarray}
Z_\perp^{(d=1,\ell)}
\propto
\frac{1}{m^2-\ell^2/4},
\end{eqnarray}
and therefore
\begin{eqnarray}
-\log Z_\perp^{(d=1,\ell)}
=
\log\!\left(m^2-\frac{\ell^2}{4}\right)
+\text{const}
=
\log\!\left(m-\frac{\ell}{2}\right)
+
\log\!\left(m+\frac{\ell}{2}\right)
+\text{const}.
\end{eqnarray}
Near the endpoint threshold, the singular part is therefore
\begin{eqnarray}
-\log Z_\perp^{(d=1,\ell)}
=
\log m_{\rm eff}
+\text{smooth terms},
\qquad
m_{\rm eff}=m-\frac{\ell}{2}.
\end{eqnarray}
Thus the logarithmic coefficient is reduced from \(2\log m\) to
\(\log m_{\rm eff}\), in the same sense that the BFSS\(_4\) and BFSS\(_3\)
boundary logarithms were halved by the Bessel endpoint prefactor. Consequently,
if the thermal fugacity dependence enters through
\begin{eqnarray}
m_{\rm eff}(x)
=
m_{\rm eff}(0)\Big(1+2x^2+O(x^4)\Big),
\end{eqnarray}
then
\begin{eqnarray}
\log\frac{Z_\perp^{(d=1,\ell)}(x)}
{Z_\perp^{(d=1,\ell)}(0)}
=
-\log\frac{m_{\rm eff}(x)}{m_{\rm eff}(0)}
=
-2x^2+O(x^4).
\end{eqnarray}
Therefore the BFSS\(_2\) radial endpoint coefficient is
\begin{eqnarray}
{\cal C}_{2,D}^{\rm rad}=-2.
\end{eqnarray}

\medskip
\noindent
The lower-dimensional sequence therefore exhibits three distinct degenerations.
For \(d=3\), the rank--\(2\) angular sector collapses onto a boundary stratum in
the \(2\times2\) block picture, leaving a rank--one, one-principal-angle
boundary geometry. For \(d=2\), the relative Grassmannian itself collapses, but
the singular Stiefel/Wishart measure leaves a nontrivial boundary layer. For
\(d=1\), the rank itself collapses: there is only one endpoint eigenvalue in
each sector, no Vandermonde, and no principal-angle variable. Thus BFSS\(_2\)
is a pure rank--one radial endpoint problem. The HCIZ factor is simply
\eqref{I_O1_exact}, while the Bessel convolution reduces to the rank--one
factor \(I_0(\ell\sqrt{vw})\). Its effect is again twofold: it shifts the
endpoint mass to \(m_{\rm eff}=m-\ell/2\), and it reduces the radial logarithmic
mass coefficient from \(2\log m\) to the threshold logarithm
\(\log m_{\rm eff}\). This logarithm is not Grassmannian in origin; it is the
rank--one radial remnant of the endpoint measure.

\section{Emergent Grassmannian geometry}\label{section4}

\subsection{Relative Grassmannian modes and principal angles}

\medskip
\noindent
It is useful to clarify the geometry behind the coefficient
\(-2(d-2)\) which appears in the symmetric--Wishart entropy. The endpoint
formulation does not merely produce two principal angles. Rather, it produces a relative Grassmannian angular sector. The principal
angles are the invariant singular-value coordinates on this sector, while the
remaining Grassmannian variables describe the angular degeneracies of these
invariants.

\medskip
\noindent
The relevant geometric object is the space of two-planes in \(\mathbb{R}^d\).
The space of orthonormal two-frames is the Stiefel manifold
\begin{eqnarray}
V_2(\mathbb{R}^d)
=
\frac{O(d)}{O(d-2)},
\qquad
\dim V_2(\mathbb{R}^d)=2d-3.
\end{eqnarray}
Quotienting by the internal \(O(2)\) rotation of the frame gives the
Grassmannian
\begin{eqnarray}
Gr(2,d)
=
\frac{O(d)}{O(2)\times O(d-2)},
\qquad
\dim Gr(2,d)=2(d-2).
\label{Gr2d_def}
\end{eqnarray}
Equivalently, for oriented planes one may replace \(O\) by \(SO\); the local
dimension is unchanged.

\medskip
\noindent
In the two-endpoint problem one should think of two two-planes,
say \(P_V\) and \(P_W\), embedded in \(\mathbb{R}^d\). The full pair belongs to
\begin{eqnarray}
Gr(2,d)\times Gr(2,d).
\end{eqnarray}
However, the global \(O(d)\) symmetry allows us to fix one of them, say
\begin{eqnarray}
P_V=P_0=\mathrm{span}(e_1,e_2).
\end{eqnarray}
Then the second plane \(P_W\) is described by a point of \(Gr(2,d)\). Thus,
after removing the common global orientation, \(Gr(2,d)\) is the relative
configuration space of the two endpoint two-planes. This is analogous to the
reduction of a two-body problem to a relative coordinate.

\medskip
\noindent
Near the aligned saddle \(P_W=P_V=P_0\), a nearby two-plane may be written as a
graph over \(P_0\),
\begin{eqnarray}
e_1\longrightarrow e_1+u,
\qquad
e_2\longrightarrow e_2+v,
\qquad
u,v\in P_0^\perp .
\end{eqnarray}
Since
\begin{eqnarray}
\dim P_0^\perp=d-2,
\end{eqnarray}
the two transverse vectors \(u\) and \(v\) contain \(2(d-2)\) real components.

\medskip
\noindent 
In invariant language,
\begin{eqnarray}
T_{P_0}Gr(2,d)
\simeq
\mathrm{Hom}(P_0,P_0^\perp),
\end{eqnarray}
where \(\mathrm{Hom}(P_0,P_0^\perp)\) denotes the vector space of linear maps
from \(P_0\) to its orthogonal complement. Thus an infinitesimal tilt of
\(P_0\) is a map
\begin{eqnarray}
A:P_0\longrightarrow P_0^\perp .
\end{eqnarray}
If \(P_0=\mathrm{span}(e_1,e_2)\), then
\begin{eqnarray}
A(e_1)=u,
\qquad
A(e_2)=v,
\qquad
u,v\in P_0^\perp .
\end{eqnarray}
Hence an element of the tangent space may be represented as a
\((d-2)\times2\) matrix
\begin{eqnarray}
A=
\begin{pmatrix}
| & | \\
u & v \\
| & |
\end{pmatrix}
\in \mathbb{R}^{(d-2)\times2}.
\label{A_grassmannian_tangent}
\end{eqnarray}
This is the origin of the number \(2(d-2)\): it counts the entries of the
linear deformation map \(A\), or equivalently the local relative angular
fluctuation modes around the aligned saddle.

\medskip
\noindent
The principal angles are not all the entries of \(A\). They are the singular
values of \(A\). For \(d\geq4\), the matrix \(A\) can have rank two, and one may
write locally
\begin{eqnarray}
A
=
U
\begin{pmatrix}
\theta_1 & 0 \\
0 & \theta_2 \\
0 & 0 \\
\vdots & \vdots
\end{pmatrix}
R,
\end{eqnarray}
where \(\theta_1,\theta_2\) are the two principal angles. The remaining variables in \(U\in SO(d-2)\) and \(R\in SO(2)\) determine how
the principal-angle directions are embedded. The matrix \(U\) selects the
transverse directions in \(P_0^\perp\), while \(R\) selects the orthonormal
directions inside the reference plane \(P_0\). These variables are not new
principal angles; they are the angular degeneracies of the principal-angle data
inside the full relative Grassmannian measure.

\medskip
\noindent
It is useful to make this explicit in BFSS\(_6\), where \(d=5\). In this case
\begin{eqnarray}
Gr(2,5)
=
\frac{O(5)}{O(2)\times O(3)},
\qquad
\dim Gr(2,5)=2(5-2)=6.
\end{eqnarray}
Fix the reference plane
\begin{eqnarray}
P_0=\mathrm{span}(e_1,e_2),
\end{eqnarray}
so that
\begin{eqnarray}
P_0^\perp=\mathrm{span}(e_3,e_4,e_5).
\end{eqnarray}
A nearby plane is described by a linear map
\begin{eqnarray}
A:P_0\longrightarrow P_0^\perp,
\qquad
A\in\mathbb{R}^{3\times2}.
\end{eqnarray}
Using a singular-value decomposition, we may write locally
\begin{eqnarray}
A
=
U
\begin{pmatrix}
\tan\vartheta_1 & 0 \\
0 & \tan\vartheta_2 \\
0 & 0
\end{pmatrix}
R,
\label{A_BFSS6_SVD}
\end{eqnarray}
where \(U\in SO(3)\) acts in the transverse space \(P_0^\perp\), and
\(R\in SO(2)\) acts inside the reference plane \(P_0\). The two quantities
\(\vartheta_1,\vartheta_2\) are the two principal angles. Infinitesimally,
\(\tan\vartheta_i\simeq\vartheta_i\).

\medskip
\noindent
Thus the six Grassmannian variables may be counted as
\begin{eqnarray}
\underbrace{\vartheta_1,\vartheta_2}_{2\ \hbox{\scriptsize principal angles}}
\quad+\quad
\underbrace{U\in SO(3)}_{3\ \hbox{\scriptsize transverse orientation angles}}
\quad+\quad
\underbrace{R\in SO(2)}_{1\ \hbox{\scriptsize in-plane orientation angle}}.
\end{eqnarray}
Equivalently,
\begin{eqnarray}
6=2+3+1.
\end{eqnarray}
The three angles in \(U\) choose the oriented two-frame in the transverse
three-dimensional space \(P_0^\perp\). The angle in \(R\) chooses which
orthonormal directions inside the original plane \(P_0\) are tilted by
\(\vartheta_1\) and \(\vartheta_2\). These four variables are therefore not new
principal angles. They are the angular degeneracies specifying how the two
principal tilts are embedded inside \(\mathbb{R}^5\).

\medskip
\noindent
In this explicit BFSS\(_6\) example, the endpoint/Wishart angular sector is
therefore a six-dimensional relative Grassmannian geometry:
\begin{eqnarray}
Gr(2,5):
\qquad
6
=
\underbrace{2}_{\hbox{\scriptsize invariant principal angles}}
+
\underbrace{4}_{\hbox{\scriptsize orientation degeneracies}}.
\end{eqnarray}
The symmetric--Wishart entropy counts all six local angular fluctuation modes,
not only the two principal-angle invariants.

\medskip
\noindent
Thus one should distinguish between
\begin{eqnarray}
\hbox{principal-angle invariants}
\qquad\hbox{and}\qquad
\hbox{Grassmannian angular modes}.
\end{eqnarray}
For two two-planes there are at most two principal angles, but the relative
Grassmannian tangent space has dimension \(2(d-2)\). For \(d\geq4\), this
dimension may be decomposed as
\begin{eqnarray}
2(d-2)
=
\underbrace{2}_{\hbox{\scriptsize principal angles}}
+
\underbrace{1}_{\hbox{\scriptsize in-plane }SO(2)}
+
\underbrace{(2d-7)}_{\hbox{\scriptsize transverse orientation angles}}.
\end{eqnarray}
The universal in-plane \(SO(2)\) angle specifies which orthonormal directions
inside the reference plane \(P_0\) realize the two principal angles. The
remaining \(2d-7\) variables are transverse: they specify the orthonormal
two-frame inside
\begin{eqnarray}
P_0^\perp\simeq \mathbb{R}^{d-2},
\end{eqnarray}
namely a point of
\begin{eqnarray}
V_2(\mathbb{R}^{d-2})
=
\frac{SO(d-2)}{SO(d-4)}.
\end{eqnarray}
These variables are angular degeneracies, not additional principal angles.

\medskip
\noindent
The coefficient \(2(d-2)\) in the entropy is therefore not the number of
principal angles. It is the number of real Gaussian angular modes around the
aligned saddle.

\medskip
\noindent
This also clarifies the relation to the endpoint/Wishart variables. The reduced
endpoint data split as
\begin{eqnarray}
2d
=
\underbrace{4}_{v_1,v_2,w_1,w_2}
+
\underbrace{2(d-2)}_{\hbox{\scriptsize relative Grassmannian modes}}.
\label{endpoint_grassmannian_count}
\end{eqnarray}
The four variables \(v_1,v_2,w_1,w_2\) are the Wishart radial eigenvalues, while
the remaining \(2(d-2)\) variables form the relative angular Grassmannian
sector. The two principal angles are the invariant "radial" coordinates on this
relative Grassmannian; the remaining variables are their transverse angular
degeneracies.

%\medskip
%\noindent
%The reduced-mass analogy is also helpful. If both planes are allowed to move
%near an aligned reference plane \(P_0\), we may write
%\begin{eqnarray}
%P_1=P_0+A_1,
%\qquad
%P_2=P_0+A_2,
%\qquad
%A_1,A_2\in \mathrm{Hom}(P_0,P_0^\perp).
%\end{eqnarray}
%Then
%\begin{eqnarray}
%A_{\rm cm}=\frac{A_1+A_2}{2},
%\qquad
%A_{\rm rel}=A_2-A_1.
%\end{eqnarray}
%The variable \(A_{\rm cm}\) is the common orientation of the pair inside
%\(\mathbb{R}^d\). It is the analogue of the center-of-mass motion and is removed
%by the global \(O(d)\) symmetry, up to an overall group volume. The variable
%\(A_{\rm rel}\) is the relative angular motion. It is precisely the
%Grassmannian fluctuation that controls the Wishart entropy.

%\medskip
%\noindent
%At the aligned saddle,
%\begin{eqnarray}
%A_{\rm rel}=0,
%\qquad
%P_V=P_W.
%\end{eqnarray}
%The saddle itself is a point in the relative Grassmannian. However, the entropy
%is controlled by the Gaussian volume of fluctuations around that point. Since
%there are \(2(d-2)\) real angular modes, the Gaussian determinant gives
%\begin{eqnarray}
%Z_{\rm W}
%\sim
%M^{-\frac{1}{2}\,2(d-2)}
%=
%M^{-(d-2)}.
%\end{eqnarray}
%Using
%\begin{eqnarray}
%M(x)=M(0)\Big(1+2x^2+O(x^4)\Big),
%\end{eqnarray}
%one obtains
%\begin{eqnarray}
%\log\frac{Z_{\rm W}(x)}{Z_{\rm W}(0)}
%=
%-(d-2)\log\frac{M(x)}{M(0)}
%=
%-2(d-2)x^2+O(x^4).
%\label{Wishart_entropy_from_Grassmannian}
%\end{eqnarray}
%Thus
%\begin{eqnarray}
%{\cal C}_{2,D}^{\rm W}=-2(d-2).
%\end{eqnarray}
%This coefficient is the entropy of the relative Grassmannian tangent space
%around the aligned saddle.

\subsection{Lower-dimensional degenerations of the endpoint geometry}

\medskip
\noindent
The case \(d=3\), corresponding to BFSS\(_4\), is degenerate. Then
\begin{eqnarray}
A\in \mathbb{R}^{1\times2},
\qquad
A=(a\;\;b),
\end{eqnarray}
so \(A\) has only one singular value,
\begin{eqnarray}
\sigma=\sqrt{a^2+b^2}.
\end{eqnarray}
Equivalently, two planes in \(\mathbb{R}^3\) always intersect in a line, and
only one principal angle can be nonzero. Thus the relative Grassmannian
\(Gr(2,3)\) is two-dimensional, but its invariant principal-angle data is only
one-dimensional. The second local variable is the angular degeneracy of the
single tilt direction.

\medskip
\noindent
The case \(d=2\), corresponding to BFSS\(_3\), is even more degenerate. In this
case the endpoint two-plane fills the whole ambient space,
\begin{eqnarray}
Gr(2,2)=\{\mathrm{pt}\},
\qquad
\dim Gr(2,2)=0.
\end{eqnarray}
Thus there is no nontrivial relative Grassmannian motion and no principal angle
between two endpoint two-planes. Equivalently,
\begin{eqnarray}
T_{P_0}Gr(2,2)
\simeq
\mathrm{Hom}(P_0,P_0^\perp)=0,
\qquad
P_0^\perp=0.
\end{eqnarray}
The angular variable which appears in the \(O(2)\) HCIZ integral is therefore
not a Grassmannian principal angle. It is instead the relative in-plane
eigenbasis angle between the two endpoint spectra \(\Lambda_V\) and
\(\Lambda_W\).

\medskip
\noindent
For \(d=1\), corresponding to BFSS\(_2\), the degeneration is complete: the
rank itself collapses. There is no endpoint two-plane at all. Each endpoint
Gram matrix has only one nonzero eigenvalue,
\begin{eqnarray}
\Lambda_V=(v),
\qquad
\Lambda_W=(w),
\end{eqnarray}
and the relative Grassmannian language no longer applies. There is no
Vandermonde factor, no principal-angle variable, and no Grassmannian angular
entropy. The endpoint problem becomes purely radial.

\medskip
\noindent
For \(d=4\),
\begin{eqnarray}
A\in\mathbb{R}^{2\times2},
\end{eqnarray}
and the full rank--two principal-angle geometry appears for the first time.
This is why \(d=4\) is the first dimension in which two nonzero principal
angles are genuinely independent. For \(d>4\), the number of principal angles
remains two, but their transverse degeneracy grows with \(d\).

\medskip
\noindent
The important point is that the reduction of the entropy in \(d=3\) is not
accidental. It reflects the degeneration of the relative Grassmannian geometry.
For \(d>3\), the relevant continuum saddle is the interior symmetric--Wishart
saddle. The relative angular space is \(Gr(2,d)\), whose tangent space has
dimension \(2(d-2)\), and whose invariant data contain two nontrivial principal
angles. This gives the universal Wishart entropy
\begin{eqnarray}
{\cal C}_{2,D}^{\rm W}
=
-2(d-2).
\end{eqnarray}

\medskip
\noindent
For \(d=3\), however, the relative Grassmannian is \(Gr(2,3)\). Since two
planes in \(\mathbb{R}^3\) always intersect in a line, there is only one
nontrivial principal angle. In the present formulation this one-angle geometry
is realized dynamically: the HCIZ sector first drives the endpoint problem to
the boundary branch,
\begin{eqnarray}
v_2=w_2=0,
\qquad
v_1=w_1=r,
\end{eqnarray}
and the Bessel endpoint asymptotics then produces the dressed boundary entropy
\begin{eqnarray}
{\cal C}_{2,D}^{\rm bdry,+}
=
-1.
\end{eqnarray}
Thus the coefficient \(-1\) is the \(d=3\) one-principal-angle analogue of the
higher-dimensional coefficient \(-2(d-2)\).

\medskip
\noindent
For \(d=2\), the analytic continuation of the large--\(d\) Wishart entropy gives
\begin{eqnarray}
{\cal C}_{2,D}^{\rm W}\big|_{d=2}=0,
\end{eqnarray}
because the interior relative Grassmannian sector has collapsed. BFSS\(_3\) is
therefore not governed by the interior Wishart saddle. Instead, the negative
Stiefel/Wishart exponent,
\begin{eqnarray}
\frac{d-3}{2}=-\frac12,
\end{eqnarray}
produces an integrable boundary singularity, and the endpoint dynamics is
controlled by a boundary layer in which one eigenvalue collapses,
\begin{eqnarray}
v_2=w_2=0,
\qquad
v_1=w_1=u.
\end{eqnarray}
Before the Bessel convolution, this boundary layer produces
\begin{eqnarray}
V_{\rm bdry}(u)\supset -2\log u .
\end{eqnarray}
After the Bessel endpoint prefactor is included, the source-dominated boundary
problem gives instead
\begin{eqnarray}
V_{\rm bdry}^{(\ell)}(u)\supset -\log u .
\end{eqnarray}
Thus the BFSS\(_3\) logarithmic entropy is also halved by the Bessel endpoint
prefactor, but now relative to its own boundary-layer saddle rather than
relative to the smooth large--\(d\) Grassmannian entropy.

\medskip
\noindent
For \(d=1\), the rank--two endpoint structure itself has disappeared. The HCIZ
factor becomes trivial,
\begin{eqnarray}
{\cal I}_{O(1)}=e^{\kappa_2vw},
\end{eqnarray}
and the Bessel convolution reduces to the rank--one factor
\begin{eqnarray}
I_0\!\Big(\ell\sqrt{vw}\Big).
\end{eqnarray}
Without the Bessel factor, the two radial endpoint integrations give
\begin{eqnarray}
-\log Z_\perp^{(d=1)}
=
2\log m+\text{const}
+\cdots .
\end{eqnarray}
After the Bessel convolution is included, the exact rank--one radial integral
at \(\kappa_2=0\) gives the threshold logarithm
\begin{eqnarray}
-\log Z_\perp^{(d=1,\ell)}
=
\log\!\left(m^2-\frac{\ell^2}{4}\right)
+\text{const}.
\end{eqnarray}
Thus, near the endpoint threshold \(m_{\rm eff}=m-\ell/2\), the singular part is
\begin{eqnarray}
-\log Z_\perp^{(d=1,\ell)}
=
\log m_{\rm eff}
+
\text{smooth terms}.
\end{eqnarray}
The Bessel convolution therefore again shifts the mass and halves the
logarithmic coefficient. The difference is that in BFSS\(_2\) this logarithm is
purely radial, not Grassmannian in origin.

\medskip
\noindent
The resulting hierarchy may be summarized as
\begin{eqnarray}
d>3:
\qquad
\hbox{interior rank--two Grassmannian saddle},
\qquad
{\cal C}_{2,D}^{\rm W}=-2(d-2),
\end{eqnarray}
\begin{eqnarray}
d=3:
\qquad
\hbox{one-principal-angle boundary degeneration},
\qquad
{\cal C}_{2,D}^{\rm bdry,+}=-1,
\end{eqnarray}
\begin{eqnarray}
d=2:
\qquad
\hbox{collapsed Grassmannian and boundary-layer saddle},
\qquad
{\cal C}_{2,D}^{\rm bdry}=-2,
\end{eqnarray}
and
\begin{eqnarray}
d=1:
\qquad
\hbox{rank collapse and radial endpoint saddle},
\qquad
{\cal C}_{2,D}^{\rm rad}=-2 .
\end{eqnarray}

\medskip
\noindent
In this sense the HCIZ/Bessel interplay realizes the same geometric
endpoint-entropy mechanism differently in different dimensional regimes. For
\(d>3\), the interior symmetric--Wishart saddle is governed by the full
rank--two relative Grassmannian geometry, and the entropy counts the
corresponding angular fluctuation modes. For \(d=3\), the HCIZ endpoint geometry
fixes the boundary branch first, and the Bessel factor dresses the resulting
rank--one, one-principal-angle problem. For \(d=2\), the relative Grassmannian
has collapsed completely, and the relevant entropy comes instead from the
boundary layer of the singular Stiefel/Wishart measure. For \(d=1\), the rank
itself collapses, and the endpoint entropy becomes purely radial: the Bessel
convolution still shifts the effective mass and halves the logarithmic
coefficient, but the logarithm is no longer Grassmannian. Thus the same
geometric endpoint-entropy mechanism appears as full relative Grassmannian
geometry for \(d>3\), one-principal-angle boundary geometry for \(d=3\),
collapsed Grassmannian boundary-layer geometry for \(d=2\), and a rank--one
radial endpoint geometry for \(d=1\).

\medskip
\noindent
The endpoint/Wishart formulation provides therefore a mechanism for the
emergence of a relative Grassmannian geometry of dimension \(2(d-2)\) from
\(\mathbb{R}^d\), together with its lower-dimensional degenerations. The two
principal angles are the invariant singular-value coordinates of this relative
geometry when the rank--two Grassmannian sector exists. In the degenerate
low-dimensional cases, this geometry successively reduces to a one-angle
boundary geometry at \(d=3\), to a boundary-layer remnant at \(d=2\), and finally
to a purely radial rank--one endpoint geometry at \(d=1\).

\subsection{Dimensional regimes from the SIS matrix--ball picture}

\medskip
\noindent
The endpoint formulation exposes a simple geometric structure behind the
lower--dimensional reductions. For \(d\ge2\), the two endpoint vectors
\(V^\mu\), \(\mu=1,2\), span a two-plane in \(\mathbb{R}^d\), and similarly for
\(W^\mu\). Thus the relative angular sector is naturally described by the
relative position of two elements of
\begin{eqnarray}
Gr(2,d)
=
\frac{O(d)}{O(2)\times O(d-2)},
\qquad
\dim Gr(2,d)=2(d-2).
\end{eqnarray}
\medskip
\noindent
In our HCIZ/Bessel reduction, only the \emph{relative} orientation
\begin{eqnarray}
O:=O_W O_V^T\in O(d)
\end{eqnarray}
matters, and because $\Lambda_V,\Lambda_W$ have rank $2$, the integrand depends
only on the $2\times 2$ block $\Omega$ obtained by projecting $O$ between the two
planes.  Concretely, if $U_V,U_W\in V_2(\mathbb{R}^d)$ are orthonormal frames
spanning the $V$- and $W$-planes, then
\begin{eqnarray}
\Omega := U_W^T U_V \in \mathbb{R}^{2\times 2},
\qquad
\Omega^T\Omega\le I_2.
\end{eqnarray}
This matrix \(\Omega\) encodes all \(O(d)\)-invariant information about the
relative position of the two endpoint two-planes. Equivalently, its singular
values are precisely the cosines of the principal angles,
\begin{eqnarray}
s_i=\cos\theta_i.
\end{eqnarray}
A detailed derivation of this identification, together with its relation to the
Schur--Ingham--Siegel auxiliary eigenvalues, is given in Appendix \ref{appendix2}.

\medskip
\noindent
The number of independent principal angles is
\begin{eqnarray}
r(d)=\min\{2,d-2\},
\qquad d\ge2.
\end{eqnarray}
Hence
\begin{eqnarray}
r(2)=0,\qquad r(3)=1,\qquad r(d\ge4)=2.
\end{eqnarray}
Accordingly, the Grassmannian dimension decomposes as
\begin{eqnarray}
\dim Gr(2,d)
=
\underbrace{r(d)}_{\hbox{principal angles}}
+
\underbrace{\big(2d-4-r(d)\big)}_{\hbox{angular degeneracies}} .
\label{Gr_decomp_correct}
\end{eqnarray}
Thus the familiar residual factor \(2d-6\) has a literal Grassmannian meaning
only in the stable regime \(d\ge4\), where two independent principal angles
exist. For \(d=3\) the residual degeneracy is instead \(1\), and for \(d=2\) it
is \(0\). Below \(d=4\), any continuation of the large--\(d\) formula must
therefore be interpreted as an analytic continuation of the Wishart/Stiefel
Jacobian, not as an honest Grassmannian dimension count.

\medskip
\noindent
The four dimensional regimes are summarized as follows:
\begin{center}
\small
\renewcommand{\arraystretch}{1.2}
\begin{tabularx}{\textwidth}{c|X|X|X|c}
dimension & geometry & angular data & saddle type & entropy coefficient \\
\hline
\(d\ge4\)
&
generic two-planes
&
two principal angles
&
interior Wishart
&
\({\cal C}_{2,D}^{\rm W}=-2(d-2)\)
\\
\(d=3\)
&
planes intersect in a line
&
one principal angle
&
boundary HCIZ
&
\({\cal C}_{2,D}^{\rm bdry,+}=-1\)
\\
\(d=2\)
&
planes fill \(\mathbb{R}^2\)
&
no principal angle; in-plane angle only
&
boundary layer
&
\({\cal C}_{2,D}^{\rm bdry}=-2\)
\\
\(d=1\)
&
rank collapse
&
no angular sector
&
radial endpoint
&
\({\cal C}_{2,D}^{\rm rad}=-2\)
\end{tabularx}
\end{center}

\medskip
\noindent
In the stable regime \(d\ge4\), the \(2\times2\) overlap block \(\Omega\) lies
in the interior of the matrix-ball domain,
\begin{eqnarray}
\Omega\Omega^T<{\bf 1}_2,
\label{matrix_ball_condition}
\end{eqnarray}
and the Haar pushforward is absolutely continuous:
\begin{eqnarray}
\rho_{d,2}(\Omega)
\propto
{\bf 1}_{\Omega\Omega^T<{\bf 1}_2}
\det\!\big({\bf 1}_2-\Omega\Omega^T\big)^{\frac{d-5}{2}}.
\label{rho_matrix_ball_synthesis}
\end{eqnarray}
Both singular values of \(\Omega\) vary independently, so the two principal
angles are genuine invariants. At the saddle/geometric level, the
Schur--Ingham--Siegel (SIS) auxiliary matrix is conjugate to
\(\Omega\Omega^T\), and its eigenvalues encode the same invariant data as
\begin{eqnarray}
s_i^2=\cos^2\theta_i .
\end{eqnarray}
Thus \(d\ge4\) is the interior symmetric--Wishart regime. The full SIS
formulation of this matrix-ball problem, including the auxiliary-matrix
representation and its saddle-point structure, is discussed in detail
elsewhere \cite{Ydri2025}. Here we only use its geometric consequence:
the angular entropy of the interior saddle is controlled by the relative
Grassmannian dimension
\begin{eqnarray}
\dim Gr(2,d)=2(d-2),
\end{eqnarray}
which gives the universal coefficient
\begin{eqnarray}
{\cal C}_{2,D}^{\rm W}=-2(d-2).
\end{eqnarray}

\medskip
\noindent
The low-dimensional cases arise by degeneration of this matrix-ball picture.
At \(d=3\), the pushforward is no longer supported in the interior
matrix-ball domain \eqref{matrix_ball_condition}. Instead, the image collapses
onto a boundary stratum. Explicitly, the Euler-angle reduction gives
\begin{eqnarray}
\Omega
=
R(\alpha)
\left(
\begin{array}{cc}
c&0\\
0&1
\end{array}
\right)
R(\gamma),
\qquad
c=\cos\beta,
\end{eqnarray}
and hence
\begin{eqnarray}
\mathrm{singval}(\Omega)=\{1,|c|\}.
\end{eqnarray}
One singular value is frozen at \(1\), or equivalently one principal angle is
forced to vanish. In SIS language, the conjugate auxiliary eigenvalue associated
with \(1-s_1^2\) runs to a singular limit. This is why the \(d=3\) problem must
be treated by the real \(c\)-resolved HCIZ reduction derived in Appendix
\ref{appendix1}, rather than by a smooth interior matrix-ball saddle.

\medskip
\noindent
At \(d=2\), the relative Grassmannian collapses completely:
\begin{eqnarray}
Gr(2,2)=\{\mathrm{pt}\}.
\end{eqnarray}
The two endpoint two-planes fill the whole ambient space, so there is no
Grassmannian principal angle. In terms of the overlap block, the singular values
are frozen at the corner of the matrix ball,
\begin{eqnarray}
\Omega\Omega^T={\bf 1}_2.
\end{eqnarray}
The angle that remains in the \(O(2)\) HCIZ integral is therefore not a
principal angle; it is the relative in-plane eigenbasis angle between
\(\Lambda_V\) and \(\Lambda_W\). The dominant effect in BFSS\(_3\) is instead
the second mechanism: the Stiefel/Wishart measure develops an integrable
boundary singularity, and the endpoint dynamics is controlled by a boundary
layer.

\medskip
\noindent
At \(d=1\), even the \(2\times2\) overlap block no longer exists. The endpoint
rank itself collapses:
\begin{eqnarray}
\Lambda_V=(v),
\qquad
\Lambda_W=(w).
\end{eqnarray}
There is no matrix ball, no principal-angle sector, no Vandermonde, and no
Grassmannian entropy. The HCIZ factor is simply
\begin{eqnarray}
{\cal I}_{O(1)}=e^{\kappa_2vw},
\end{eqnarray}
and the remaining structure is purely radial. The Bessel convolution still
produces the endpoint mass threshold, but the associated logarithm is radial,
not Grassmannian.

\medskip
\noindent
The dimensional regimes may alternatively be summarized as follows:

\begin{center}
\small
\renewcommand{\arraystretch}{1.2}
\begin{tabularx}{\textwidth}{c|X|X|X}
dimension & SIS / matrix-ball picture & angular data & endpoint mechanism \\
\hline
\(d\ge4\)
&
interior of the \(2\times2\) matrix ball
&
two independent singular values \(s_1,s_2\)
&
interior symmetric--Wishart saddle
\\
\(d=3\)
&
boundary stratum of the matrix ball
&
one frozen singular value \(s_1=1\), one variable \(s_2=|c|\)
&
marginal case; \(c\)-resolved HCIZ selects the boundary branch
\\
\(d=2\)
&
singular-value corner of the matrix ball 
&
no principal angle, \(s_1=s_2=1\); only an in-plane eigenbasis angle
&
Stiefel/Wishart boundary layer
\\
\(d=1\)
&
no \(2\times2\) block
&
no angular sector
&
pure radial endpoint saddle
\end{tabularx}
\end{center}

\medskip
\noindent
Thus two mechanisms must be distinguished.

\begin{enumerate}
\item The first is the \emph{geometric degeneration} of the SIS matrix-ball
support. In the stable regime \(d\ge4\), and in particular starting at
\(d=4\), the \(2\times2\) overlap block has two independent singular values and
lies in the interior of the matrix ball. This is the regime in which
\begin{eqnarray}
{\cal C}_{2,D}^{\rm W}=-2(d-2)
\end{eqnarray}
is the honest Grassmannian/Wishart entropy of the interior matrix-ball saddle.
As the dimension is lowered, the support degenerates as
\begin{eqnarray}
d\ge4&:&\hbox{ interior ball},\nonumber\\
d=3&:&\hbox{ boundary stratum},\nonumber\\
d=2&:&\hbox{ collapsed corner},\nonumber\\
d=1&:&\hbox{ rank collapse}.
\end{eqnarray}
The case \(d=3\) is the marginal geometric transition: one singular value is
frozen at \(1\), one principal angle vanishes, and the SIS interior
matrix-ball description is replaced by the \(c\)-resolved HCIZ reduction.

\item The second is \emph{measure-driven endpoint dominance}. This refers to
the dominance of boundary or radial endpoint saddles induced by the
low-dimensional measure and by the Bessel endpoint asymptotics. At \(d=3\), the
endpoint measure is marginal: the factor \((v_1v_2)^{(d-3)/2}\) neither
suppresses nor enhances the boundary, so the boundary branch is selected
dynamically by the HCIZ sector. At \(d=2\), the relative Grassmannian has
collapsed completely, but the Stiefel/Wishart measure develops an integrable
boundary singularity and creates a genuine boundary layer. At \(d=1\), the
angular geometry disappears altogether and only the radial endpoint threshold
remains. Thus the large--\(d\) coefficient \(-2(d-2)\) should not be
extrapolated blindly to low dimensions; one must first identify the appropriate
endpoint geometry.
\end{enumerate}

\medskip
\noindent
Consequently, the large--\(d\) SIS/Grassmannian picture explains the origin of
the stable rank--two entropy in the interior matrix-ball regime. The
lower-dimensional models are obtained by tracking two departures from this
regime: the degeneration of the matrix-ball support and the takeover of the
endpoint measure.

\subsection{The relative Grassmannian as a relational four-dimensional geometry}

\medskip
\noindent
We finally emphasize the geometric interpretation of the relative Grassmannian
sector. The Grassmannian variables should not be thought of merely as auxiliary
angular variables. Rather, they describe the relative configuration of two
endpoint two-planes inside the ambient space \(\mathbb{R}^d\).

\medskip
\noindent
Let
\begin{eqnarray}
P_V,\;P_W\subset\mathbb{R}^d
\end{eqnarray}
be the two endpoint two-planes. In the stable regime \(d\ge4\), if the two
planes are fixed in generic relative position, their span is four-dimensional:
\begin{eqnarray}
\dim(P_V+P_W)=4.
\end{eqnarray}
Thus a generic pair of endpoint two-planes defines a relational
four-dimensional geometry inside \(\mathbb{R}^d\). The point, however, is not
only that such a four-dimensional span exists. The point is that this
four-dimensional relational geometry can move, tilt, rotate, and be embedded in
different ways inside the ambient \(d\)-dimensional space. The role of the
relative Grassmannian is precisely to describe this configuration space.

\medskip
\noindent
This is analogous to the reduction of a two-particle system. For two particles
in \(\mathbb{R}^3\), removing the center-of-mass motion leaves a relative
position vector in \(\mathbb{R}^3\). In the present problem, removing the common
global \(O(d)\) rotation allows us to fix one plane,
\begin{eqnarray}
P_V=P_0=\mathrm{span}(e_1,e_2).
\end{eqnarray}
The remaining degree of freedom is the relative position of the second plane,
\begin{eqnarray}
P_W\in Gr(2,d),
\qquad
Gr(2,d)=\frac{O(d)}{O(2)\times O(d-2)}.
\end{eqnarray}
Thus \(Gr(2,d)\) plays the role of the relative configuration space of the two
endpoint planes, just as \(\mathbb{R}^3\) plays the role of the relative
configuration space of two particles.

\medskip
\noindent
Locally near the aligned configuration, the second plane may be written as a
graph over \(P_0\),
\begin{eqnarray}
e_1\longrightarrow e_1+u,
\qquad
e_2\longrightarrow e_2+v,
\qquad
u,v\in P_0^\perp .
\end{eqnarray}
Equivalently, the relative motion is encoded in a linear map
\begin{eqnarray}
A:P_0\longrightarrow P_0^\perp,
\qquad
A\in\mathbb{R}^{(d-2)\times2}.
\end{eqnarray}
The two transverse vectors \(u\) and \(v\) specify how the second plane leaves
the first one. Since each lives in the \((d-2)\)-dimensional transverse space,
the relative motion contains
\begin{eqnarray}
2(d-2)
\end{eqnarray}
real degrees of freedom, in agreement with
\begin{eqnarray}
\dim Gr(2,d)=2(d-2).
\end{eqnarray}

\medskip
\noindent
The principal angles now acquire a simple meaning. They are the invariant
singular-value data of the relative map \(A\). They measure the magnitudes of
the independent tilts between the two planes. Equivalently, they specify the
invariant shape of the pair of two-planes, namely its congruence class under
rigid \(O(d)\) rotations. They do not, however, specify the full relative
configuration. The remaining Grassmannian variables record how this shape is
oriented and embedded inside \(\mathbb{R}^d\).

\medskip
\noindent
Thus the distinction is
\begin{eqnarray}
\hbox{principal angles}
&=&
\hbox{shape invariants of the pair }(P_V,P_W),
\nonumber\\
\hbox{full Grassmannian}
&=&
\hbox{relative motion, orientation, and embedding data}.
\end{eqnarray}
The four-dimensional object is the generic span \(P_V+P_W\). The Grassmannian
is the space of ways in which its two two-plane constituents are placed
relative to one another inside the ambient space.

\medskip
\noindent
The simplest illustration is \(d=3\). Two planes in \(\mathbb{R}^3\)
necessarily intersect in a line, so one principal angle is zero and the other
is the ordinary angle between the planes. Fixing this angle fixes the relative
shape: two planes meeting, for example, at \(30^\circ\). But it does not fix the
full Grassmannian configuration. After fixing one plane \(P_V=P_0\), the second
plane is specified by two data: the angle \(\theta\) at which it meets \(P_0\),
and the orientation of the intersection line
\begin{eqnarray}
L=P_0\cap P_W
\end{eqnarray}
inside \(P_0\). Fixing \(\theta\) fixes the shape, while the line \(L\) may
still rotate inside \(P_0\). This remaining orientation is the angular
degeneracy. Thus \(Gr(2,3)\) contains one principal-angle variable and one
angular-degeneracy variable.

\medskip
\noindent
The lower-dimensional degenerations then acquire a simple interpretation. For
\(d\ge4\), two endpoint two-planes can generically span a four-dimensional
subspace, and the relative Grassmannian describes the motion of this relational
four-dimensional geometry inside \(\mathbb{R}^d\). At \(d=3\), this generic
four-dimensional span is no longer possible: two planes in \(\mathbb{R}^3\)
always intersect in a line, their span is only three-dimensional, and only one
principal angle survives. At \(d=2\), both planes fill the ambient space and the
relative Grassmannian collapses. At \(d=1\), the rank--two plane picture itself
disappears.

\medskip
\noindent
Thus the emergent Grassmannian sector may be viewed as the configuration space
of a relational four-dimensional geometry generated by two endpoint two-planes.
The number \(2(d-2)\) is not the dimension of the four-dimensional span itself;
it is the number of ways in which this relational four-dimensional geometry can
move, orient, and embed itself inside the ambient \(d\)-dimensional space.

\section{Conclusion}\label{conclusion}

\medskip
\noindent
In the present paper we developed a Gram--Wishart--Stiefel formulation of the
\(N=2\), large--\(d\) BFSS/BMN endpoint theory. The shifted planar boundary
model was rewritten in terms of radial Wishart eigenvalues and relative
Stiefel angular variables, with the large--\(R\) aligned contribution \(-A\)
kept inside the Gaussian kernel. This formulation separates the radial endpoint
measure, the aligned Bessel sector, and the transverse
Harish--Chandra--Itzykson--Zuber (HCIZ) angular interaction.

\medskip
\noindent
A central result is that the nonperturbative Bessel kernel cannot be replaced
by the naive aligned saddle alone. Its leading effect is twofold: it shifts the
endpoint mass and supplies an inverse Wishart--Stiefel prefactor. This inverse
prefactor removes the spurious doubled entropy and leads, in the stable
rank--two regime, to the symmetric--Wishart saddle with logarithmic coefficient
\begin{eqnarray}
{\cal C}_{2,D}^{\rm W}=-2(d-2).
\label{conclu2}
\end{eqnarray}
Thus the continuum \(D\)-channel entropy is not controlled merely by the radial
Wishart variables, but by the full rank--two angular geometry generated by the
endpoint construction.

\medskip
\noindent
The lower-dimensional BFSS models were then analyzed directly. For BFSS\(_4\)
(\(d=3\)), the \(O(3)\) angular integral reduces to a \(c\)-resolved HCIZ
problem. The HCIZ sector dynamically selects a boundary branch before the
Bessel factor is included, and the Bessel endpoint asymptotics dresses this
rank--one boundary geometry, giving
\begin{eqnarray}
{\cal C}_{2,D}^{\rm bdry,+}=-1.
\end{eqnarray}
For BFSS\(_3\) (\(d=2\)), the relative Grassmannian collapses, but the
Stiefel/Wishart measure develops an integrable boundary singularity. The
resulting boundary layer is again dressed by the Bessel endpoint prefactor,
leading to
\begin{eqnarray}
{\cal C}_{2,D}^{\rm bdry}=-2.
\end{eqnarray}
For BFSS\(_2\) (\(d=1\)), the rank--two endpoint structure disappears
altogether. The problem becomes purely radial, and the Bessel convolution
produces the threshold logarithm associated with
\(m_{\rm eff}=m-\ell/2\), giving
\begin{eqnarray}
{\cal C}_{2,D}^{\rm rad}=-2.
\end{eqnarray}

\medskip
\noindent
The geometric content of these results is the emergence of a relative
Grassmannian sector from the rank--two endpoint variables. Each endpoint
configuration determines a two-plane in the ambient space \(\mathbb{R}^d\).
Denoting these planes by
\begin{eqnarray}
P_V,\;P_W\subset\mathbb{R}^d,
\end{eqnarray}
the angular problem is the problem of their relative position. After the common
global \(O(d)\) orientation is removed, one may fix \(P_V=P_0\), and the
remaining angular variable is the position of \(P_W\). Thus the relative
angular sector is
\begin{eqnarray}
Gr(2,d)
=
\frac{O(d)}{O(2)\times O(d-2)},
\qquad
\dim Gr(2,d)=2(d-2).
\end{eqnarray}
This is the geometric origin of the factor \(2(d-2)\). It counts the full set
of relative Grassmannian fluctuation modes, not merely the invariant
principal-angle variables.

\medskip
\noindent
Near the aligned configuration this statement becomes explicit. A nearby plane
may be written as a graph over
\begin{eqnarray}
P_0=\mathrm{span}(e_1,e_2),
\end{eqnarray}
namely
\begin{eqnarray}
e_1\longrightarrow e_1+u,
\qquad
e_2\longrightarrow e_2+v,
\qquad
u,v\in P_0^\perp .
\end{eqnarray}
Equivalently, the infinitesimal relative motion is a linear map
\begin{eqnarray}
A:P_0\longrightarrow P_0^\perp,
\qquad
A\in\mathbb{R}^{(d-2)\times2}.
\end{eqnarray}
Since \(P_0^\perp\) has dimension \(d-2\), the two transverse vectors \(u\) and
\(v\) contain \(2(d-2)\) real components. The principal angles are obtained
only after taking the singular values of this map. Thus the principal angles
are invariant coordinates on the relative Grassmannian, while the remaining
Grassmannian variables specify the orientation and embedding degeneracies of
these tilts inside \(\mathbb{R}^d\). The symmetric--Wishart entropy is
therefore not the entropy of two angles; it is the entropy of the full relative
Grassmannian tangent space around the aligned saddle.

\medskip
\noindent
The same geometry admits an equivalent Schur--Ingham--Siegel (SIS)
matrix--ball description. Since the endpoint spectra have rank two, the HCIZ
integrand depends only on the \(2\times2\) overlap block
\begin{eqnarray}
\Omega=U_W^TU_V
\end{eqnarray}
between orthonormal frames spanning \(P_W\) and \(P_V\). This block satisfies
\begin{eqnarray}
\Omega\Omega^T\le {\bf 1}_2,
\end{eqnarray}
and its singular values are the cosines of the principal angles,
\begin{eqnarray}
s_i=\cos\theta_i .
\end{eqnarray}
For \(d\ge4\), the Haar pushforward to \(\Omega\) is supported in the interior
of the \(2\times2\) matrix ball,
\begin{eqnarray}
\Omega\Omega^T<{\bf 1}_2.
\end{eqnarray}
Both singular values vary independently, the two principal angles are genuine
invariants, and the SIS/Wishart saddle is an honest interior matrix--ball
saddle. This is the stable rank--two regime in which the
Grassmannian/Wishart entropy is well defined and gives \eqref{conclu2}.

\medskip
\noindent
The lower-dimensional theories are therefore not obtained by blindly
substituting small values of \(d\) into the large--\(d\) expression. They arise
by successive degenerations of the same relative Grassmannian, or equivalently
of the same SIS matrix--ball support. At \(d=3\), the \(2\times2\) overlap block
no longer fills the interior matrix ball. Its image is supported on a boundary
stratum: one singular value is frozen at \(1\), one principal angle is forced
to vanish, and only one nontrivial principal angle remains. This is the
geometric reason why BFSS\(_4\) must be treated by the real \(c\)-resolved HCIZ
reduction rather than by the smooth interior SIS saddle.

\medskip
\noindent
At \(d=2\), the degeneration is stronger. Both endpoint two-planes fill the
ambient space \(\mathbb{R}^2\), so the relative Grassmannian collapses to a
point. In matrix--ball language both singular values are frozen at the
singular-value corner,
\begin{eqnarray}
s_1=s_2=1,
\qquad
\Omega\Omega^T={\bf 1}_2.
\end{eqnarray}
There is therefore no Grassmannian principal angle. The angular variable which
survives in the \(O(2)\) HCIZ integral is only the relative in-plane eigenbasis
angle between the endpoint spectra. At \(d=1\), even the rank--two plane
picture disappears: there is no \(2\times2\) overlap block, no Vandermonde, no
principal-angle sector, and no Grassmannian entropy. The endpoint problem is
purely radial.

\medskip
\noindent
This hierarchy separates two mechanisms. The first is the geometric
degeneration of the SIS matrix--ball support:
\begin{eqnarray}
d\ge4&:&\hbox{ interior matrix ball},\nonumber\\
d=3&:&\hbox{ boundary stratum},\nonumber\\
d=2&:&\hbox{ singular-value corner},\nonumber\\
d=1&:&\hbox{ rank collapse}.
\end{eqnarray}
The second is measure-driven endpoint dominance. In low dimensions, the
Wishart/Stiefel measure and the Bessel endpoint asymptotics can move the
dominant contribution away from the smooth interior saddle and toward boundary
or radial endpoint configurations. Thus \(-2(d-2)\) is the entropy of the
stable interior rank--two Grassmannian/Wishart saddle, and should not be
extrapolated into regimes where the endpoint geometry has already degenerated.

\medskip
\noindent
The final geometric interpretation is relational. For \(d\ge4\), two generic
endpoint two-planes span a four-dimensional subspace,
\begin{eqnarray}
\dim(P_V+P_W)=4.
\end{eqnarray}
This span is a relational four-dimensional geometry generated by the two
two-plane constituents. The Grassmannian \(Gr(2,d)\) is not the dimension of
this four-dimensional object itself. Rather, it is the configuration space of
the ways in which the two two-planes are placed relative to one another inside
\(\mathbb{R}^d\). The principal angles specify the invariant shape of the pair
of planes, while the remaining Grassmannian variables specify the orientation,
motion, and embedding of this shape in the ambient space.

\medskip
\noindent
Overall, the endpoint formulation reveals a unified geometric mechanism behind
the \(D\)-channel entropy. In the stable regime, this entropy is the
Grassmannian/Wishart entropy of an interior rank--two saddle. In low dimensions,
the same mechanism survives only through boundary or radial degenerations: a
one-principal-angle boundary geometry in BFSS\(_4\), a boundary-layer geometry
in BFSS\(_3\), and a rank--one radial endpoint geometry in BFSS\(_2\). Thus the
HCIZ/Bessel reductions, the SIS matrix--ball picture, and the dimensional
hierarchy of endpoint entropies are all different manifestations of the same
underlying relative Grassmannian geometry.

\section{Historical note}\label{history}

\medskip
\noindent
The work presented in this paper was carried out within a broader
programme combining large-\(d\) Gaussian reduction, Molien--Weyl
singlet projection, endpoint/Wishart formulations and Monte Carlo
methods. The central concern of this programme is the structure
of matrix quantum mechanics and its interpretation as matrix
quantum gravity; see
\cite{Ydri:2026gsy,Ydri:2026smo,YdriUnpublished0,Ydri:2026rke,
YdriUnpublished1,YdriUnpublished3}
and \cite{Ydri2025}.

\medskip
\noindent
Two working assumptions underlie the developments presented
here and in the preceding papers. First, matrix quantum mechanics
constitutes the primary dynamical framework: its holographic
interpretation leads to quantum black-hole dynamics, while its
noncommutative matrix degrees of freedom provide a microscopic
route to emergent geometry and gravity. From this perspective,
zero-dimensional matrix models, including the IKKT
model~\cite{Ishibashi:1996xs}, may be regarded as reductions,
limits or approximations of an underlying one-dimensional matrix
quantum theory. Second, Gaussian structures are ``unreasonably''
effective in matrix quantum mechanics, capturing a substantial
part of its nontrivial quantum dynamics even before the full
interacting theory is restored. The results obtained in the
present paper provide a further illustration of this viewpoint.

\medskip
\noindent
Historically, the present programme in matrix quantum mechanics,
matrix quantum gravity and matrix quantum computation grew out
of our earlier programme on emergent geometry in Yang--Mills
and multitrace matrix models
\cite{GeomTransition,MatrixGaugeEmergent,TopologyChange,
MultitraceApproach,EmergentMultitrace,MultitracePhase},
which itself grew out of our earlier investigations of
noncommutative geometry, fuzzy spaces and noncommutative
field theory
\cite{NCRegulator,ScalingFuzzy,OriginUVIR,
GaugeUVIR,NCGaugeMC,NCphi4Phase}.
The connection with the matrix models of string theory
was developed further in \cite{MatrixTheoryReview}.

\section{Acknowledgments}

\medskip
\noindent
The author also acknowledges the use of ChatGPT-5.5, as well as previous versions, in several auxiliary capacities: 
(1) as a language editor; 
(2) as a LaTeX generator; 
(3) as a Mathematica-like symbolic tool; 
(4) as an assistant in searching for and reviewing references; 
and, more importantly, 
(5) as an ``artificial'' sounding board for testing, organizing, and refining ideas, effectively replacing in this role the function often played by human collaborators. However, the scientific vision, concept, design, direction, final scientific and mathematical editing, and all intellectual responsibility for this work remain solely with the author.

\appendix
\section{HCIZ reduction for BFSS$_4$ (the $O(3)$ angular integral)}\label{appendix1}

\subsection{Reduction through Euler angles}

\medskip
\noindent
Since the endpoint Gram matrices in  \(d=3\) have rank at most $2$
(because the endpoint vectors are planar), we may take
\begin{eqnarray}
\Lambda_V=\mathrm{diag}(v_1,v_2,0),\qquad
\Lambda_W=\mathrm{diag}(w_1,w_2,0),
\qquad v_i,w_i\ge 0.
\end{eqnarray}
The angular invariant is
\begin{eqnarray}
{\cal T}:=\mathrm{Tr}\!\big(\Lambda_W O\Lambda_V O^T\big)
=\sum_{i=1}^2\sum_{j=1}^2 w_i v_j\,O_{ij}^2,
\label{T_O3_def}
\end{eqnarray}
so only the upper--left $2\times2$ block of $O\in O(3)$ enters.

\medskip
\noindent First, because ${\cal T}$ depends only on the squared matrix elements $O_{ij}^2$,
the integrand is invariant under $O\mapsto -O$ and under the disconnected
component of $O(3)$.  Thus, up to an overall $v,w$--independent constant,
\begin{eqnarray}
{\cal I}_{d=3}(\Lambda_W,\Lambda_V)
:=
\int_{O(3)} d\mu(O)\;e^{\kappa_2{\cal T}}
\;\propto\;
\int_{SO(3)} d\mu(O)\;e^{\kappa_2{\cal T}}.
\label{O3_to_SO3}
\end{eqnarray}

\medskip
\noindent Next, we employ Euler parameterization, i.e. we parameterize $SO(3)$ by Euler angles
\begin{eqnarray}
O(\alpha,\beta,\gamma)=R_z(\alpha)\,R_y(\beta)\,R_z(\gamma),
\qquad
0\le\alpha,\gamma<2\pi,\quad 0\le\beta\le\pi,
\end{eqnarray}
with Haar measure
\begin{eqnarray}
d\mu(O)=\frac{1}{8\pi^2}\,\sin\beta\; d\alpha\, d\beta\, d\gamma.
\label{Haar_SO3}
\end{eqnarray}
Recall
\begin{eqnarray}
R_z(\theta)=
\left(\!\begin{array}{ccc}
\cos\theta&-\sin\theta&0\\
\sin\theta&\cos\theta&0\\
0&0&1
\end{array}\!\right),
\qquad
R_y(\beta)=
\left(\!\begin{array}{ccc}
\cos\beta&0&\sin\beta\\
0&1&0\\
-\sin\beta&0&\cos\beta
\end{array}\!\right).
\end{eqnarray}
Let $A$ denote the upper--left $2\times2$ block of $O$. A direct multiplication gives
\begin{eqnarray}
A(\alpha,\beta,\gamma)
=
\left(\!\begin{array}{cc}
\cos\alpha\,\cos\beta\,\cos\gamma-\sin\alpha\,\sin\gamma
&
-\cos\alpha\,\cos\beta\,\sin\gamma-\sin\alpha\,\cos\gamma
\\[1mm]
\sin\alpha\,\cos\beta\,\cos\gamma+\cos\alpha\,\sin\gamma
&
-\sin\alpha\,\cos\beta\,\sin\gamma+\cos\alpha\,\cos\gamma
\end{array}\!\right).\nonumber\\
\label{A_explicit}
\end{eqnarray}
Introduce the planar rotation
\begin{eqnarray}
R(\theta)=
\left(\!\begin{array}{cc}
\cos\theta&-\sin\theta\\
\sin\theta&\cos\theta
\end{array}\!\right).
\end{eqnarray}
Then the matrix in \eqref{A_explicit} factorizes as
\begin{eqnarray}
A(\alpha,\beta,\gamma)
=
R(\alpha)\,
\left(\!\begin{array}{cc}
c&0\\
0&1
\end{array}\!\right)\,
R(\gamma),
\label{A_block_correct}
\end{eqnarray}
with $c:=\cos\beta$ (and $s:=\sin\beta$). In particular, the singular values of $A$ are $\{1,|\cos\beta|\}$, and $A$ becomes rank--one at
$\beta=\pi/2$.

Writing $D_V=\mathrm{diag}(v_1,v_2)$ and $D_W=\mathrm{diag}(w_1,w_2)$, the invariant
reduces to
\begin{eqnarray}
{\cal T}(\alpha,\beta,\gamma)
=
\mathrm{Tr}\!\big(D_W\,A(\alpha,\beta,\gamma)\,D_V\,A(\alpha,\beta,\gamma)^T\big).
\label{T_block_trace_again}
\end{eqnarray}

\medskip
\noindent
Define the combinations
\begin{eqnarray}
\varphi_1:=\alpha+\gamma,\qquad
\varphi_2:=\alpha-\gamma.
\label{phi_psi_def}
\end{eqnarray}
The Jacobian is $|\partial(\alpha,\gamma)/\partial(\varphi_1,\varphi_2)|=\frac12$, and the flat measure becomes
\begin{eqnarray}
\frac{d\alpha}{2\pi}\frac{d\gamma}{2\pi}
=
\frac12\,
\frac{d\varphi_1}{2\pi}\frac{d\varphi_2}{2\pi}.
\label{measure_phi_psi}
\end{eqnarray}
On the torus $(\alpha,\gamma)\in[0,2\pi)^2$ one may equivalently integrate $(\varphi_1,\varphi_2)$ over $[0,2\pi)^2$
since the map is a $2$--to--$1$ covering; \eqref{measure_phi_psi} already accounts for this.

\medskip
\noindent A straightforward trigonometric computation shows that
${\cal T}$ is affine in $\cos(2\varphi_1)$, $\cos(2\varphi_2)$, $\cos(\varphi_1-\varphi_2)$ and $\cos(\varphi_1+\varphi_2)$:
\begin{eqnarray}
{\cal T}(\alpha,\beta,\gamma)
=
T_0(c)\;+\;T_{1}(c)\cos(2\varphi_1)\;+\;T_{2}(c)\cos(2\varphi_2)\;+\;T_{-}(c)\cos(\varphi_1-\varphi_2)\;+\;T_{+}(c)\cos(\varphi_1+\varphi_2),\nonumber\\
\label{T_affine_alpha_correct}
\end{eqnarray}
with coefficients (where $v_{\pm}=v_1\pm v_2$ and $w_{\pm}=w_1\pm w_2$)

\begin{eqnarray}
  &&  T_0=\frac{w_+v_+}{4}(1+c^2),\label{T0}\\
  &&T_{1}=\frac{w_-v_-}{8}(1+c)^2,\quad T_{2}=\frac{w_-v_-}{8}(1-c)^2,\label{Tuv}\\
  &&T_-=-\frac{w_+v_-}{4}(1-c^2),\quad T_+=-\frac{w_-v_+}{4}(1-c^2)\label{T-+}.
\end{eqnarray}
Equations \eqref{T0}--\eqref{Tuv}--\eqref{T-+} are the key
reduction identities for $d=3$.

\subsection{Angular Fourier--Bessel expansion and the small-\(\kappa_2\) check}

\medskip
\noindent
Using \eqref{measure_phi_psi} and \eqref{T0}--\eqref{Tuv}--\eqref{T-+}, we find 
\begin{eqnarray}
{\cal I}_{d=3}(\kappa_2)
&=&
\int_{-1}^{1}\frac{dc}{2}\;
e^{\kappa_2 T_0}
\int_0^{2\pi}\frac{d\varphi_1}{2\pi}\int_0^{2\pi}\frac{d\varphi_2}{2\pi}\;
\exp\!\Big(\kappa_2T_{1}\cos(2\varphi_1)\;+\;\kappa_2T_{2}\cos(2\varphi_2)\nonumber\\
&+&\kappa_2T_{-}\cos(\varphi_1-\varphi_2)\;+\;\kappa_2T_{+}\cos(\varphi_1+\varphi_2)\Big).
\label{I_O3_phi_psi_form}
\end{eqnarray}

\medskip
\noindent
Now use the Jacobi--Anger expansions
\begin{eqnarray}
e^{x\cos\theta}=\sum_{m\in\mathbb Z} I_m(x)\,e^{im\theta}.
\label{Jacobi_Anger}
\end{eqnarray}
Apply \eqref{Jacobi_Anger} four times:
\begin{eqnarray}
e^{\kappa_2 T_1\cos(2\varphi_1)}
&=&
\sum_{m\in\mathbb Z} I_m(\kappa_2 T_1)\,e^{i2m\varphi_1},
\\
e^{\kappa_2 T_2\cos(2\varphi_2)}
&=&
\sum_{n\in\mathbb Z} I_n(\kappa_2 T_2)\,e^{i2n\varphi_2},
\\
e^{\kappa_2 T_{-}\cos(\varphi_1-\varphi_2)}
&=&\sum_{p\in\mathbb Z} I_p(\kappa_2 T_-)\,e^{ip (\varphi_1-\varphi_2)},
\\
e^{\kappa_2 T_{+}\cos(\varphi_1+\varphi_2)}
&=&\sum_{q\in\mathbb Z} I_q(\kappa_2 T_+)\,e^{iq (\varphi_1+\varphi_2)},
\label{mixed_term_expansion}
\end{eqnarray}
Multiplying these and integrating over $u,v$ projects to the zero modes:
\begin{eqnarray}
\int_0^{2\pi}\frac{d\varphi_1}{2\pi}\int_0^{2\pi}\frac{d\varphi_2}{2\pi}\;
e^{i(2m+p+q)\varphi_1}\,e^{i(2n-p+q)\varphi_2}
=
\delta_{2m+p+q,\,0}\;\delta_{2n-p+q,\,0}.
\end{eqnarray}
Hence the \emph{exact} angle integrals are done, but in a way that keeps the coupling honest:
\begin{eqnarray}
{\cal I}_{d=3}(\kappa_2)=
\int_{-1}^{1}\frac{dc}{2}\;
e^{\kappa_2 T_0(c)}
\sum_{\substack{p,q\in\mathbb Z\\ p\equiv q\ (\mathrm{mod}\ 2)}}
I_{-\frac{p+q}{2}}(\kappa_2 T_1(c))\;
I_{\frac{p-q}{2}}(\kappa_2 T_2(c))\;
I_{p}\!(\kappa_2 T_{-}(c))\;
I_{q}\!(\kappa_2 T_{+}(c)).\nonumber\\
\label{I_O3_double_Bessel_series}
\end{eqnarray}
Using $I_{-m}=I_m$, this can be written more symmetrically as
\begin{eqnarray}
{\cal I}_{d=3}(\kappa_2)=
\int_{-1}^{1}\frac{dc}{2}\;
e^{\kappa_2 T_0(c)}
\sum_{\substack{p,q\in\mathbb Z\\ p\equiv q\ (\mathrm{mod}\ 2)}}
I_{\frac{p+q}{2}}(\kappa_2 T_1(c))\;
I_{\frac{p-q}{2}}(\kappa_2 T_2(c))\;
I_{p}\!(\kappa_2 T_{-}(c))\;
I_{q}\!(\kappa_2 T_{+}(c)).\nonumber\\
\label{I_O3_double_Bessel_series_symmetric}
\end{eqnarray}
The notation $ p\equiv q$ means that $p\pm q$ must be even. This completes the $O(3)$ HCIZ reduction for the rank--$2$ endpoint spectrum.

\medskip
\noindent
To obtain the small-\(\kappa_2\) expansion up to \(O(\kappa_2^2)\), we start from the quadruple--Bessel representation
\eqref{I_O3_double_Bessel_series_symmetric}. Using
\begin{eqnarray}
I_0(x)=1+\frac{x^2}{4}+O(x^4),\qquad I_{\pm1}(x)=\frac{x}{2}+O(x^3),
\end{eqnarray}
one checks that up to $O(\kappa_2^2)$ only the $(p,q)=(0,0)$ mode contributes.
Hence
\begin{eqnarray}
{\cal I}_{d=3}(\kappa_2)
&=&
\int_{-1}^{1}\frac{dc}{2}\;
e^{\kappa_2 T_0(c)}
I_0(\kappa_2 T_1)
I_0(\kappa_2 T_2)
I_0(\kappa_2 T_-)
I_0(\kappa_2 T_+)
+O(\kappa_2^3).
\end{eqnarray}
Expanding, we obtain
\begin{eqnarray}
{\cal I}_{d=3}(\kappa_2)
&=&
\int_{-1}^{1}\frac{dc}{2}
\left[
1+\kappa_2 T_0
+\kappa_2^2
\left(
\frac{T_0^2}{2}
+\frac{T_1^2+T_2^2+T_-^2+T_+^2}{4}
\right)
\right]
+O(\kappa_2^3).
\label{I_O3_expanded_integrand}
\end{eqnarray}
Using  \eqref{T0}--\eqref{Tuv}--\eqref{T-+}, and performing the $c$--integrals,
\begin{eqnarray}
\int_{-1}^{1}\frac{dc}{2}\,T_0
&=&
\frac{w_+v_+}{3},
\\
\int_{-1}^{1}\frac{dc}{2}\,\frac{T_0^2}{2}
&=&
\frac{7}{120}\,w_+^2v_+^2,
\\
\int_{-1}^{1}\frac{dc}{2}\,\frac{T_1^2+T_2^2+T_-^2+T_+^2}{4}
&=&
\frac{1}{40}\,w_-^2v_-^2
+\frac{1}{120}\,w_+^2v_-^2
+\frac{1}{120}\,w_-^2v_+^2,
\end{eqnarray}
we finally obtain
\begin{eqnarray}
{\cal I}_{d=3}(\kappa_2)
&=&
1
+\kappa_2\,\frac{w_+v_+}{3}
\\
&&
+\kappa_2^2
\left[
\frac{7}{120}w_+^2v_+^2
+\frac{1}{40}w_-^2v_-^2
+\frac{1}{120}w_+^2v_-^2
+\frac{1}{120}w_-^2v_+^2
\right]
+O(\kappa_2^3).\nonumber\\
\label{I_O3_small_kappa_final}
\end{eqnarray}
\medskip
\noindent
Now we rewrite this equation in terms of the invariants
\begin{eqnarray}
S_1^V=v_1+v_2,\qquad S_2^V=v_1^2+v_2^2,
\qquad
S_1^W=w_1+w_2,\qquad S_2^W=w_1^2+w_2^2.
\end{eqnarray}
Using
\begin{eqnarray}
(v_1-v_2)^2=2S_2^V-(S_1^V)^2,
\qquad
(w_1-w_2)^2=2S_2^W-(S_1^W)^2,
\end{eqnarray}
one finds that the coefficient of $\kappa_2^2$ in \eqref{I_O3_small_kappa_final} becomes
\begin{eqnarray}
\frac{1}{60}\Big[
4(S_1^W)^2(S_1^V)^2
-2(S_1^W)^2 S_2^V
-2(S_1^V)^2 S_2^W
+6\,S_2^W S_2^V
\Big].
\label{IO3_kappa2_in_S}
\end{eqnarray}
Therefore, at $d=3$,
\begin{eqnarray}
{\cal I}_{d=3}(\Lambda_W,\Lambda_V)
&=&
1+\kappa_2\,\frac{S_1^W S_1^V}{3}
+\frac{\kappa_2^2}{2}\;
\frac{
4(S_1^W)^2(S_1^V)^2
-2(S_1^W)^2 S_2^V
-2(S_1^V)^2 S_2^W
+6\,S_2^W S_2^V
}{
3\cdot 2\cdot 5
}
+O(\kappa_2^3).\nonumber\\
\label{IO3_matches_zonalprime_d3}
\end{eqnarray}
This agrees precisely with the zonal-moment formula obtained elsewhere
\cite{Ydri2025}, once that formula is specialized to \(d=3\). Indeed,
in this case
\begin{eqnarray}
d(d-1)(d+2)=3\cdot2\cdot5,
\qquad
d+1=4,
\qquad
2d=6.
\end{eqnarray}

\subsection{Summary of the HCIZ reduction of the BFSS$_4$}

\medskip
\noindent
Using the Euler reduction and setting $c=\cos\beta\in[-1,1]$, one may write
\begin{eqnarray}
{\cal I}_{d=3}(\Lambda_W,\Lambda_V)=
\int_{-1}^{1}\frac{dc}{2}\;{\cal Z}_{\rm ang}(c;\Lambda_W,\Lambda_V),
\label{Id3_c_resolved}
\end{eqnarray}
where ${\cal Z}_{\rm ang}$ is the exact two--torus average
\begin{eqnarray}
{\cal Z}_{\rm ang}(c;\Lambda_W,\Lambda_V)
:=
\int_{0}^{2\pi}\frac{d\varphi_1}{2\pi}\int_{0}^{2\pi}\frac{d\varphi_2}{2\pi}\;
\exp\!\Big(\kappa_2\,{\cal T}(c;\varphi_1,\varphi_2)\Big).
\label{Zang_c_def}
\end{eqnarray}

\medskip
\noindent
It is then convenient to define the \emph{$c$--resolved holonomy potential}
\begin{eqnarray}
V_{\rm hol}(c;\Lambda_W,\Lambda_V)
:=
-\log {\cal Z}_{\rm ang}(c;\Lambda_W,\Lambda_V),
\label{Vhol_c_def}
\end{eqnarray}
so that \eqref{Id3_c_resolved} becomes
\begin{eqnarray}
{\cal I}_{d=3}(\Lambda_W,\Lambda_V)
\;\propto\;
\int_{-1}^{1}\frac{dc}{2}\;
\exp\!\Big(-V_{\rm hol}(c;\Lambda_W,\Lambda_V)\Big).
\label{Id3_as_1d_potential}
\end{eqnarray}
Equation \eqref{Id3_as_1d_potential} is an \emph{exact} representation of the holonomy sector
as a one--dimensional statistical integral with potential $V_{\rm hol}(c)$.

\medskip
\noindent
Equivalently, by Jacobi--Anger expansions and integrating over the two torus angles exactly,
one finds the \emph{exact} $c$--integral with a discrete Bessel sum:
\begin{eqnarray}
{\cal Z}_{\rm ang}(c;\Lambda_W,\Lambda_V)=
e^{\kappa_2 T_0(c)}
\sum_{\substack{p,q\in\mathbb Z\\ p\equiv q\ (\mathrm{mod}\ 2)}}
I_{\frac{p+q}{2}}(\kappa_2 T_1(c))\;
I_{\frac{p-q}{2}}(\kappa_2 T_2(c))\;
I_{p}\!(\kappa_2 T_{-}(c))\;
I_{q}\!(\kappa_2 T_{+}(c)),\nonumber\\
\label{Id3_quadruple_bessel}
\end{eqnarray}
with $T_0,T_1,T_2,T_\pm$ given explicitly in \eqref{T0}--\eqref{Tuv}--\eqref{T-+}.
This provides an explicit generator of the full $\kappa_2$ expansion.

\subsection{Failure of the four-dimensional reduced effective potential in BFSS$_4$}

\medskip
\noindent
In the reduced endpoint formulation, the planar endpoint partition function
factorizes into a Wishart--type eigenvalue measure times the holonomy factor.
For general \(d\), we have
\begin{eqnarray}
Z_\perp
&\propto&
\int dv_1\,dv_2\;
e^{-m(v_1+v_2)}
(v_1v_2)^{\frac{d-3}{2}}
|v_1-v_2|
\int dw_1\,dw_2\;
e^{-m(w_1+w_2)}
(w_1w_2)^{\frac{d-3}{2}}
|w_1-w_2|
\nonumber\\
&&\times
{\cal I}_{d}(\Lambda_W,\Lambda_V).
\label{Zperp_general_form}
\end{eqnarray}
Specializing to \(d=3\) removes the factors
\((v_1v_2)^{(d-3)/2}\) and \((w_1w_2)^{(d-3)/2}\). Thus
\begin{eqnarray}
Z_\perp^{(d=3)}
\propto
\int dv_1\,dv_2\;
e^{-m(v_1+v_2)}
|v_1-v_2|
\int dw_1\,dw_2\;
e^{-m(w_1+w_2)}
|w_1-w_2|\;
{\cal I}_{d=3}(\Lambda_W,\Lambda_V).
\label{Zperp_d3_form}
\end{eqnarray}

\medskip
\noindent
If the angular variable \(c=\cos\beta\) is integrated out, the holonomy
contribution is defined by
\begin{eqnarray}
V_{\rm hol}^{(d=3)}
:=
-\log {\cal I}_{d=3}(\Lambda_W,\Lambda_V).
\end{eqnarray}
The resulting four-dimensional reduced effective potential is therefore
\begin{eqnarray}
V_{\rm eff}(v,w)
&=&
m(v_1+v_2)-\log|v_1-v_2|
+
m(w_1+w_2)-\log|w_1-w_2|
\nonumber\\
&&
-\log {\cal I}_{d=3}(\Lambda_W,\Lambda_V).
\label{Veff_4d_reduced}
\end{eqnarray}
The angular integral \({\cal I}_{d=3}\) may be expanded in \(\kappa_2\) either
by computing Haar moments directly, or equivalently by expanding the
quadruple--Bessel representation
\eqref{I_O3_double_Bessel_series_symmetric} and performing the \(c\)-integral
term by term. To quadratic order one recovers
\eqref{IO3_matches_zonalprime_d3}, which gives
\begin{eqnarray}
V_{\rm hol}^{(d=3)}
&=&
-\log {\cal I}_{d=3}
\nonumber\\
&=&
-\frac{\kappa_2}{3}\,S_1^W S_1^V
-\frac{\kappa_2^2}{180}
\Big[
2(S_1^W)^2(S_1^V)^2
-6(S_1^W)^2 S_2^V
-6(S_1^V)^2 S_2^W
+18\,S_2^W S_2^V
\Big]
+O(\kappa_2^3).\nonumber\\
\label{Vhol_smallk_simplified}
\end{eqnarray}
Higher orders in \(\kappa_2\) can be generated equivalently by the zonal
\((\)Jack, \(\alpha=2)\) expansion of \({\cal I}_d\) specialized to \(d=3\),
or by the systematic expansion of the quadruple--Bessel representation followed
by the explicit \(c\)-integrals. In both routes, the result can be organized
entirely in terms of the spectral invariants of \(\Lambda_V\) and
\(\Lambda_W\).

\medskip
\noindent
However, the reduced potential \eqref{Veff_4d_reduced} is not a reliable saddle
potential in the weak-\(\kappa_2\) truncation. To see the problem, introduce
\begin{eqnarray}
v_\pm=\frac{v_1\pm v_2}{2},
\qquad
w_\pm=\frac{w_1\pm w_2}{2}.
\end{eqnarray}
The classical plus Vandermonde part becomes
\begin{eqnarray}
V_{\rm cl}
=
2m v_+ - \log|2v_-|
+
2m w_+ - \log|2w_-|.
\label{Vcl_vpm_wpm}
\end{eqnarray}
The logarithmic terms decrease without bound as \(|v_-|\) or \(|w_-|\) grow,
so this part of the potential is non-confining in the eigenvalue splitting
directions.

\medskip
\noindent
At quadratic order in \(\kappa_2\), the reduced holonomy correction has the same
qualitative problem: its contribution to the splitting directions is also
negative. Schematically,
\begin{eqnarray}
\frac{\partial V_{\rm eff}}{\partial v_-}
=
-\frac{1}{v_-}
-\kappa_2^2\,(\hbox{positive})\,v_-
+\cdots ,
\label{dVdvminus_failure}
\end{eqnarray}
and therefore no real finite solution for \(v_-\) is obtained in this truncated
four-dimensional potential. The same statement holds in the \(w_-\) direction.
Thus the reduced potential has no finite interior saddle in the eigenvalue
splitting directions.

\medskip
\noindent
This does not mean that the BFSS\(_4\) saddle is absent. Rather, it shows that
the four-dimensional reduced potential is dynamically incomplete. Once the
angular variable \(c\) is integrated out perturbatively, the angular
backreaction that could control the eigenvalue splittings is no longer visible.
Consequently, \eqref{Veff_4d_reduced} by itself is not a reliable saddle
potential. Its runaway behavior in the eigenvalue splitting directions reflects
the premature integration over \(c\) and the weak-\(\kappa_2\) truncation, rather
than a genuine statement about the BFSS\(_4\) saddle.

\medskip
\noindent
A consistent saddle analysis must therefore keep the full \(c\)-resolved angular
partition function, so that the radial eigenvalues and the residual angular
coordinate remain coupled. This is the formulation adopted in the main text.

\section{The relation between principal angles and auxiliary eigenvalues}\label{appendix2}

\medskip
\noindent
\emph{Principal angles and the singular values of the $2\times 2$ block.}
Let $\Pi_U,\Pi_Z\subset\mathbb R^d$ be two oriented $2$--planes.
Choose orthonormal frames
\begin{eqnarray}
U=(u_1,u_2)\in \mathbb R^{d\times 2},\qquad
Z=(z_1,z_2)\in \mathbb R^{d\times 2},
\qquad
U^TU=Z^TZ=\mathbf 1_2.
\end{eqnarray}
The \emph{principal angles} $(\theta_1,\theta_2)$ between $\Pi_U$ and $\Pi_Z$
are defined by the singular values of the $2\times 2$ overlap matrix
\begin{eqnarray}
S:=U^T Z\in\mathbb R^{2\times 2},
\qquad
\mathrm{singval}(S)=\{\cos\theta_1,\cos\theta_2\},\qquad 0\le \theta_1\le \theta_2\le \frac{\pi}{2}.
\label{principal_angles_def}
\end{eqnarray}
Equivalently, the eigenvalues of $SS^T$ are $\cos^2\theta_i$.

\medskip
\noindent
In the rank--$2$ HCIZ problem one works in a basis where the supports of
$\Lambda_V$ and $\Lambda_W$ are on the coordinate plane
\begin{eqnarray}
\Pi=\mathrm{span}\{e_1,e_2\}.
\end{eqnarray}
The orthogonal matrix $O$ rotates this plane as
\begin{eqnarray}
\Pi'=O\Pi.
\end{eqnarray}
Define then $\Pi=\Pi_U$ and $\Pi'=\Pi_Z$, i.e. 
\begin{eqnarray}
U=(u_1,u_2)=(e_1,e_2),\qquad Z=(z_1,z_2)=(Oe_1,Oe_2)\qquad\Rightarrow\qquad Z=OU.
\end{eqnarray}
The overlap matrix in this case is
\begin{eqnarray}
S=U^TZ=U^TOU \qquad \Leftrightarrow \qquad S_{ia} = e_i^T (O e_a) = O_{ia},\qquad i,a=1,2.
\end{eqnarray}
This means that the relevant overlap is precisely the $2\times 2$ principal block
\begin{eqnarray}
S\equiv X:=\big(O_{ia}\big)_{i,a=1,2}.
\end{eqnarray}
Then the singular values of $X$ are exactly the cosines of the principal angles
between the two $2$--planes picked out by the source supports:
\begin{eqnarray}
\mathrm{singval}(X)=\{s_1,s_2\}=\{\cos\theta_1,\cos\theta_2\}.
\label{singvals_equal_cos}
\end{eqnarray}
This is the \emph{precise} sense in which $(s_1,s_2)$ and $(\theta_1,\theta_2)$
carry the same geometric degrees of freedom.

\bigskip
\noindent
\emph{What do the auxiliary eigenvalues $\lambda_i$ encode?}
The Schur--Ingham--Siegel (SIS) step introduces an auxiliary symmetric matrix $M$
to represent the determinant weight in the pushforward measure.
Schematically (rank--$2$ case),
\begin{eqnarray}
\det(\mathbf 1_2-XX^T)^{p}
\ \propto\
\int_{\mathrm{Sym}_2^+} dM\;
(\det M)^{-(p+\frac{3}{2})}\;
\exp\!\Big(-\mathrm{Tr}M(\mathbf 1_2-XX^T)\Big),
\quad p=\frac{d-5}{2}.\nonumber\\
\label{IS_identity_rank2}
\end{eqnarray}
The crucial structural fact is:

\medskip
\begin{center}
\emph{$M$ is conjugate to $XX^T$, hence its eigenvectors align with those of $XX^T$ at saddle level,
and its eigenvalues are functions of the singular values $s_i^2$ (hence of $\cos^2\theta_i$).}
\end{center}

\medskip
\noindent
Indeed, the exponent in \eqref{IS_identity_rank2} is
\begin{eqnarray}
-\mathrm{Tr}M(\mathbf 1_2-XX^T),
\end{eqnarray}
so the steepest--descent condition in $M$ forces it to probe the inverse of
$\mathbf 1_2-XX^T$. Concretely, at the level of eigenvalues one expects
the conjugate variables to scale as
\begin{eqnarray}
\mu_i \sim \frac{1}{1-s_i^2}=\frac{1}{1-\cos^2\theta_i}=\frac{1}{\sin^2\theta_i},
\label{mu_vs_angles}
\end{eqnarray}
up to $\kappa_2$--dependent deformations once the source term
$\exp(\kappa_2\mathrm{Tr}(W X V X^T))$ is included.
After the oscillatory rotation used in our SIS normalization,
one works with eigenvalues $\lambda_i$ of $iM$, and thus:
\begin{center}
\emph{$\lambda_i$ are not angles, but they are conjugate to $s_i^2$ and therefore encode the same geometric invariants as $\theta_i$.}
\end{center}

\medskip
\noindent
For $d=3$, one singular value is frozen: $s_1=1$ (equivalently $\theta_1=0$).
Then \eqref{mu_vs_angles} predicts
\begin{eqnarray}
1-s_1^2=0
\qquad\Rightarrow\qquad
\mu_1\to +\infty.
\end{eqnarray}
In other words, one auxiliary eigenvalue is driven to a singular limit.
Only the second invariant survives, corresponding to the single
nontrivial principal angle.

\medskip
\noindent
This is the consistent reconciliation:
\begin{itemize}
\item For $d\ge 4$, $(\lambda_1,\lambda_2)$ encode two independent invariants
      $(s_1^2,s_2^2)=(\cos^2\theta_1,\cos^2\theta_2)$.
\item At $d=3$, the geometry enforces $s_1=1$ (one principal angle $0$),
      so one conjugate eigenvalue runs to a singular limit; only one nontrivial
      invariant remains.
\item The Euler description keeps this manifestly real: the single principal variable
      is $c=\cos\beta=s_2$.
\end{itemize}

\medskip
\noindent In this precise sense SIS does not smoothly extend to $d=3$:
the pushforward measure collapses onto a boundary stratum,
and the auxiliary eigenvalue description develops a matching singular direction.

\end{document}